\documentclass[aps,prb,amsmath,amssymb,amsfonts,floatfix,nofootinbib,superscriptaddress,longbibliography,12pt,reprint,noeprint]{revtex4-2}
\usepackage[utf8]{inputenc}
\usepackage{geometry}
\usepackage[english]{babel}
\usepackage{blindtext}
\usepackage{amsmath,amsfonts,amssymb}

\usepackage{graphicx}
\usepackage{float}
\usepackage{times}
\usepackage{color}
\usepackage{subfigure}
\usepackage{indentfirst}
\usepackage{setspace}
\usepackage{bm}

\usepackage[colorlinks,linkcolor=blue,anchorcolor=blue,citecolor=green]{hyperref}
\usepackage{CJKutf8}  

\usepackage{hyperref}

\allowdisplaybreaks
\newcounter{fan}

\newcounter{amg}

\begin{document}

    \title{Entanglement transitions and multifractality in monitored free-fermions with random long-range hopping}
	\author{Bo Fan (\begin{CJK*}{UTF8}{gbsn}范波\end{CJK*})}
	\email{bo.fan@kit.edu}
    \affiliation{\mbox{Institute for Quantum Materials and Technologies, Karlsruhe Institute of Technology, 76131 Karlsruhe, Germany}}
    \affiliation{\mbox{Institut f\"ur Theorie der Kondensierten Materie, Karlsruhe Institute of Technology, 76131 Karlsruhe, Germany}}
	\affiliation{Shanghai Center for Complex Physics, School of Physics and Astronomy, 
		\\Shanghai Jiao Tong University, Shanghai 200240, China}
	\author{Antonio M. Garc\'ia-Garc\'ia}
	\email{amgg@sjtu.edu.cn}
	\affiliation{Shanghai Center for Complex Physics, School of Physics and Astronomy, 
		\\Shanghai Jiao Tong University, Shanghai 200240, China}
		
	\vspace{0.3cm}
	\date{\today}
	\vspace{0.3cm}
	
	\begin{abstract}
		We study the entanglement dynamics of a one-dimensional chain of monitored non-interacting complex fermions with random power-law hopping characterized by a decay exponent $\alpha$. 
		For $\alpha \lesssim 1$, in stark contrast with the case of hopping to nearest neighbors, the scaling of the entanglement entropy (EE) of the steady state with system size $L$ is faster than logarithmic for any monitoring or disorder strength and it tends towards a linear (volume-law) scaling for sufficiently small $\alpha \lesssim 1/2$.
		For $\alpha > 3/2$, the EE is in the area-law phase, namely, no scaling with $L$, for any monitoring strength. For $1 < \alpha \lesssim 3/2$, we identify an $\alpha$-dependent measurement-induced phase transition (MIPT) at a critical value of the monitoring strength separating the mentioned area-law and sub-volume-law phases. At this critical point, the EE scales logarithmically with system size, and the density-density correlation function, closely related to the EE, exhibits multifractal features. These results highlight the importance of superdiffusive classical hopping in the entanglement dynamic of quantum many-body systems and also help differentiate its role with respect to conventional sources of entanglement such as genuine quantum non-locality.  
	\end{abstract}
	
	\maketitle

	\section{Introduction}
    Even for non-interacting quantum systems, the dynamics of disordered systems is highly sensitive to dimensionality. For any amount of disorder, in one and two dimensions, quantum coherence effects induce destructive interference and ultimately Anderson localization \cite{anderson58,abrahams1979} that completely arrests diffusion. 
    In higher dimensions, quantum coherence effects are weaker, so the dynamics remains diffusive for weak disorder, but an metal-insulator transition \cite{abrahams1979,evers08,garcia2008a}, usually referred to as Anderson transition, occurs at a dimension-dependent critical disorder strength, characterized by multifractal eigenstates \cite{castellani1986,schreiber1991,rodriguez2011multifractal,garcia2007dimensional,wegner1979}, intermediate level statistics \cite{shapiro1993,bogomolny1996,kravtsov1994,garcia2003} and non mean-field critical exponents \cite{garcia2008a}. 
    However, this result is only valid if hopping is sufficiently short-range. Random long-range power-law hopping \cite{mirlin1996,mirlin00,evers08} characterized by a decay exponent $\alpha$, strongly suppresses localization effects. Metallic features such as superdiffusive dynamics and random-matrix spectral correlations occur even in one spatial dimension provided that $\alpha < 1$.
    For $1 < \alpha < 3/2$, eigenstates are power-law localized and spectral correlations are given by Poisson statistics. For $\alpha = 1$, the system share all the features of a disordered system with short-range hopping in higher dimensions at the mentioned Anderson transition. For $\alpha > 3/2$, features are similar to those of a system with short-range hopping which in one dimension behaves as an insulator. Note that the phase diagram for long-range hopping depends only on the decay exponent $\alpha$ and not on the strength of disorder.

    In recent years, there have been a growing interest in the entanglement dynamics of monitored quantum many-body systems \cite{cao2019a,turkeshi2021,schiro2024,buchhold2021a,fava2023,fava2024,poboiko2023,chahine2023entanglement,poboiko2023a,poboiko2025,altland2022,lunt2020a,lunt2022,alberton2021a,Jian2020a,Jian2023,Fisher2022,poboiko2024}, due to its relevance in both quantum information, especially quantum computation, and the very foundations of quantum mechanics. Of particular importance in this research area is the identification and characterization of measurement-induced phase transitions (MIPTs) \cite{Li2018a,Skinner2019a,nahum2021a,ippoliti2021a,Li2019a,carisch2023,Szyniszewski2019a,Szyniszewski2020,Turkeshi2020a,turkeshi2021,Turkeshi2022a,legal2023,Soares2025,poboiko2025}.
    Due to technical challenges, an important part of the research has focused on the entanglement dynamics of monitored one-dimensional (1d) non-interacting fermions, either Majorana or complex fermions. The use of the non-linear sigma model (NLSM) \cite{wegner1979,efetov1983supersymmetry} in this context \cite{poboiko2023,fava2024}, together with large scale numerical simulations \cite{garcia2025,garcia2026} based on GPU acceleration, have been instrumental to demonstrate that \cite{poboiko2023,garcia2025,garcia2026} both clean and disordered non-interacting 1d complex fermions are in the area-law phase for any finite monitoring strength, namely, there is no MIPT in the system. By contrast, 1d non-interacting Majorana fermions exhibit a MIPT \cite{fava2023,Jian2023} in the clean limit. A MIPT also occurs in other situations such as non-interacting complex fermions in higher spatial dimensions \cite{poboiko2023a,poboiko2025,chahine2023entanglement} or if interactions are turned on \cite{poboiko2024}.  
    
    \begin{figure}[!htbp]
    	\begin{center}
    		\includegraphics[width=8.0cm]{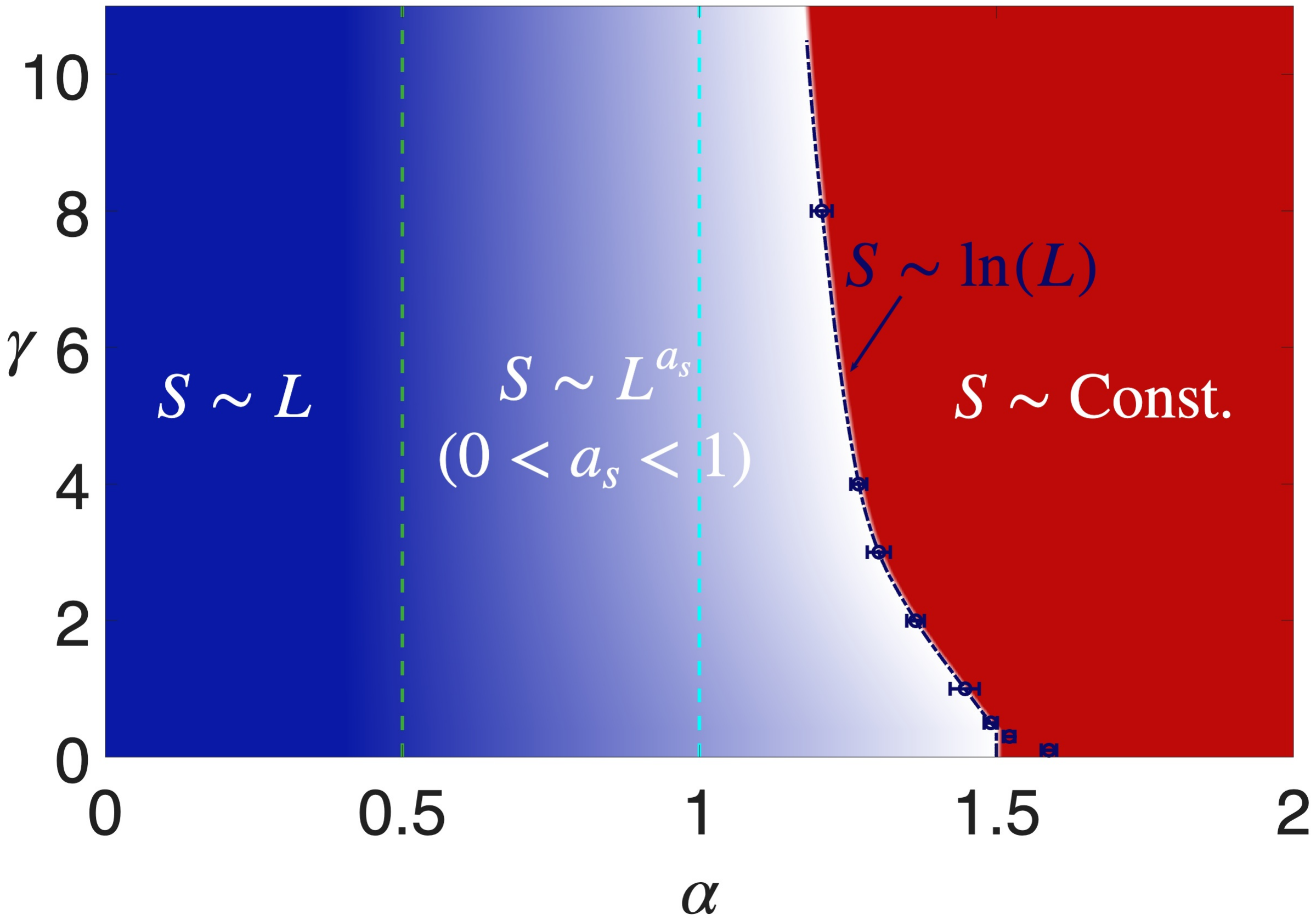}
    		\caption{
    			Phase diagram showing the scaling of the steady-state entanglement entropy $S$ with system size $L$, as a function of the power-law hopping exponent $\alpha$ and the monitoring strength $\gamma$.
    			For $\alpha \lesssim 1$, the EE exhibits sub-volume-law scaling, $S \propto L^{a_s}$ with $0 < a_s <1 $ for any $\gamma$, and tends towards $a_s \to 1$, volume law, for $\alpha \lesssim 1/2$.  For $\alpha > 3/2$, the system is in the area-law phase, $a_s = 0$ for any finite monitoring strength $\gamma > 0$, while in the region $1 < \alpha \lesssim 3/2$, it experiences a line of MIPT (navy dash-dotted line) between sub-volume-law and area-law phases at $\alpha_c(\gamma)$. At this critical point, the EE scales logarithmically, $S \propto \ln(L)$, and certain correlation functions become multifractal, see Fig.~\ref{Fig:f_alpha_q}. 
    			Numerically, $\alpha_c(\gamma)$ (circles with error bars) along the critical line is determined via a finite-size analysis of the mutual information, see Fig.~\ref{Fig:I2_vs_alpha_QSD} and Fig.~\ref{Fig:I2_vs_alpha_QSD_app} for details. 
    		}\label{Fig:phase_diagram}
    	\end{center}
    	\vspace{-0.5cm}
    \end{figure}  
    Given this rich phenomenology, a natural question to ask is how the entanglement dynamics changes when nearest-neighbor hopping is replaced by long-range hopping. Previous studies have addressed this question in several complementary settings. The free fermions with long-range non-unitary dynamic \cite{Zhang2022universal} can give rise to (sub-)volume-law entanglement for $\alpha <3/2$, and logarithmic scaling for $\alpha >3/2$, which is qualitatively the short-range limit. 
    The interplay between monitoring and long-range hopping in the entanglement dynamics has already been studied for 1d complex fermions, but in the limit of non-random hopping \cite{minato2022,mueller2022} where it was found that for $\alpha \lesssim 3/2$ the EE grows with the system size independently on the monitoring strength. More generally, recent works have demonstrated that the long-range coupling enters through either the unitary dynamics \cite{passarelli2026range} or in the measurement operators \cite{russomanno2023entanglement,piccitto2025entanglement}, leading to the intermediate sub-volume-law phase entanglement scaling behavior, which is absent in the monitored free fermion system with only short-range hopping.
    These results demonstrate that long-range couplings can fundamentally modify the competition between unitary dynamics and measurements.
    Here, we shall unveil a more intricate dynamical behavior by considering random long-range hopping plus a diagonal random potential. More specifically, we study a 1d chain of non-interacting spinless complex fermions with random power-law hopping subjected to continuous monitoring of the occupation number via the quantum state diffusion (QSD) protocol. This protocol preserves the Gaussian nature of fermionic states and enables efficient numerical simulation.

    We study the entanglement dynamics by a numerical calculation of the entanglement entropy (EE), mutual information (MI), and density-density correlation functions in the dynamical steady state, after saturation of the EE, as a function of $\alpha$ and the measurement strength $\gamma$. For $\alpha \lesssim 1$, we find that the EE grows faster than logarithmically with system size for any monitoring strength and approaches a linear growth for $\alpha \lesssim 1/2$. In contrast, for $\alpha > 3/2$, the EE follows an area-law, namely, no scaling with system size, for any $\gamma$. We note that accurately identifying the critical transition in the region of small $\gamma \lesssim 0.3$ and $\alpha \gtrsim 3/2$ is numerically challenging because large system sizes are required. However, we provide theoretical arguments supporting that for $\alpha > 3/2$ the system is always in the area-law phase.
    For $1 < \alpha \lesssim 3/2$, a MIPT between sub-volume-law and area-law phases occur at a certain $\gamma = \gamma_c(\alpha)$ that depends on $\alpha$ and the monitoring strength $\gamma$. Around this critical point, the growth of EE is logarithmic with the system size, and the density-density correlation function is multifractal. These findings, summarized in Fig.~\ref{Fig:phase_diagram}, illustrate the profound impact of random long-range hopping on the entanglement properties of monitored non-interacting quantum many-body systems.  
    
    We now introduce the model, outline the quantum trajectory method used to simulate the continuously monitored dynamics and define the employed observables.  
       
	\section{Model, Measurement Protocol and Observables}
	
    We consider a one-dimensional chain of $L$ sites at half filling, containing $N = L/2$ complex fermions, with random power-law hopping, described by the Hamiltonian \cite{mirlin00,MirlinEvers00,evers08,vega2019multifractality}:	
	\begin{equation}
		H = \sum_{i,j}^{L} t_{ij} \left[ \hat{c}_{i}^\dagger \hat{c}_j + \hat{c}_j^\dagger \hat{c}_{i} \right], \label{eq:H}
	\end{equation}
	where $\hat{c}_j$ and $\hat{c}_j^\dagger$ are the annihilation and creation operators for fermions at site $ j $. The long-range hopping between sites $i$ and $j$ are given by $t_{ij} = J_{ij} \left[1+\left(\frac{\sin(\pi (i-j)/L)}{\pi/L}\right)^{2\alpha}\right]^{-0.5}$, where $J_{ij}$ are independent random variables drawn from a normal distribution of zero mean and unit variance. Periodic boundary conditions are imposed via the sine function in $t_{ij}$ to minimize boundary effects. At large separation, the variance of the hopping decays as $\overline{t_{ij}^2}\sim |i-j|^{-2\alpha}$, and the exponent $\alpha$ therefore controls the range of the random hopping. This model is the one-dimensional power-law random banded matrix (PRBM) problem, whose single-particle states are known to display extended, multifractal, and localized regimes depending on $\alpha$ \cite{mirlin00,MirlinEvers00,evers08,vega2019multifractality}.
    
    The monitored dynamics is modeled by the quantum trajectory formalism \cite{molmer93,zoller1987}. Since the Hamiltonian is quadratic in the fermion operators, we choose to weakly monitor the local occupation number at each site, so the resulting evolution preserves the Gaussian nature of the state, enabling efficient numerical simulations. Specifically, we employ the QSD protocol, which describes the continuous stochastic evolution through the Stochastic Schr\"odinger equation \cite{cao2019a,alberton2021a,szyniszewski2023}:
	\begin{equation}
		\begin{aligned}
			d|\psi(t)\rangle = &-i H |\psi(t)\rangle dt - \frac{\gamma dt}{2} \sum_j (\hat{n}_j - \langle \hat{n}_j \rangle)^2 |\psi(t)\rangle\\
			& + \sum_j (\hat{n}_j - \langle \hat{n}_j \rangle) dW_j^t |\psi(t)\rangle,
		\end{aligned}
		\label{eq.SSE}
	\end{equation}
	where $\hat{n}_j = \hat{c}_{j}^\dagger \hat{c}_j$, $\gamma$ is the measurement strength, and the noise increments are defined as $dW_j^t = \sqrt{\gamma} dB_j(t)$, with $dB_j(t)$ the standard independent Wiener processes. It is then follows that the zero mean $\overline{dW_j^t} = 0$, and variance $\overline{dW_i^tdW_j^t} = \gamma dt \delta_{ij}$.
	The QSD equation describes the time evolution of the quantum state under continuous monitoring of local the occupation number, which can be approximated by trotterization \cite{cao2019a}:
	\begin{equation}
		\begin{aligned}
			| \psi_{t + dt} \rangle &\approx C e^{\sum_j [dW_j^t + (2 \langle \hat{n}_j \rangle_{t} - 1) \gamma dt] \hat{n}_j} e^{-iH dt} | \psi_t \rangle \\
			&= C | V_{t + dt} \rangle,
		\end{aligned}
		\label{eq.trotterization}
	\end{equation}
	where $C$ normalizes the state, and $ | V_{t + dt} \rangle $ is orthonormalized using QR decomposition to preserve the purity of the Gaussian state. This scheme not only captures the interplay between unitary evolution and weak stochastic measurements, but also allows for a relatively large time step $dt$ without introducing significant numerical errors. The dynamical evolution is initialized with the N\'{e}el state: $\psi(t=0) \rangle = \prod_{i=1}^{L/2} \hat{c}_{2i-1}^\dagger | \text{vac} \rangle $. To ensure the accuracy of the numerical simulation, we use different time steps $1 \times 10^{-3} \le dt \le 0.05$ depending on $\gamma$. 
	
    The central object in the analysis is the equal-time correlation matrix
    \begin{equation}
        D_{ij}(t)= \langle \hat c_i^\dagger \hat c_j\rangle_t  \equiv  \langle\psi(t)|\hat c_i^\dagger \hat c_j|\psi(t)\rangle .
        \label{eq:di}
    \end{equation}
    For a subsystem $A$ of length $\ell_A$, let $\{\lambda_j(t)\}$ be the eigenvalues of $D_A(t)$, the restriction of $D(t)$ to $A$. The EE of subsystem $A$ is then obtained from
    \begin{equation} \label{eq:sl}
        S(\ell_A,t)= -\sum_{j=1}^{\ell_A} \left[ \lambda_j(t)\ln\lambda_j(t) + (1-\lambda_j(t))\ln(1-\lambda_j(t)) \right].
    \end{equation}
    
	In order to suppress statistical fluctuations, all observables are averaged over both disorder realizations and quantum trajectories. The number of disorder realizations ranges from $64$ to $1280$, depending on $\alpha$ and the system size $L$. Each disorder realization is further averaged over $3 \sim 24$ quantum trajectories depending on the measurement strength $\gamma$. The dynamical evolution is continued until the ensemble-averaged EE $S(\ell_A,t)$ saturates, which means $S(\ell_A,t) \approx S(\ell_A,t_f)$ for all $t \ge t_f$, signaling that the system has reached its dynamical steady state. Since the saturation time $t_f$ depends strongly on both $\alpha$ and $\gamma$, we first estimate it using smaller systems, typically $L=128$ with $\ell_A = L/2$, for each parameter set.

	To locate the MIPT more accurately, we additionally compute the MI between two distant regions. The system is divided into four equal consecutive subsystems $a, b, c$ and $d$, each with $L/4$ sites. The MI between subsystem $a$ and $c$ is then defined as, 
	\begin{equation}\label{eq:mi}
		\mathcal{I}_2 = S_a + S_c - S_{a\cup c} 
	\end{equation}
	where $S_a$, $S_c$ and $S_{a\cup c}$ denote the EE Eq.~\eqref{eq:sl} of the subsystem $a$, $c$ and $a\cup c$, respectively. The mutual information is particularly useful, because it decays to zero in the area-law phase, grows in the sub-volume-law phase, and becomes approximately scale invariant at the critical point of the MIPT.
    
	Finally, we also study the density-density correlation function $C(r)$, which is determined directly from the correlation matrix via 
	\begin{equation} \label{eq:cr}  
		C(r) = |D_{i+r,i}|^2, \qquad r > 0,
	\end{equation} 
	with $D_{i+r,i}$ given by Eq.~\eqref{eq:di}. This quantity is directly related to particle number fluctuations. For a Gaussian state, the second cumulant of the particle number $N_{\ell_A}=\sum_{i\in {\ell_A}}\hat n_i$ is
    \begin{equation}
        C_{\ell_A}^{(2)}=\langle N_{\ell_A}^2\rangle-\langle N_{\ell_A}\rangle^2  =\sum_{i\in {\ell_A},j\notin \ell_A}|D_{ij}|^2. \label{eq:na2}
    \end{equation}
    
    The entanglement entropy can be expanded in terms of the even cumulants of particle-number fluctuations \cite{levitov2009,Burmistrov2017,poboiko2023},
    \begin{equation} \label{eq.Klich_Levitov} 
		S(\ell_A) = {\pi^2\over 3}C_{\ell_A}^{(2)} + {\pi^4\over 15}C_{\ell_A}^{(4)} + {2\pi^6\over 945}C_{\ell_A}^{(6)} + \cdots  .
	\end{equation}
    In the monitored free fermion system, the leading scaling of $S(\ell_A)$ is well captured by ${\pi^2\over 3}C_{\ell_A}^{(2)}$, which has been extensively validated in previous works \cite{Burmistrov2017,poboiko2023,poboiko2023a,garcia2025,garcia2026}. We further confirm it in Appendix~\ref{app:mut} that this approximation remains accurate across a broad range of $\alpha$ and $\gamma$ in our system.

	\section{Results and discussion}
    We now characterize the steady-state entanglement phases of the system using three complementary observables: the MI $\mathcal{I}_2$, the bipartite EE scaling $S(\ell_A = L/2)$, and the density-density correlation function $C(r)$. In addition, we analyze the singularity spectrum $f(\alpha_q)$ of $C(r)$ to characterize its multifractal properties at the MIPT. For ease of notation, we do not state explicitly the time dependence of these observables. All quantities are evaluated after the saturation of the EE, so that further time evolution is negligible.  
    
    \vspace{-0.3cm}
    \subsection{Identification of the MIPT}
    
\begin{figure}[!htbp]
    	\begin{center}
    		\includegraphics[width=4.0cm]{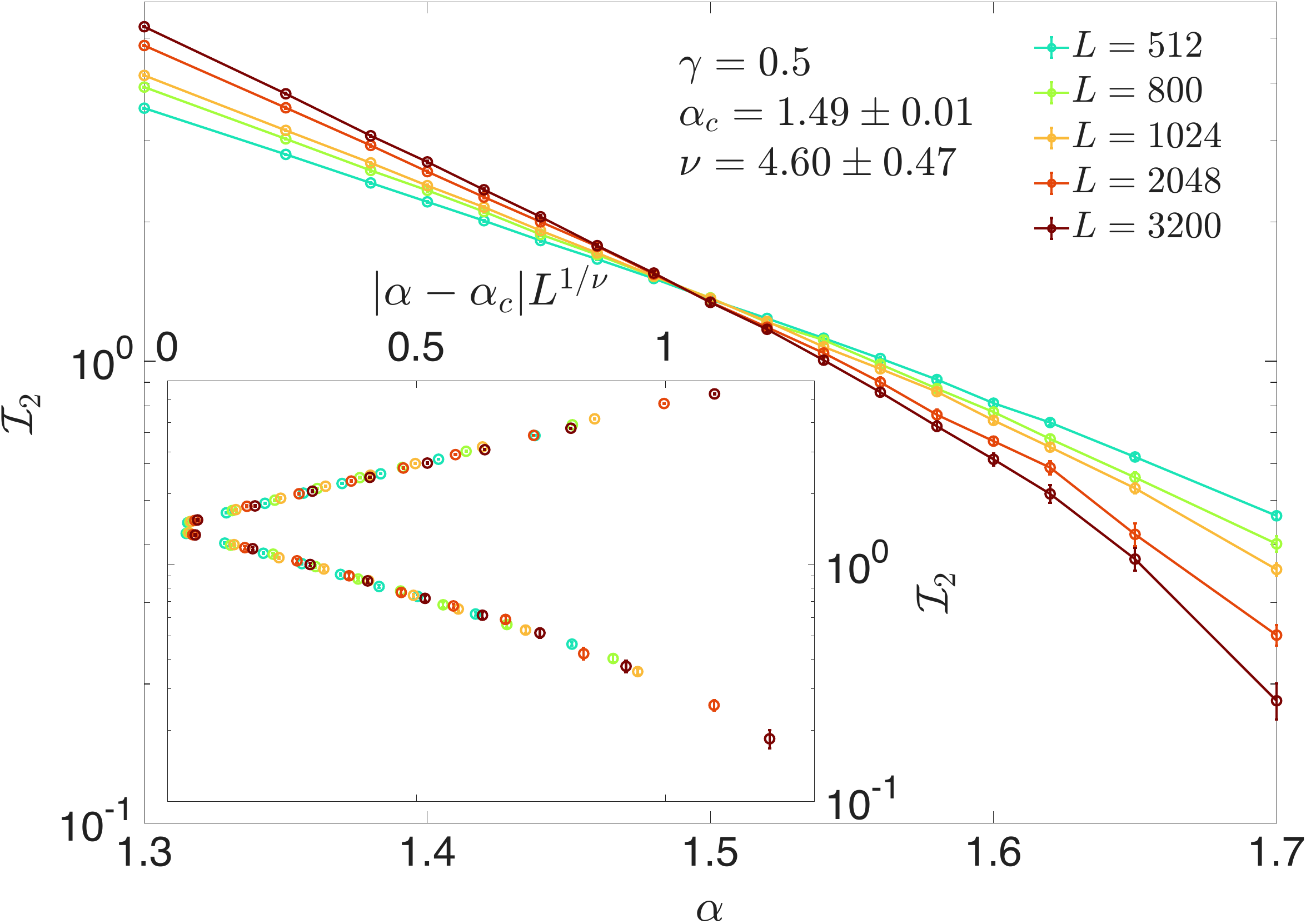} 
    		\includegraphics[width=4.0cm]{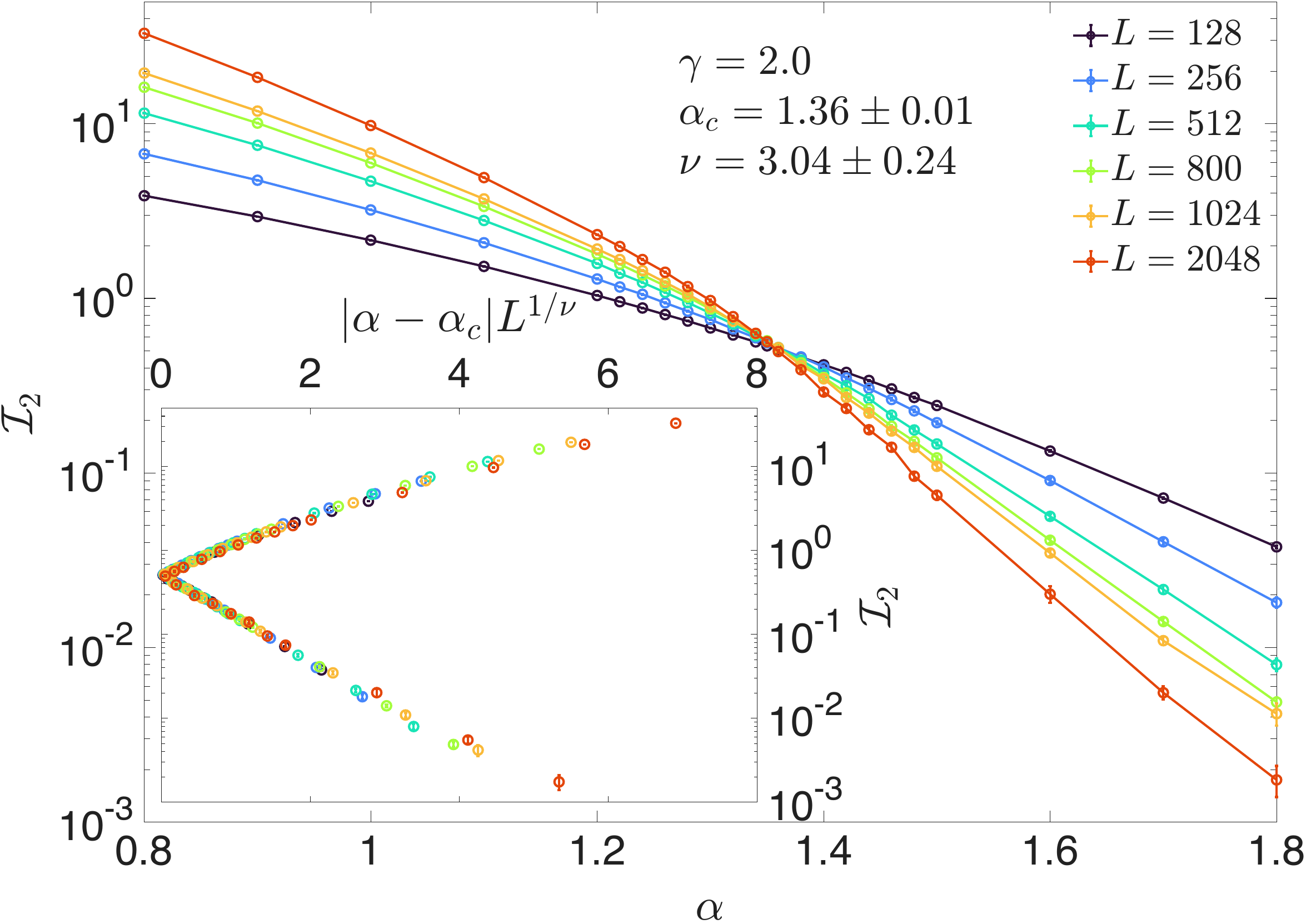} 
    		\caption{The mutual information $\mathcal{I}_2$ Eq.~\eqref{eq:mi} as a function of $\alpha$ for $\gamma = 0.5$ (left) and $\gamma = 2.0$ (right). The sharp crossing at $\alpha = \alpha_c$ indicates a MIPT, where $\mathcal{I}_2$ becomes scale-invariant. The insets show the optimal finite-size data collapse around $\alpha_c$. The estimated critical points are $\alpha_c = 1.49 \pm 0.01$ and $\alpha_c = 1.36 \pm 0.01$ for $\gamma =0.5$ and $2.0$, respectively. The critical exponent $\nu$ exhibits larger fluctuations, due to the absence of a well-defined universal critical exponent in the model we are considering. Results for additional values of $\gamma$ are presented in Appendix~\ref{app:mut},  Fig.~\ref{Fig:I2_vs_alpha_QSD_app}.
    		}\label{Fig:I2_vs_alpha_QSD}
    	\end{center}
    	\vspace{-0.5cm}
    \end{figure}

    Out first task is to identify the critical monitoring $\alpha_c$ at which the MIPT occurs for a given monitoring strength $\gamma$. We employ the MI $\mathcal{I}_2$ between two disjoint subsystems as a function of $\alpha$ for various system sizes. As shown in Fig.~\ref{Fig:I2_vs_alpha_QSD}, a sharp crossings of $\mathcal{I}_2$ for different system sizes around $\alpha_c$ signals the MIPT. At criticality, $\mathcal{I}_2$ becomes scale-invariant \cite{poboiko2023a,poboiko2025}, while away from $\alpha_c$, it increases with system size in the sub-volume-law phase and decreases in the area-law phase, allowing for a clear distinction between the two regimes. The scaling collapse of the data near $\alpha_c$ confirms the critical scaling behavior, typical of phase transitions. The obtained critical values are $\alpha_c = 1.49 \pm 0.01$ for $\gamma = 0.5$ and $\alpha_c = 1.36 \pm 0.01$ for $\gamma = 2.0$. Therefore, stronger monitoring shifts the transition to smaller $\alpha$, as expected from the enhanced measurement-induced pinning of local occupations.

    While the critical point $\alpha_c$ is determined with relatively high accuracy, the critical exponent $\nu$ exhibits larger fluctuations depending on the fitting window, likely owing to the absence of a well-defined universal critical exponent in this model. To illustrate this sensitivity, we present in Appendix~\ref{app:scaling_analysis}, Fig.~\ref{Fig:cost_function}, the heatmap of the cost function used in the data collapse procedure, highlighting the range of $\nu$ and $\alpha_c$ compatible with the data. Therefore, $\nu$ is just a parameter introduced to determine with more accuracy the location of the transition. The same procedure is employed for the rest of points depicted in the phase diagram of Fig.~\ref{Fig:phase_diagram}. We also note that finite-size effects are more pronounced at weak measurement rate ($\gamma \leq 0.5$), so for the finite-size scaling analysis in this regime we restrict to system sizes $L \geq 512$. By contrast, for stronger monitoring ($\gamma \geq 1.0$), where finite-size effects are weaker, we consider sizes $L \geq 128$.

\begin{figure}[!htbp]
    	\begin{center}
    		\includegraphics[width=4.0cm]{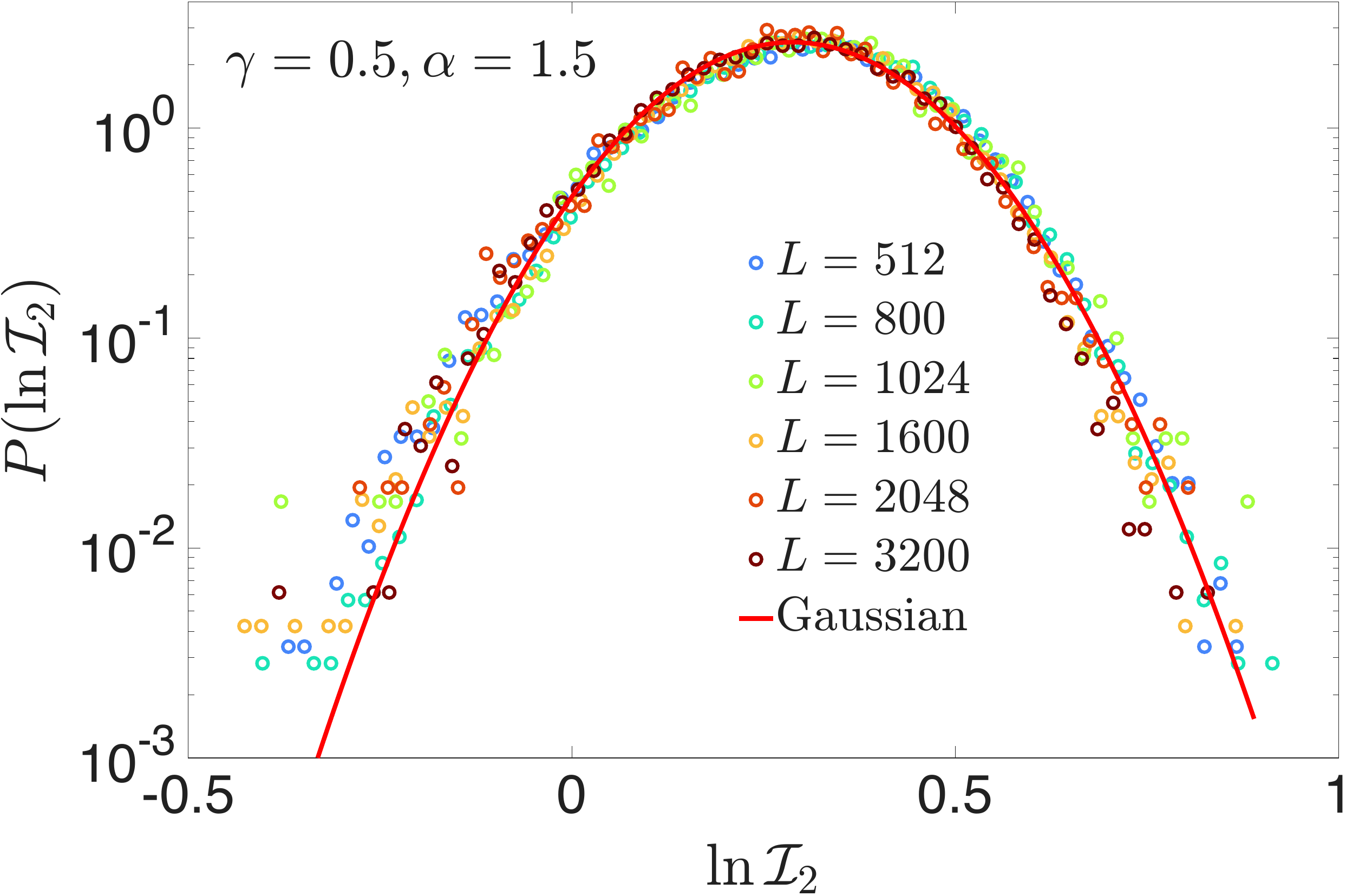} 
    		\includegraphics[width=4.0cm]{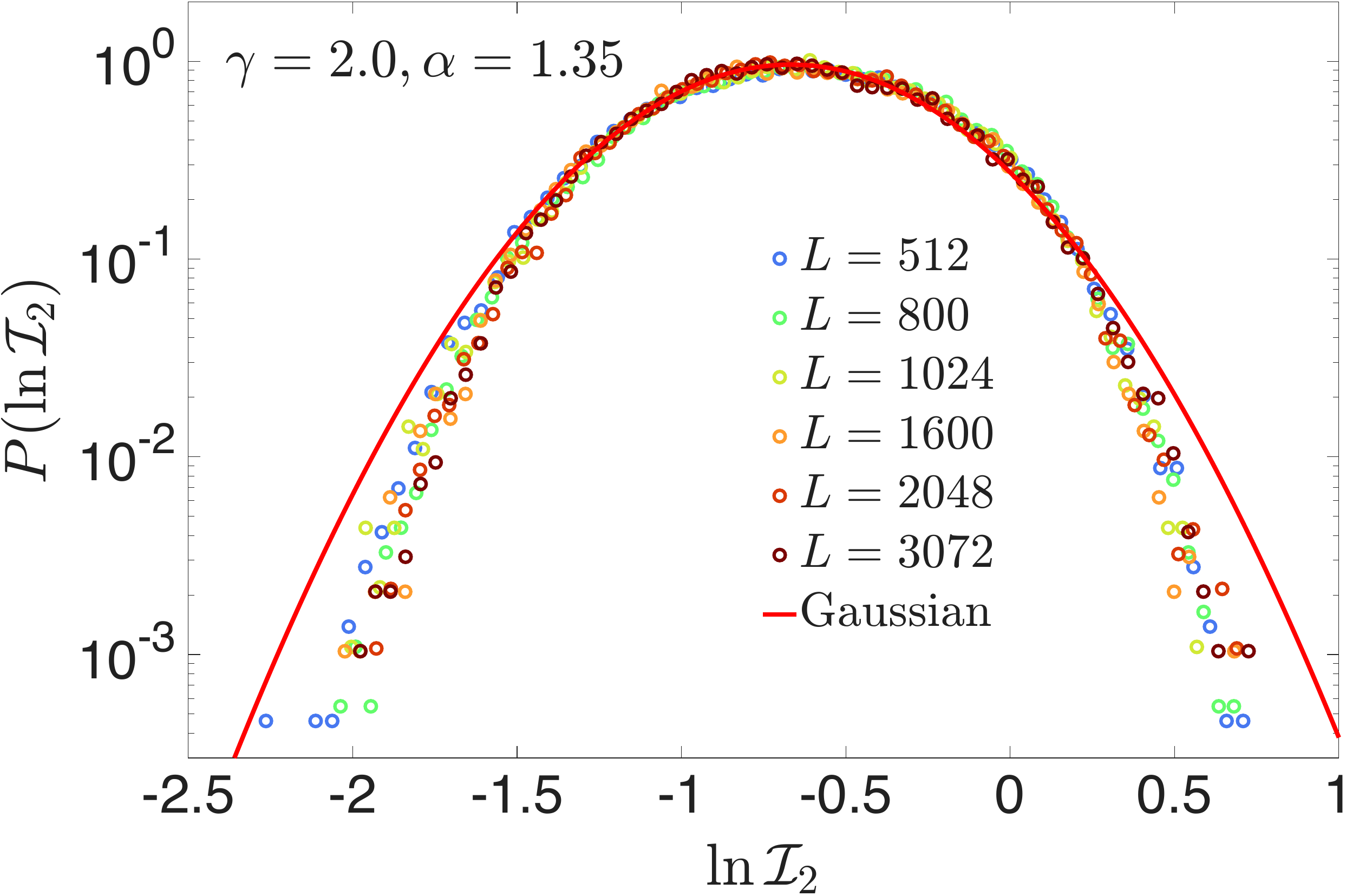} 
    		\caption{The probability distribution of the mutual information $\ln \mathcal{I}_2$ Eq.~\eqref{eq:mi} for different system size $L$ around the MIPT. Left: $\gamma =0.5, \alpha_c = 1.5$. Right: $\alpha_c = 1.35, \gamma = 2.0$. The distribution $P(\ln \mathcal{I}_2)$ is size independent, consistent with the scale invariance of $\mathcal{I}_2$ at the transition. The tails of $P(\ln \mathcal{I}_2)$ deviate from a Gaussian form, especially in the strong measurement limit $\gamma  = 2$. 
    		}\label{Fig:pro_Iac_critical}
    	\end{center}
    	\vspace{-0.8cm}
    \end{figure}

    As an additional evidence for the MIPT, we show in Fig.~\ref{Fig:pro_Iac_critical} that the probability distribution of $\ln \mathcal{I}_2$ becomes, as predicted, scale-invariant at the MIPT $\alpha \approx \alpha_c(\gamma)$. More results for other phases are shown in the Appendix~\ref{app:pro_Iac}. By contrast, in the (sub-)volume-law phase regime, the probability distribution of $\mathcal{I}_2$ agrees well with a Gaussian distribution, with a mean value that increases with system size. In the area-law phase regime, the peak of $P(\ln \mathcal{I}_2)$ shifts to $\ln \mathcal{I}_2 \to -\infty $ with increasing system size.

    \subsection{The scaling behavior across different phases}
    
\begin{figure}[!htbp]
		\begin{center}
			\includegraphics[width=4.0cm]{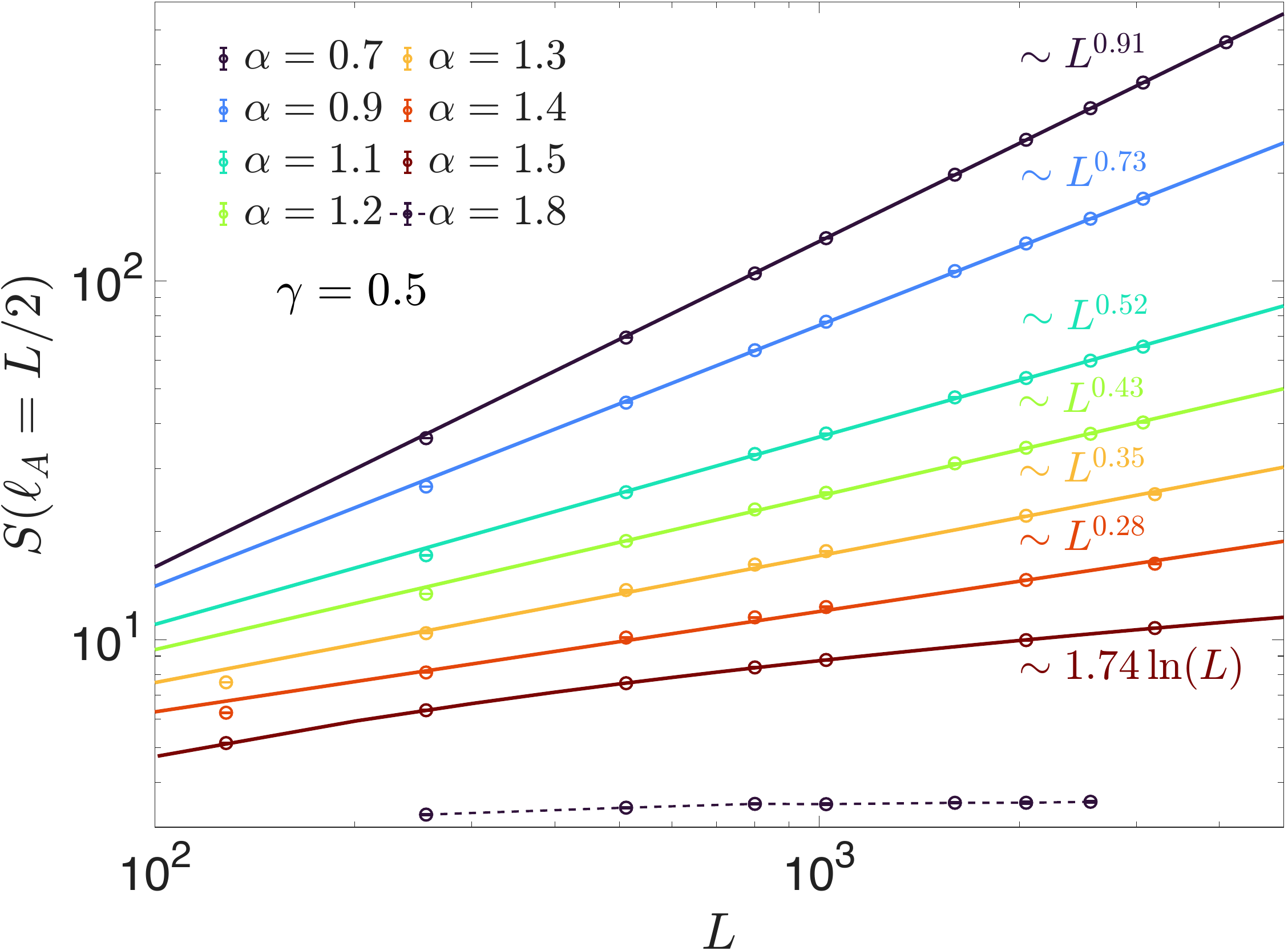} 
			\includegraphics[width=4.0cm]{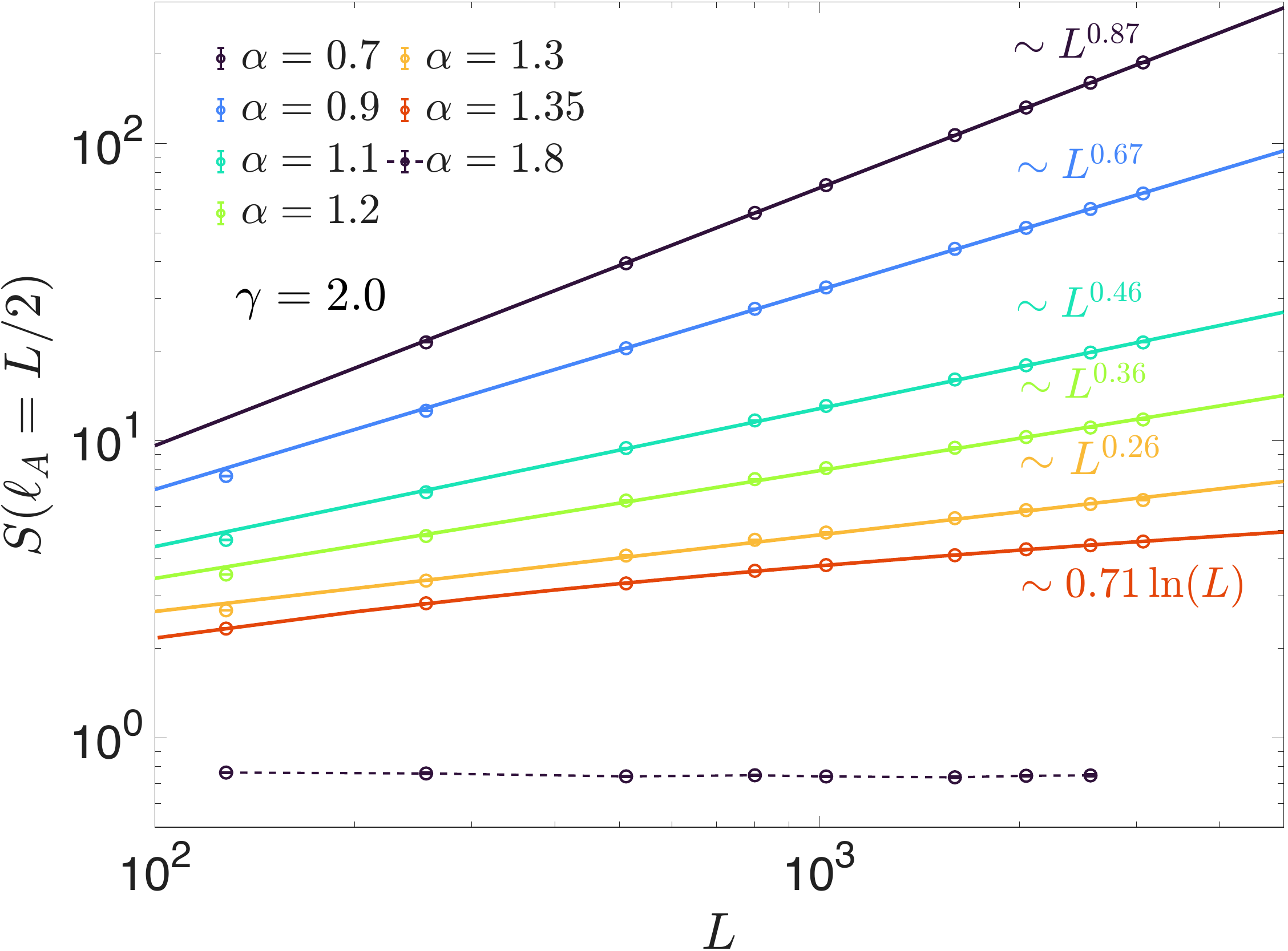} 
			\caption{
				Steady-state entanglement entropy $S(\ell_A = L/2)$ Eq.~\eqref{eq:sl} as a function of the system size $L$ for $\gamma = 0.5$ (left) and $\gamma = 2.0$ (right), and different power-law exponents $\alpha$. Fits to a power-law growth $S \sim L^{a_s(\alpha)}$ reveal a gradual increase of $a_s$ with increasing $\alpha$. For weak monitoring $\gamma = 0.5$, the system transits from $a_s \approx 1$ (volume-law) for small $\alpha$ to a logarithmic growth $\alpha \approx 3/2$, and  finally to an area-law phase for sufficiently large $\alpha$ ($\alpha = 1.8$). The difference between weak and strong monitoring is that both the logarithmic growth, and the area-law phase occur at a slightly smaller $\alpha$. These results are consistent with the mutual information analysis depicted in Fig.~\ref{Fig:I2_vs_alpha_QSD}.
				}\label{Fig:SA_vs_L_monitoring}
		\end{center}
		\vspace{-0.5cm}
	\end{figure}

    We now turn to describe the precise scaling behavior of the bipartite EE and MI with system size $L$ across different phases. Results depicted in Fig.~\ref{Fig:SA_vs_L_monitoring} indicate that for sufficiently small $\alpha$, and weak monitoring $\gamma = 0.5$, the EE exhibits a power-law growth $S(\ell_A = L/2) \sim L^{a_s}$ with $a_s > 0$. More specifically, for $\alpha \lesssim 0.5$, $a_s \sim 1$, so the scaling is volume-law. As $\alpha$ increases, the exponent $a_s$ decreases approximately linearly, following $a_s \simeq 3/2 - \alpha$ and approaches zero as $\alpha \to 3/2$, see Fig.~\ref{fig.as_vs_alpha}.  For $\alpha \approx 3/2$, the scaling becomes logarithmic $S(\ell_A = L/2) \sim \ln(L)$, which is characteristic of $(1+1)$-dimensional conformal field theories (CFTs)~\cite{calabrese2004entanglement,calabrese2009entanglement}. 
    For larger $\alpha$, the EE is independent of system size $L$, signaling the area-law phase. 
    For $\gamma = 2.0$, corresponding to the strong monitoring limit, we observe a qualitatively similar behavior, but the critical $\alpha_c$ of the MIPT, shifts to $\approx 1.35$, in agreement with the mutual information analysis in Fig.~\ref{Fig:I2_vs_alpha_QSD}.

\begin{figure}[!htbp]
		\begin{center}
			\includegraphics[width=4cm]{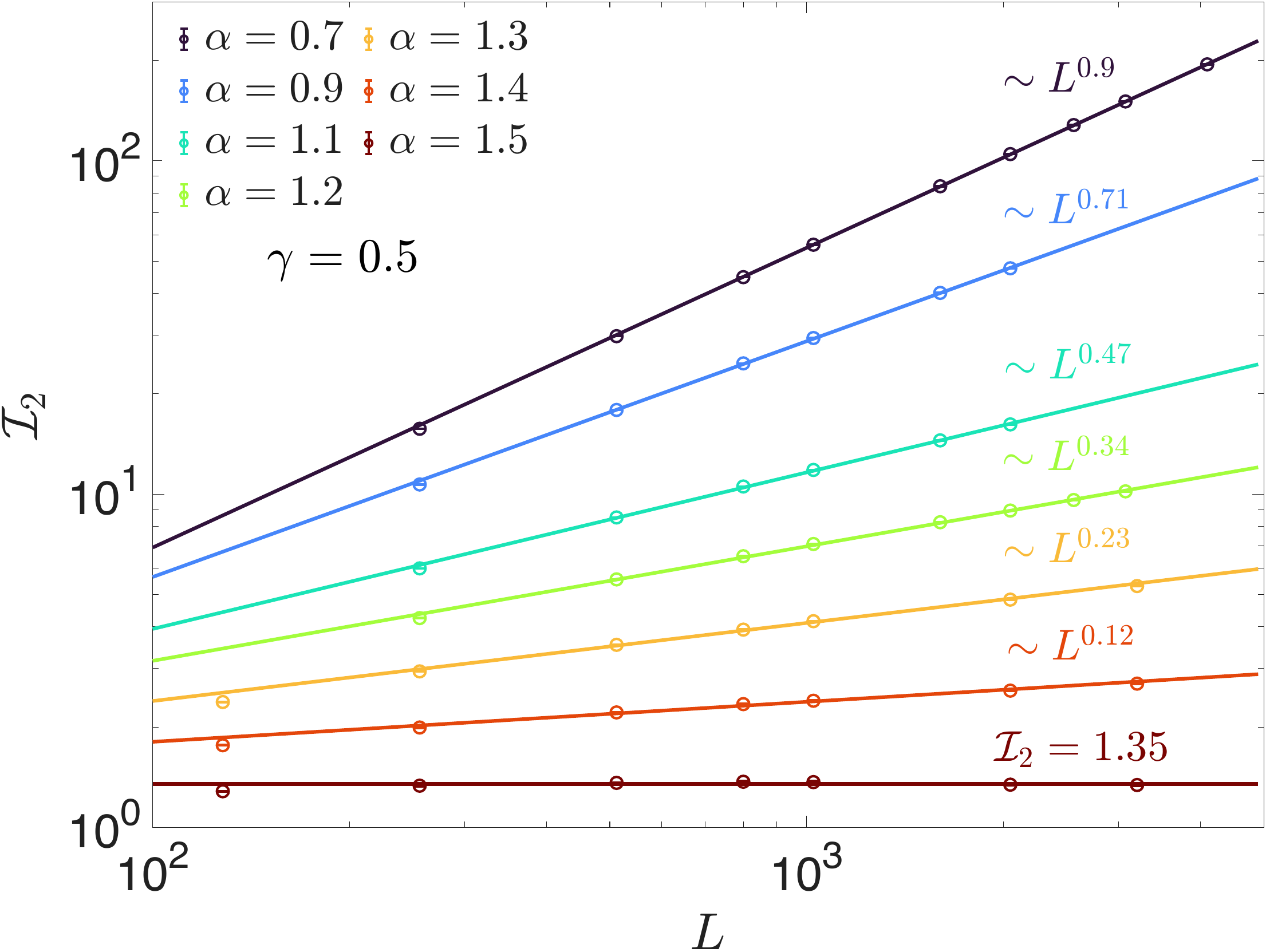} 
			\includegraphics[width=4cm]{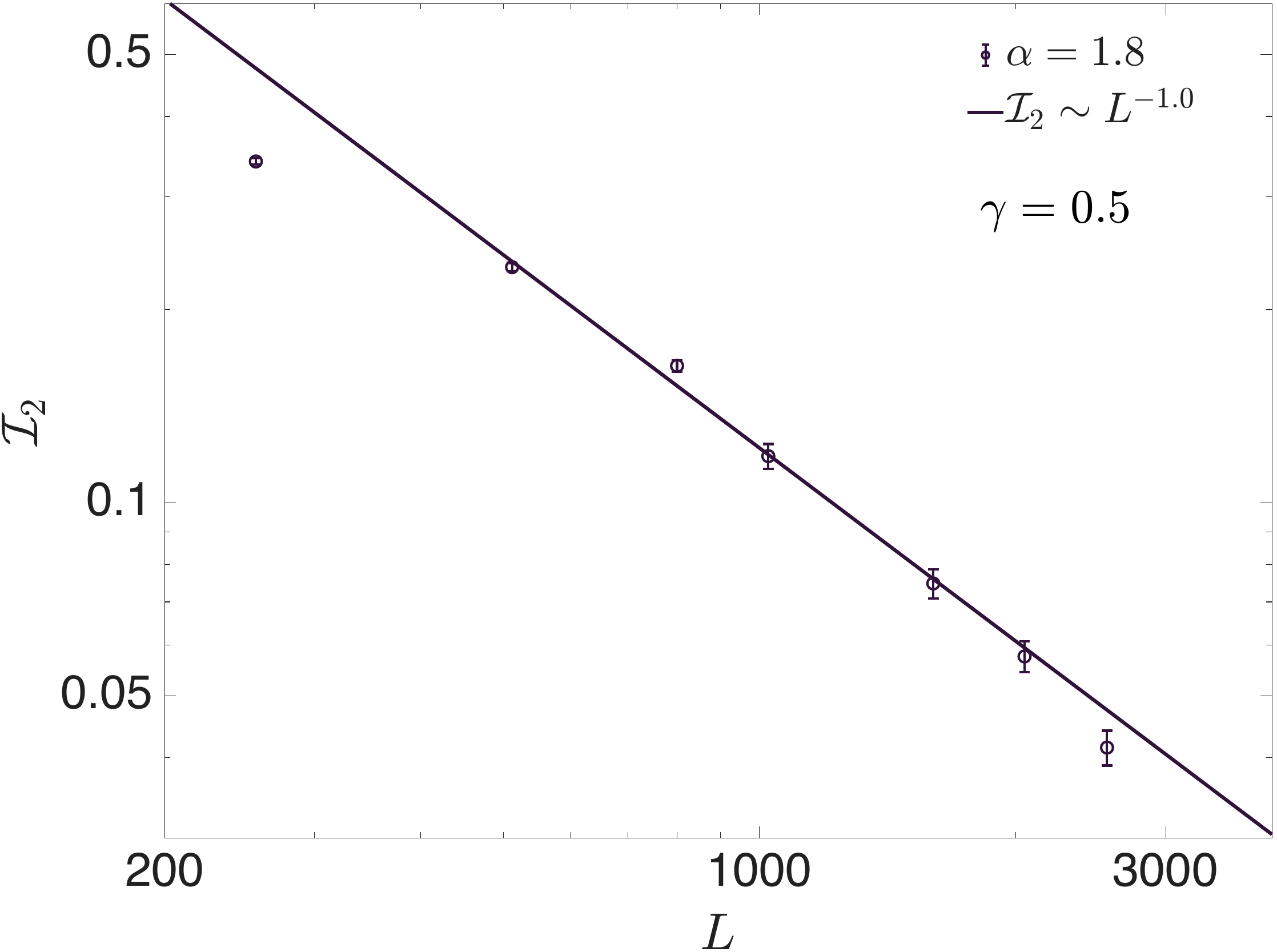} 
			\includegraphics[width=4cm]{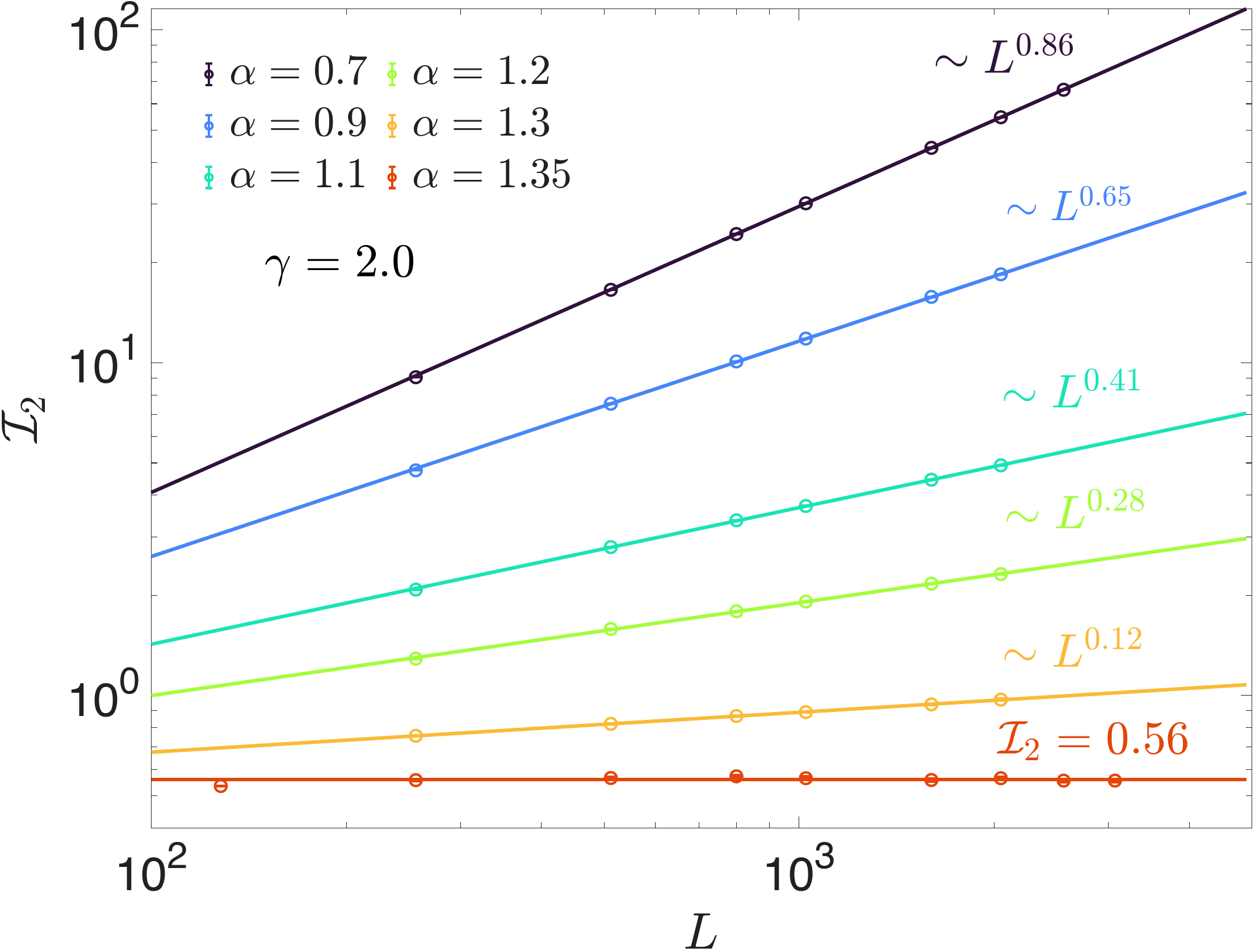}  
			\includegraphics[width=4cm]{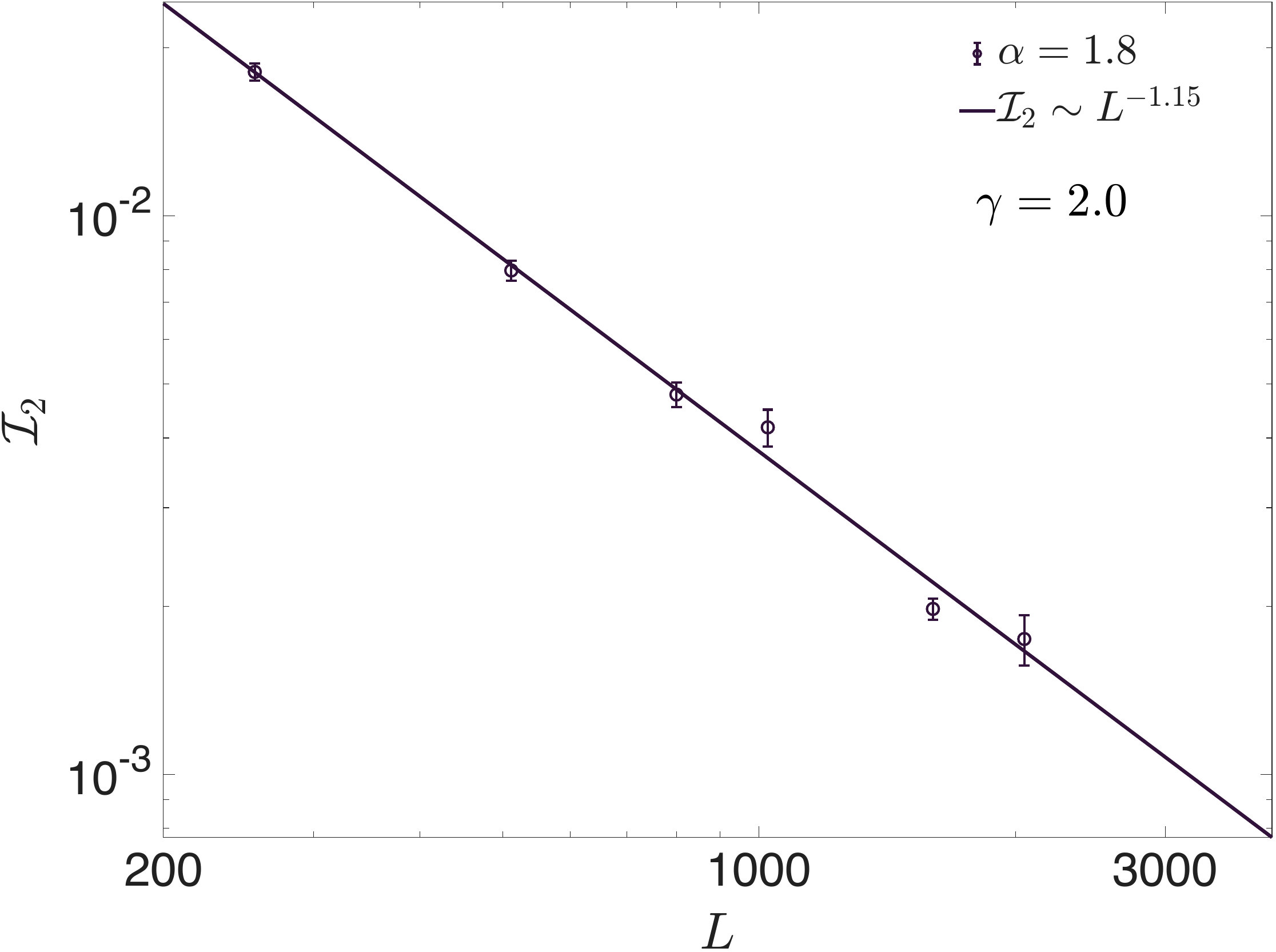} 
			\caption{ Scaling of the mutual information $\mathcal{I}_2$ with system size $L$ for $\gamma = 0.5$ (Upper panel) and $\gamma = 2.0$ (Lower panel), and different values of $\alpha$. For sufficiently small $\alpha$, $\mathcal{I}_2 \sim L^{a_p}$ with $0 < a_p \leq 1$. At the MIPT $\alpha = \alpha_c$, $\mathcal{I}_2$ is scale-invariant, and $\mathcal{I}_2$ decays as a power-law in the area-law phase. These results are consistent with the $\alpha_c$ obtained in Fig.~\ref{Fig:I2_vs_alpha_QSD} using the MI and also with those for the EE in Fig.~\ref{Fig:SA_vs_L_monitoring}. 
			}\label{Fig:I2_vs_L_monitoring}
		\end{center}
		\vspace{-0.5cm}
	\end{figure}

\begin{figure*}[!htbp]
	\begin{center}
		\subfigure[]{ \label{fig.as_vs_alpha}
			\includegraphics[width=5.2cm]{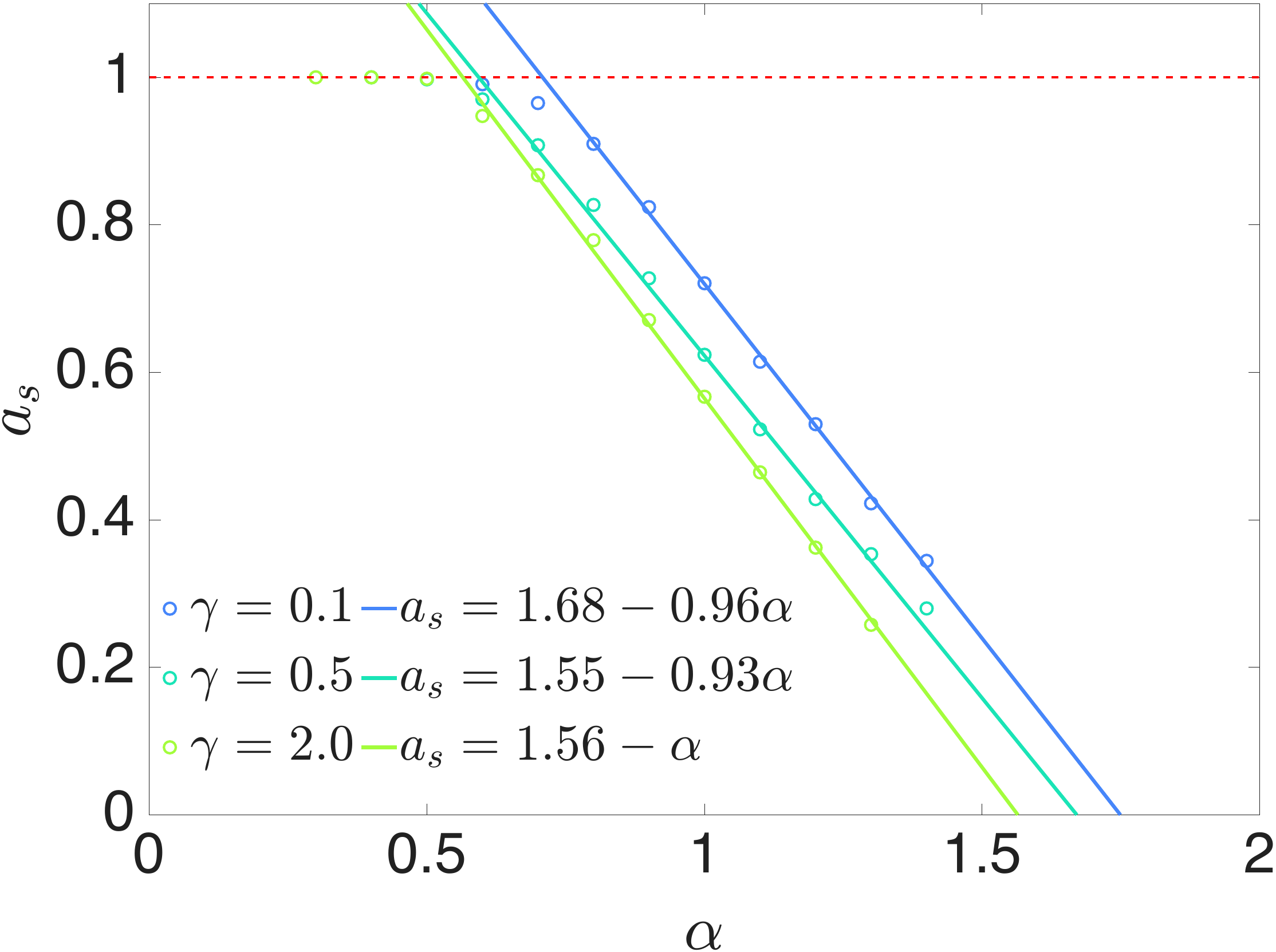} }
		\subfigure[]{ \label{fig.ap_vs_alpha}
			\includegraphics[width=5.2cm]{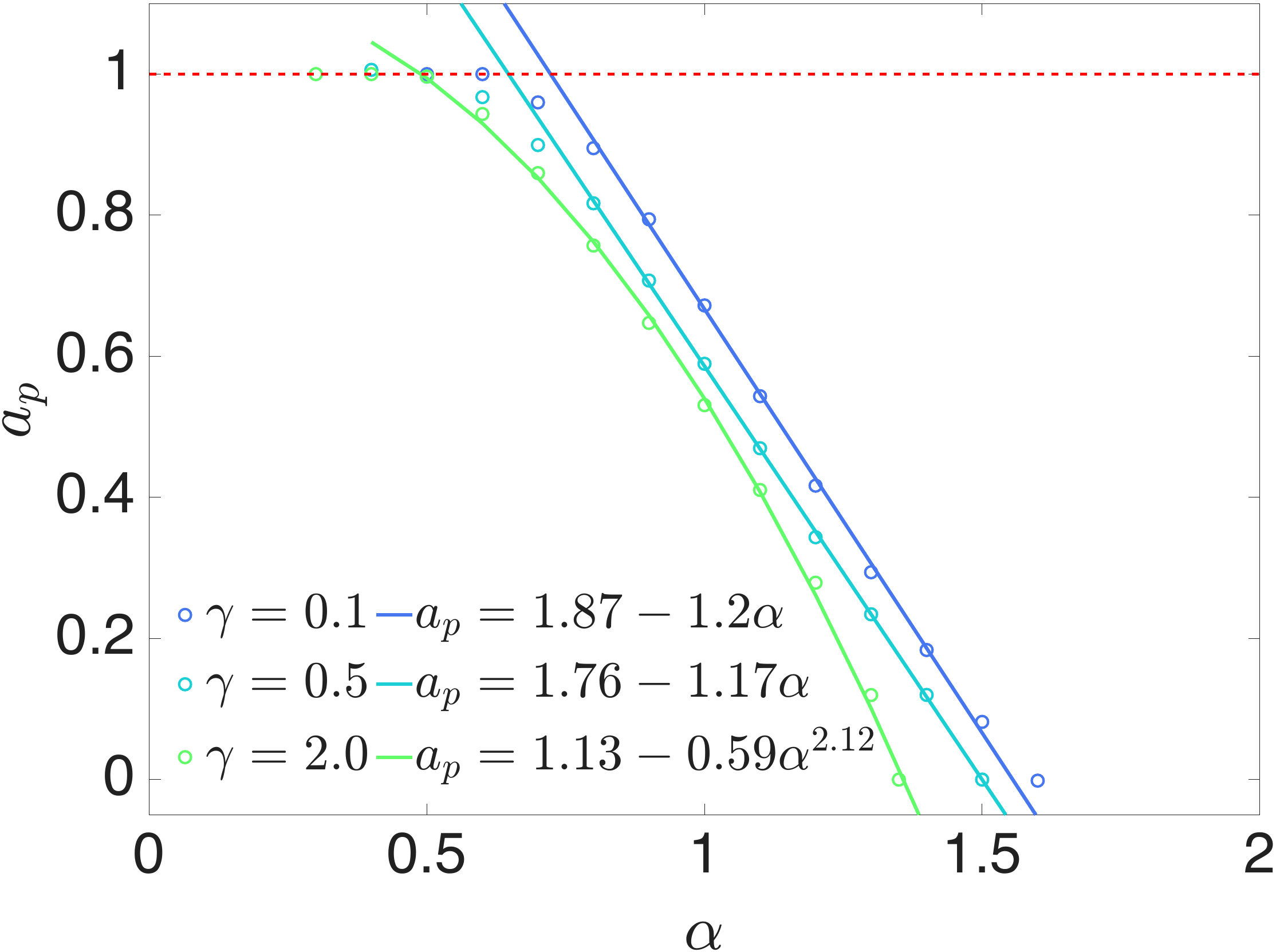} }
		\subfigure[]{ \label{fig.ar_vs_alpha}
			\includegraphics[width=5.2cm]{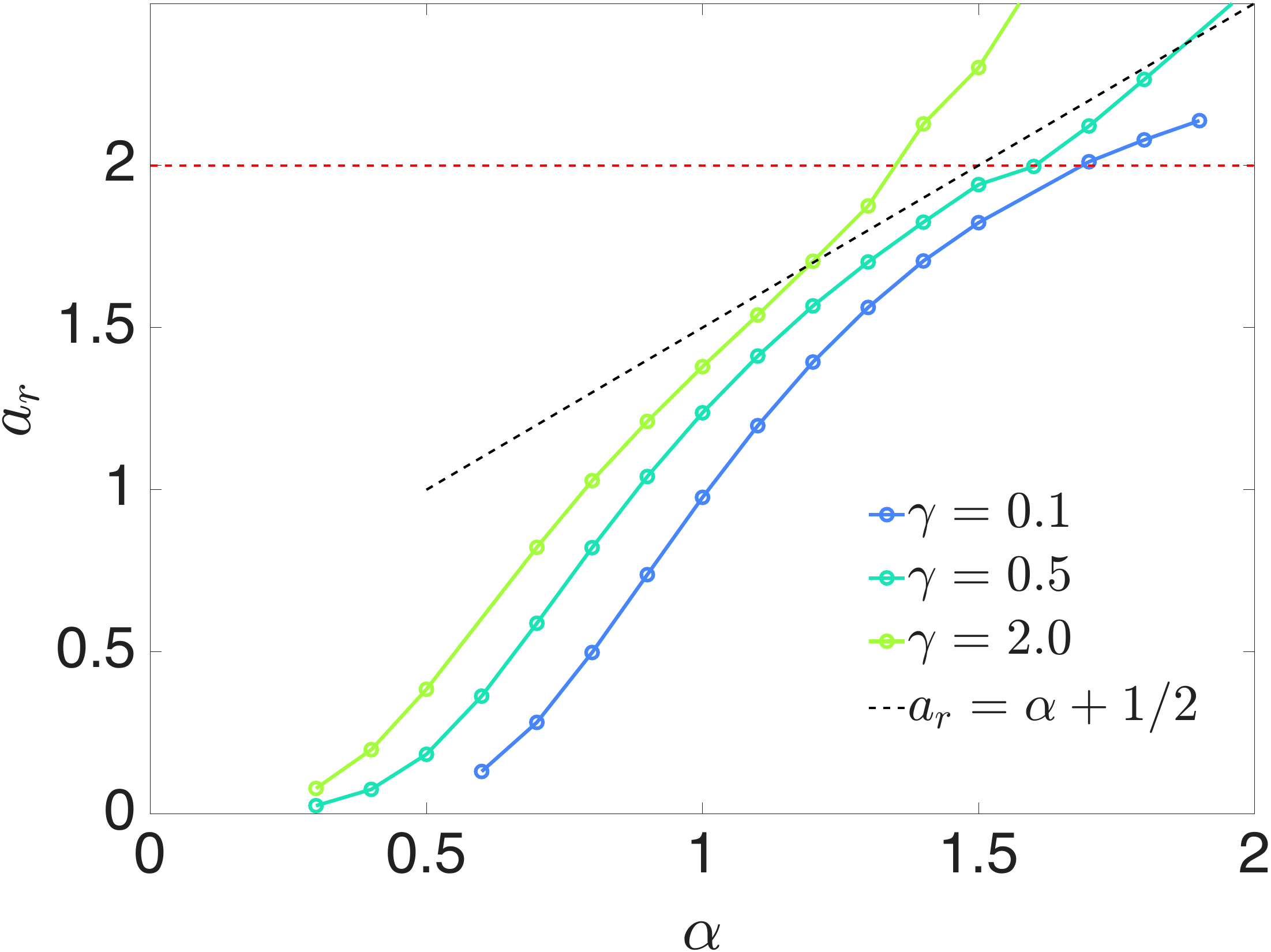} }
		\caption{ 
			\subref{fig.as_vs_alpha}. The scaling exponent $a_s$, characterizing the EE growth in the sub-volume-law phase, extracted from Fig.~\ref{Fig:SA_vs_L_monitoring}, as a function of $\alpha$. The exponent $a_s$ decreases approximately as $a_s \simeq 3/2 - \alpha$, and saturates at $a_s =1$ for sufficiently small $\alpha \lesssim 0.5$. 
			\subref{fig.ap_vs_alpha}. The power-law exponent $a_p$, characterizing the growth of $\mathcal{I}_2$, extracted from Fig.~\ref{Fig:I2_vs_L_monitoring}, as a function of $\alpha$. In the weak monitoring limit, $a_p$ decreases approximately linearly with $\alpha$. For stronger monitoring, however, $a_p$ exhibits a nonlinear dependence on $\alpha$.
			\subref{fig.ar_vs_alpha}. The power-law exponent $a_r$, characterizing the decay of $C(r)$, obtained from Fig.~\ref{Fig:Cr_full_app}, as a function of $\alpha$. The black dashed line shows the analytic prediction $a_r = \alpha + 1/2$. Following Ref.~\cite{poboiko2023}, $a_r = 2$ is the critical value of the transition from the sub-volume-law to the area-law phase. At the MIPT $\alpha = \alpha_c$, the EE scales logarithmically with system size. 
			The values of $a_s, a_r$, and $a_p$ are consistent with each other. 
		}\label{Fig:extracted_exponent_vs_alpha}
	\end{center}
	\vspace{-0.5cm}
\end{figure*}

	The scaling of MI $\mathcal{I}_2$ with the system size $L$, shown in Fig.~\ref{Fig:I2_vs_L_monitoring}, is in full agreement with that of the EE, providing further support for the phase diagram presented in Fig.~\ref{Fig:phase_diagram}. When $\alpha < \alpha_c$, $\mathcal{I}_2$ follows a power-law growth $\mathcal{I}_2 \sim L^{a_p}$ with an exponent $0 < a_p \leq 1$, characteristic of the (sub-)volume-law phase. 
	At the critical point $\alpha = \alpha_c$, $\mathcal{I}_2$ becomes scale-invariant, directly reflecting the logarithmic growth of the EE at criticality. For $\alpha > \alpha_c$, $\mathcal{I}_2$ instead decays as a power-law with $L$, signaling the area-law phase.
	
\begin{figure}[!htbp]
    	\begin{center}
    		\includegraphics[width=4cm]{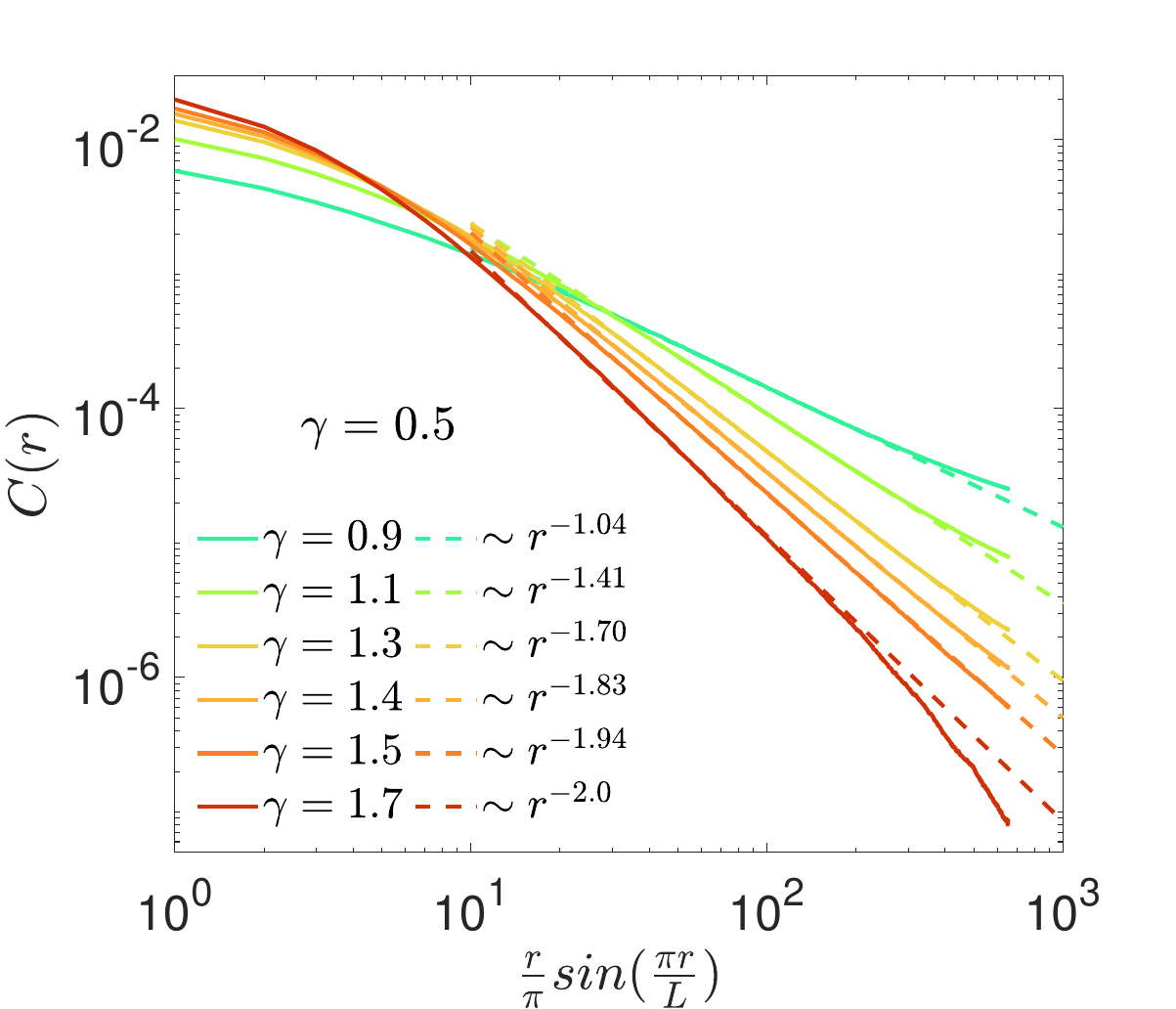} 
    		\includegraphics[width=4cm]{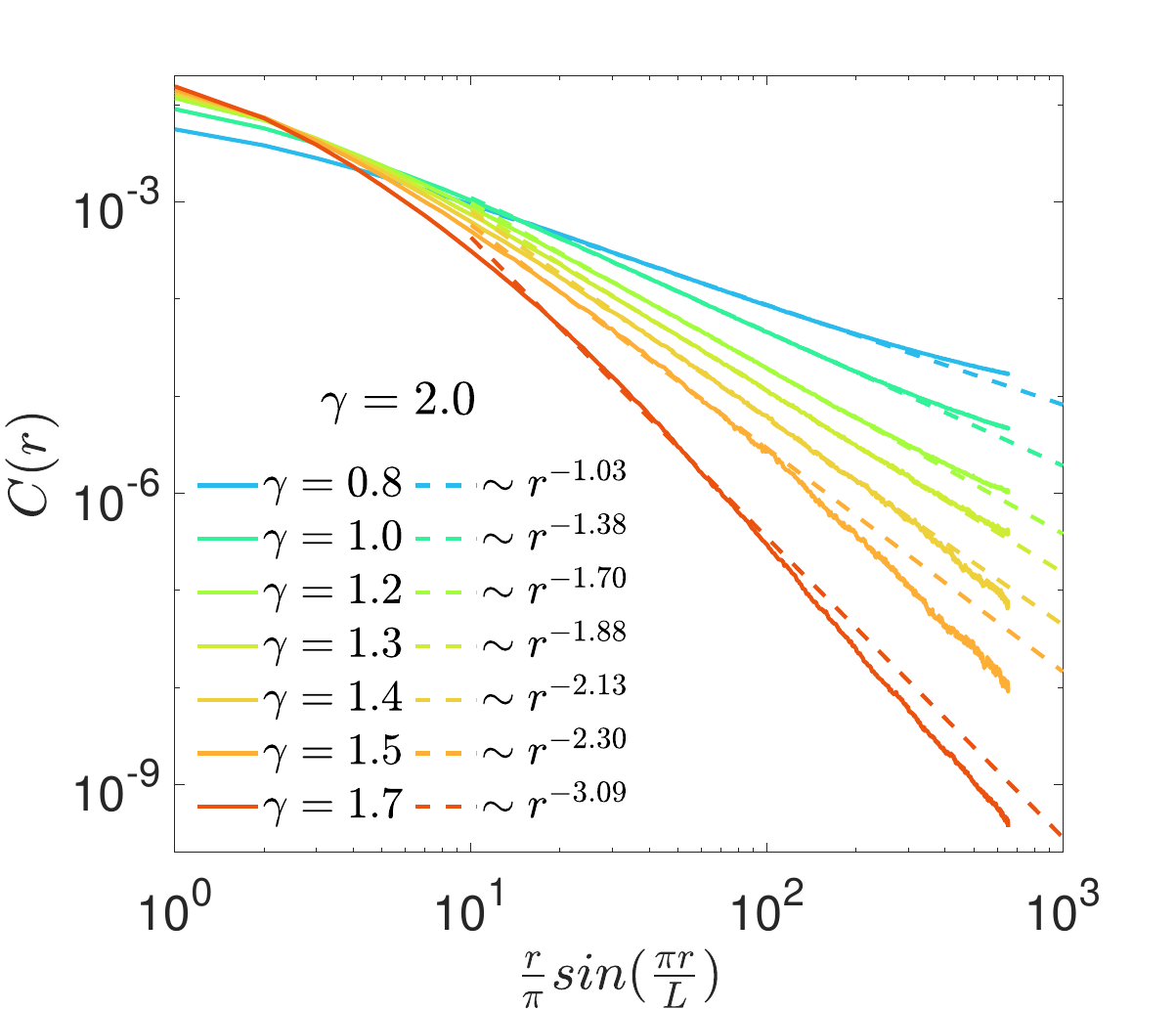}  
    		\caption{Density-density correlation function $C(r)$ Eq.~\eqref{eq:cr} as a function of distance $r$ for different values of $\alpha$. The monitoring strength is $\gamma = 0.5$ (left) and $\gamma = 2.0$ (right), with system size fixed at $L = 2048$. The dashed lines denote power-law fits $C(r) \sim 1/r^{a_r}$. 
    		The extracted power-law exponent $a_r$ is directly related to the growth of the EE with system size. Results for more values of $\alpha$ and $\gamma$ are depicted in Appendix \ref{app:mut}, Fig.~\ref{Fig:Cr_full_app}.
    		}\label{Fig:Cr_QSD}
    	\end{center}
    	\vspace{-0.5cm}
    \end{figure}

    In order to complement the description of the MIPT, we also analyze the decay of the density-density correlation function $C(r)$ Eq.~\eqref{eq:cr}, depicted in Fig.~\ref{Fig:Cr_QSD}, which is closely related to the scaling of the EE, see Eq.~\eqref{eq.Klich_Levitov} \cite{levitov2009,poboiko2023}. 
    For $\alpha < \alpha_c$, we obtain $a_r < 2$, so the decay of $C(r)$ is sufficiently slow to produce a growing particle number fluctuations Eq.~(\ref{eq:na2}) which translates into a power-law scaling of the EE with system size. 
    At the critical point $\alpha \approx \alpha_c$, $a_r = 2$, in agreement with the theoretical prediction $C(r) \sim r^{-2}$ \cite{poboiko2023}, which is consistent with a logarithmic scaling of the EE. For $\alpha > \alpha_c$, the exponent satisfies $a_r > 2$, signaling the area-law phase. 
    
	We now summarize the different scaling exponents $a_s$, $a_p$ and $a_r$ of the EE $S \sim L^{a_s}$, MI $\mathcal{I}_2 \sim L^{a_p}$ and density-density correlations $C(r)\sim r^{-a_r}$, respectively, as a function of $\alpha$ for monitoring strengths $\gamma = 0.1, 0.5$ and $2.0$.
	Results depicted in Fig.~\ref{Fig:extracted_exponent_vs_alpha} show that the increasing of $\alpha$ results in an approximately linear decrease of $a_s\simeq 3/2 - \alpha$ before saturating at $a_s = 1$ (volume-law) for $\alpha \leq 0.5$. For sufficiently weak monitoring, the decrease of $a_p$ is also approximately linear as $\alpha$ increases before saturating at $a_s \approx 1$ for $\alpha \leq 0.5$, consistent with a volume-law phase. For stronger monitoring, $a_p$ exhibits a nonlinear dependence on $\alpha$, reflecting the competition between monitoring, that tends to localize the fermions, and therefore suppressing entanglement, and long-range hopping that always enhances it. The exponent $a_r \approx 2$, describing the decay of $C(r)\propto r^{-a_r}$ is a reliable marker of the transition from the logarithmic to the area-law phase. 

	Taken together, the MI $\mathcal{I}_2$, EE $S(\ell_A = L/2)$ scaling with system size and the decay exponent of the correlation function $C(r)$ with distance, $a_s$, $a_p$ and $a_r$ respectively, provide a consistent and detailed picture of the entanglement phase diagram of the PRBM model Eq.~\eqref{eq:H} depicted in Fig.~\ref{Fig:phase_diagram}.
	
	\vspace{-0.3cm}
	\subsection{Multifractal analysis at the MIPT}

\begin{figure}[!htbp]
		\begin{center}
			\includegraphics[width=4cm]{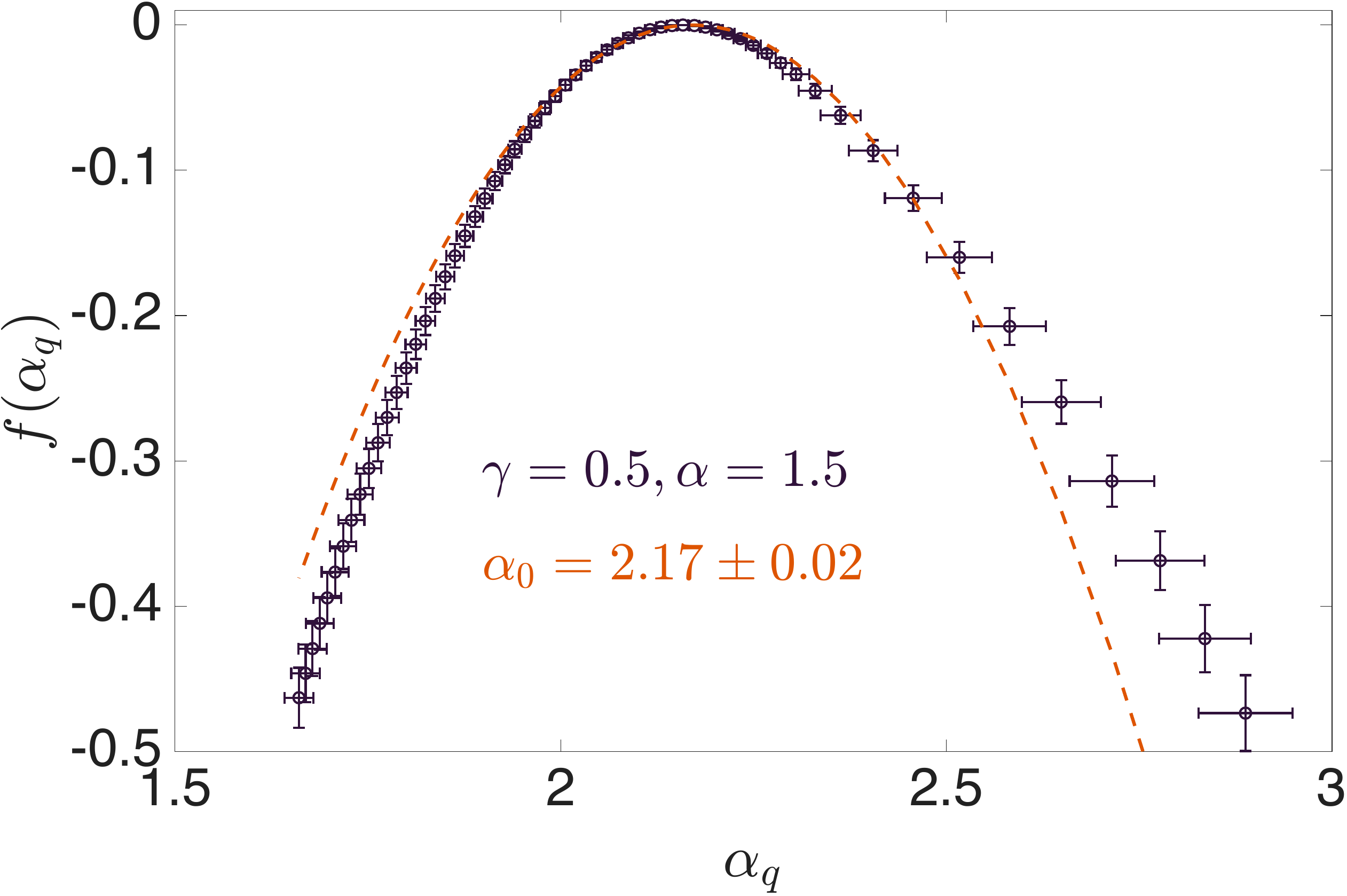} 
			\includegraphics[width=4cm]{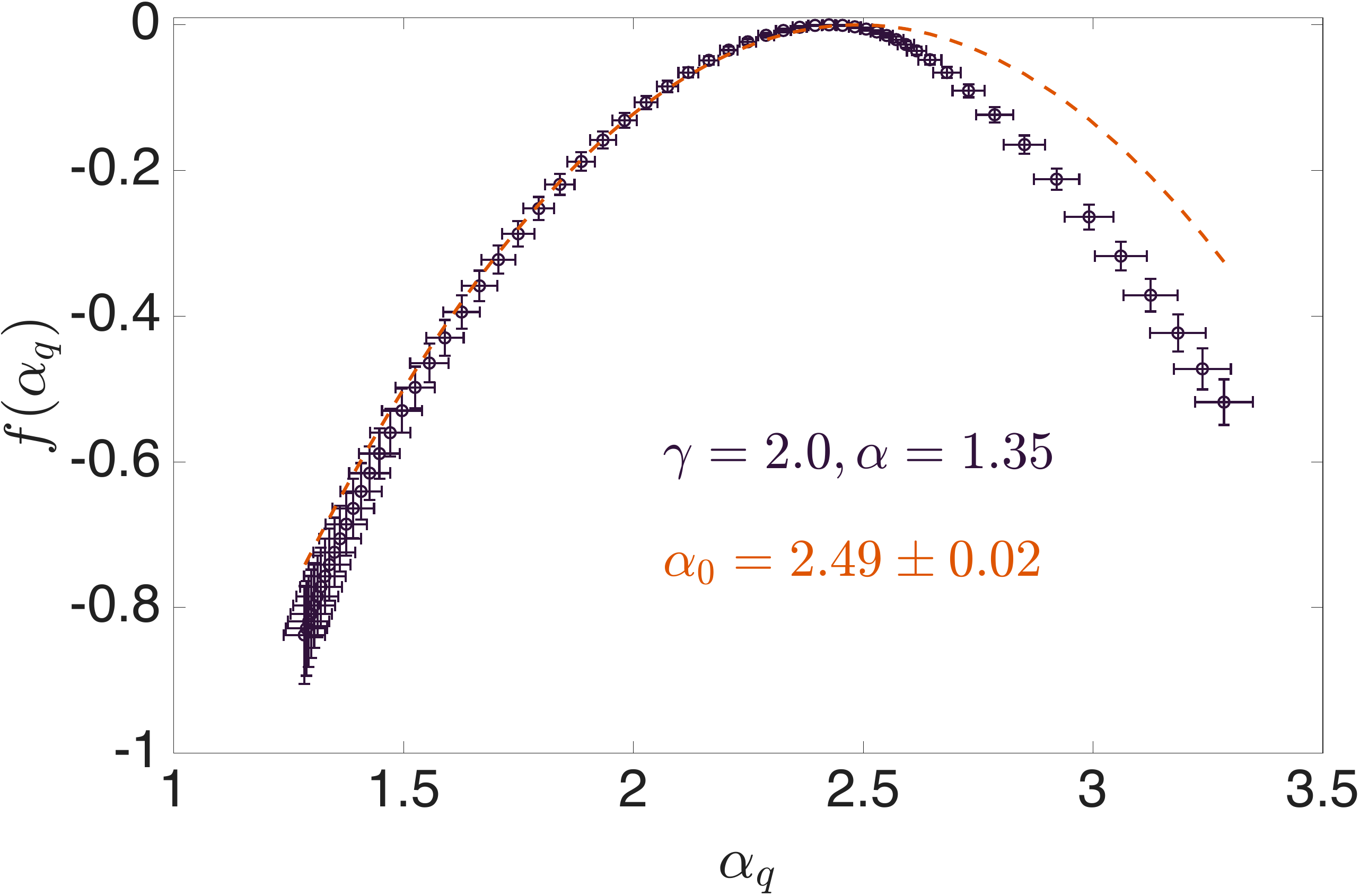} 
			\caption{
                The singularity spectrum $f(\alpha_q)$ \cite{chhabra1989direct,Jian2023,poboiko2025} at the MIPT. Dashed line: parabolic fit according to Eq.~\eqref{eq.f_alpha}. The corresponding fitted $\alpha_0$ are $2.17 \pm 0.02$ and $2.49\pm 0.02$ for $\gamma = 0.5, \alpha = 1.5$ and $\gamma = 2.0,\alpha = 1.35$, respectively.
				}\label{Fig:f_alpha_q}
		\end{center}
		\vspace{-0.8cm}
	\end{figure}

	Wavefunctions near Anderson localization transitions \cite{evers08,schreiber1991,rodriguez2011multifractal} are expected to show multifractal features. In the context of entanglement dynamics in monitored quantum systems, it has been argued \cite{Jian2023,poboiko2025} that the correlation function $C(r)$ is expected to exhibit multifractal scaling at the MIPT. In order to characterize these multifractal features of the critical states, we compute the singularity spectrum $f(\alpha_q)$ \cite{chhabra1989direct} at the MIPT. Results displayed in Fig.~\ref{Fig:f_alpha_q} reveal that the singularity spectrum is well described by the parabolic form
	\begin{equation}
		f(\alpha_q) = -\frac{(\alpha_q - \alpha_0)^2}{4(\alpha_0 - 2d)},
	\label{eq.f_alpha}
	\end{equation} 
	with $d = 1$ and $\alpha_0 = 2.17\pm 0.02$ for $\gamma = 0.5, \alpha=1.5$ and $\alpha_0 = 2.49 \pm 0.02$  for $\gamma = 2.0,\alpha = 1.35$. We note that $\alpha_0$ corresponds to the position of the peak of the singularity spectrum, $f(\alpha_0) = 0$, and determines the decay of the typical correlation function $C^{\rm typ}(r) = \exp(\overline{\ln C(r)}) \sim r^{-\alpha_0}$ \cite{Jian2023,poboiko2025}, where $\ln C(r)$ is self-averaging at MIPT. The average is taken over all disorder realizations and quantum trajectories.
	It also signals, since $\alpha_0 \sim 2d$, that the steady state at the MIPT is weakly multifractal \cite{chhabra1989direct,Jian2023,poboiko2025}.
	The dependence of $\alpha_0$ on $\gamma$ further suggests that the multifractal properties are governed by the competition between measurements and long-range hopping.

	\vspace{-0.3cm}
	\subsection{Analytical understanding of the entanglement entropy scaling}
	
	Finally, we provide a qualitative theoretical explanation of the obtained results. The monitoring of the dynamics rapidly drives the system to an infinite temperature state which typically makes it more delocalized, and therefore more entangled compared to the ground state, while the process of measurement typically involves a localization in a specific element of the basis which normally suppresses entanglement. Although for some quantum trajectories the entanglement can increase \cite{garcia2026b,schiro2024}, the combined effect is a destruction of entanglement. Regarding the role of short-range disorder, as mentioned earlier, recent results \cite{garcia2026,liao2026} have shown that a weak disordered potential does not change qualitatively the entanglement dynamics of monitored 1d complex fermions. 
	We also note that effective descriptions of the monitoring dynamics of 1d non-interacting fermions \cite{cao2019a,poboiko2023} with short-range hopping and no disorder, indicate that monitoring has the effect of inducing a finite lifetime $\sim \gamma^{-1}$ to quasiparticles. Long-distance quasiparticle pairs are therefore destroyed and no MIPT occurs because correlations are eventually suppressed. We also stress that monitoring completely suppresses any Anderson localization effect since it requires quantum coherence which is completely destroyed by measurements. 
    
    However, this quasiparticle description breaks down in the presence of random long-range hopping model, because the particles are not associated with the well-defined quasiparticle excitations propagating with a finite velocity. Instead, the long-range hopping induces direct coupling across arbitrarily large distances, leading to intrinsically nonlocal correlations. Consequently, an effective model of quasiparticles with a finite lifetime is no longer sufficient to capture the entanglement dynamics if long-range hopping is present.
    
    Following these considerations, we proceed to estimate analytically the scaling of EE with system size. 
    We focus initially on $D_{ij}$, but note that this quantity is directly related to the density-density correlations $C(r)$ Eq.~(\ref{eq:cr}) and the entanglement entropy Eq.~(\ref{eq.Klich_Levitov}). 
	From Eq.~\eqref{eq.SSE}, we obtain, following Ref.~\cite{cao2019a}, the evolution of the correlation matrix $D_{ij}$ Eq.~\eqref{eq:di},
    \begin{equation}
    	\begin{aligned}
    		dD_{ij} = &- i[H,D]_{ij}dt - \gamma dt(1-\delta_{ij})D_{ij}\\
    		&+ [D_{ij} (dW_i + dW_j) - 2\sum_{k} D_{ik} D_{kj} dW_k] 
    	\end{aligned} \label{eq.dDij}
    \end{equation}
    with $\overline{dW_i} = 0$ and $\overline{dW_idW_j} = \gamma dt \delta_{ij}$. Averaging the above dynamic equation over all trajectories, the stochastic terms vanish.
    Therefore, the average of correlation matrix $\overline{D}_{ij}$ satisfies the equation 
    \begin{equation}
    	{d\overline{D}_{ij} \over dt}= - i[H,\overline{D}]_{ij} - \gamma(1-\delta_{ij})\overline{D}_{ij} .
    \end{equation}
    
\begin{figure}[!htbp]
    	\begin{center}
    		\includegraphics[width=4cm]{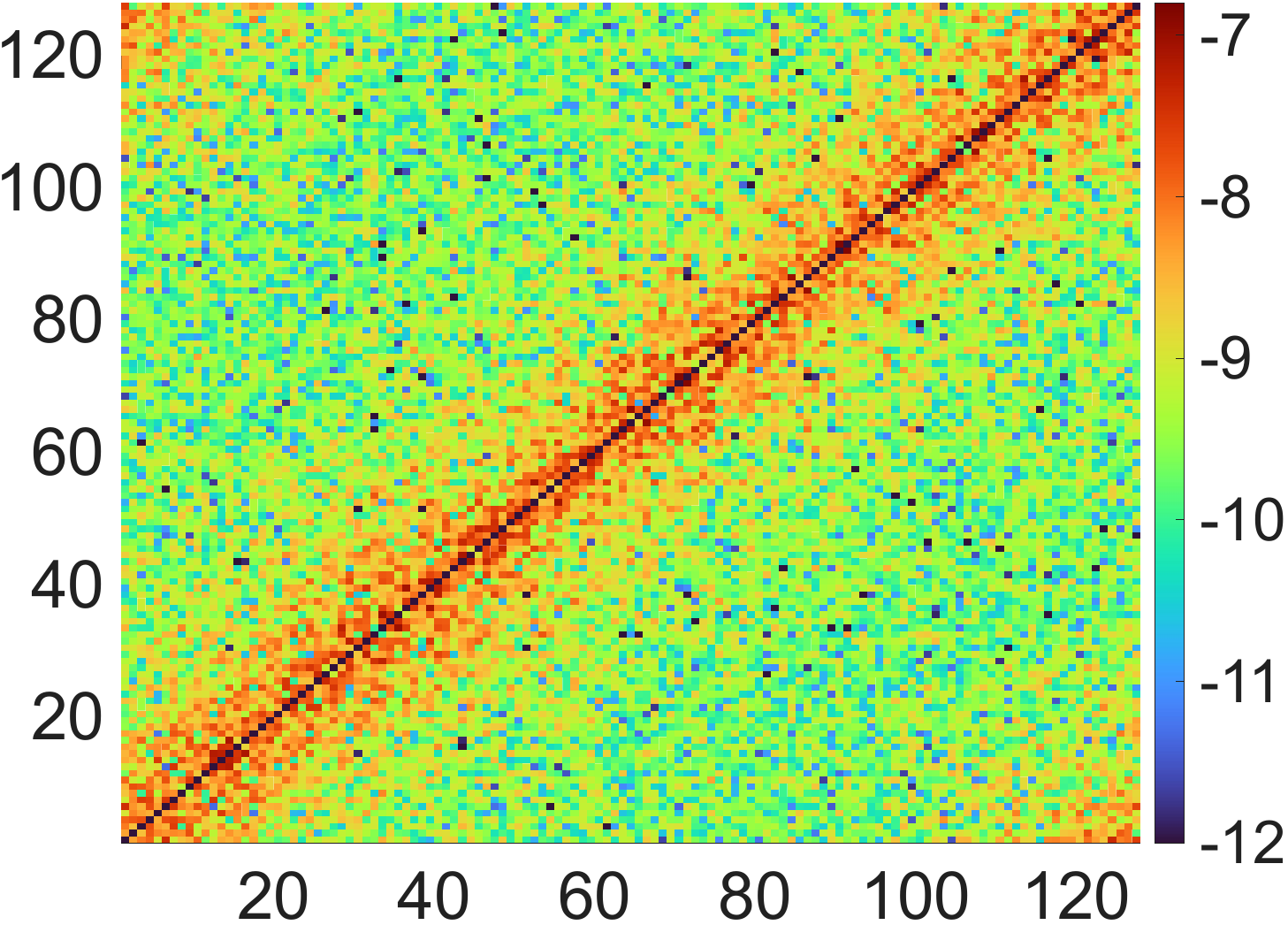} 
    		\includegraphics[width=4cm]{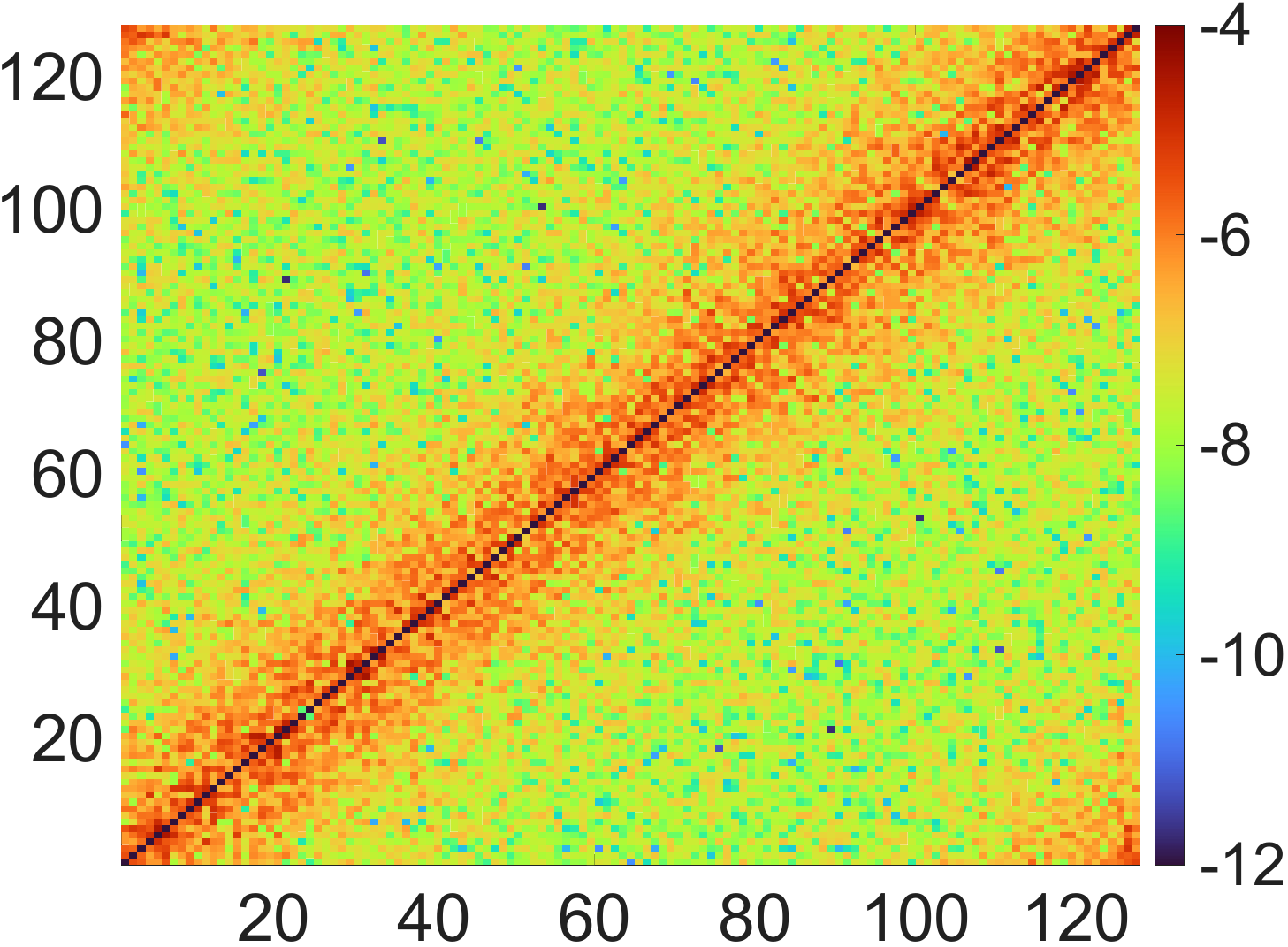} 
    		\caption{
    			The spatial distribution of $\ln {d\overline{D}_{ij} \over dt} $ (left) and $\ln \overline{D}_{ij}$ (right) for $i\neq j$ in the dynamical steady state. Note the scale of $\ln \overline{D}_{ij}$ is more than 20 times larger than $\ln {d\overline{D}_{ij} \over dt} $, even when the measurement strength is $\gamma = 0.1$ and $\alpha = 1.5$. The system size is $L = 128$. Although ($\gamma = 0.1, \alpha = 1.5$) corresponds to the critical parameter in the thermodynamic limit, considering the small system size, the results are characterizing effectively the superdiffusive regime (sub-volume-law phase).
    		}\label{Fig:dD_Dij}
    	\end{center}
    	\vspace{-0.5cm}
    \end{figure}

    The above equation can be expanded as 
    \begin{equation}
    	{d\overline{D}_{ij} \over dt}= 
    	\begin{cases}
    		- i(\sum_{k}H_{ik}\overline{D}_{kj} - \sum_{p}\overline{D}_{ip}H_{pj}) - \gamma\overline{D}_{ij}, &i\neq j\\
    		- i(\sum_{k}H_{ik}\overline{D}_{kj} - \sum_{p}\overline{D}_{ip}H_{pj}). &i=j 
    	\end{cases}
    	\label{eq.dDij_dt}
    \end{equation}    
    This expression is still too complicated to be discussed for a general value of $\alpha$ so further approximations are necessary. 
    In Fig.~\ref{Fig:dD_Dij}, we present the off-diagonal elements $\ln {d\overline{D}_{ij} \over dt} $ and $\ln \overline{D}_{ij}$ when $i\neq j$. Comparing the amplitude of $\ln {d\overline{D}_{ij} \over dt} $ and $\ln \overline{D}_{ij}$, we can safely assume that the time derivative ${d\overline{D}_{ij} \over dt}$ is negligible. With a finite measurement strength, this assumption holds particularly well in the superdiffusive and critical regions ($1/2 < \alpha \lesssim \alpha_c(\gamma)$), but it breaks down for sufficiently small $\alpha \lesssim 1/2$ due to the fast recovery of entanglement. We also assume that the off-diagonal terms cancel $\sum_{k\neq j}H_{ik}\overline{D}_{kj} - \sum_{p\ne i}\overline{D}_{ip}H_{pj} \approx 0$, so it only remains the contribution of the diagonal elements,
    \begin{equation}
    	\overline{D}_{ij} \approx {i H_{ij}\over \gamma} (\overline{D}_{ii} - \overline{D}_{jj}), \qquad i\neq j .
    	\label{eq.average_Dij}
    \end{equation}    
    
    $C(r)$ Eq.~(\ref{eq:cr}), the correlation function we are interested in, is related to $D_{ij}$ by $C_{ij} = |D_{ij}|^2 = D_{ji}D_{ij}$, where $r = |i - j|$. Since both are random quantities, their differentials are given by It\^{o}'s lemma,  
    \begin{equation}
    	\begin{aligned}
    		dC_{ij} = d|D_{ij}|^2 = D_{ji}dD_{ij} + dD_{ji}D_{ij} + dD_{ji}dD_{ij}. 
    	\end{aligned} \label{eq.dCij}
    \end{equation}
    By using Eq. \eqref{eq.dDij}, the first term of the above equation is:
    \begin{equation}
    	\begin{aligned}
    		D_{ji}dD_{ij} = -iD_{ji}[H,D]_{ij}dt -\gamma dt (1-\delta_{ij})D_{ji}D_{ij} \\
    		+ D_{ji}D_{ij} (dW_i + dW_j) - 2\sum_{k} D_{ji}D_{ik} D_{kj} dW_k ,
    	\end{aligned}
    	\label{eq.DdD}
    \end{equation}
    where the last two terms in the right side are the stochastic part. Since $dW_j$ is independent of $D_{ij}$ at the same time step, upon trajectory averaging, the last two terms cancel due to the zero mean of $dW_j$. The first term corresponds to the unitary dynamics. 
    After averaging over trajectories, we can again assume that the off-diagonal terms cancel, so only diagonal elements remain,
    \begin{equation}
    	\begin{aligned}
    		\overline{-iD_{ji} [H,D]_{ij}} \approx \overline{-i H_{ij} D_{ji}(D_{jj} - D_{ii} ) }  
    	\end{aligned}
    \end{equation}
    where $H$ and $D$ are statistically correlated because the correlation matrix $D$ evolves under the same Hamiltonian $H$. For $H_{ij}$ with Gaussian entries of zero-mean \cite{novikov1965functionals},
    \begin{equation}
    	\overline{-iH_{ij} D_{ji}(D_{jj} - D_{ii} )} \approx \overline{-i|H_{ij}|^2(D_{jj} - D_{ii} )} \overline{\partial D_{ji} \over \partial H_{ij}}. 
    \end{equation}
    Employing Eq.~\eqref{eq.average_Dij}, $\partial D_{ji}/\partial H_{ij}\sim \overline{i(D_{ii} - D_{jj})/\gamma}$. For the random power-law hopping Hamiltonian, $\overline{|H_{ij}|^2} \sim |i-j|^{-2\alpha}$, we therefore have
    \begin{equation}
    	\begin{aligned}
    		\overline{-iH_{ij} D_{ji}(D_{jj} - D_{ii} )} \sim \frac{\mu_{ij} }{|i-j|^{2\alpha}} ,
    	\end{aligned}
    \end{equation}
    where $\mu_{ij} \sim \overline{(\langle n_j \rangle- \langle n_i \rangle)^2}/\gamma$ and $\langle n_i \rangle = D_{ii}$ stands for the local occupation number at site $i$.
    Averaging over all trajectories and disorder realizations, Eq.~\eqref{eq.DdD} is rewritten as
    \begin{equation}
    	\begin{aligned}
    		\overline{D_{ji}dD_{ij} } \approx \frac{\mu_{ij} }{|i-j|^{2\alpha}} -\gamma dt  (1-\delta_{ij})\overline{|D_{ij}|^2}.
    	\end{aligned}\label{eq.average_DdD}
    \end{equation}
    
    After averaging over disorder and trajectories, the second term of Eq.~\eqref{eq.dCij} becomes equal to Eq.~\eqref{eq.average_DdD}, $\overline{dD_{ji}D_{ij} } = \overline{D_{ji}dD_{ij} }$. 
    In the last term of Eq.~\eqref{eq.dCij}, we only keep the first order $dt$ terms and neglect higher orders, which leads to  
    \begin{equation}
    	\begin{aligned}
    		dD_{ji}dD_{ij} = &|D_{ij}|^2(dW_i+dW_j)^2 + 4\sum_{k}|D_{ik}|^2|D_{kj}|^2 dW_k^2  \\
    		&- 2 D_{ji}(dW_i+dW_j) \sum_{k}D_{ik}D_{kj}dW_k \\
    		&- 2 D_{ij}(dW_i+dW_j) \sum_{k}D_{jk}D_{ki}dW_k .
    	\end{aligned}
    \end{equation}
    Averaging over all trajectories, the above equation can be simplified as
    \begin{equation}
    	\begin{aligned}
    		\overline{ dD_{ji}dD_{ij} }= &2\gamma dt (1+\delta_{ij})\overline{ |D_{ij}|^2} + 4\gamma dt\overline{ \sum_{k}|D_{ik}|^2|D_{kj}|^2 } \\
    		&- 4 \gamma dt \overline{  |D_{ij}|^2 (D_{ii} + D_{jj}) } .
    	\end{aligned}\label{eq.average_dDdD}
    \end{equation}
    
    Substituting Eq.~\eqref{eq.average_dDdD} and \eqref{eq.average_DdD} into Eq.~\eqref{eq.dCij}, we obtain the equation governing the dynamic of the  correlation function $C_{ij}$ averaged over trajectories and disorder realizations,
    \begin{equation}
    	\begin{aligned}
    		{dC_{ij} \over dt} ={ d|D_{ij}|^2\over dt} =  &\frac{\mu_{ij}}{|i-j|^{2\alpha}} + 4 \gamma\sum_{k\neq i,j}|D_{ik}|^2|D_{kj}|^2 \\
    		& - 4 \gamma |D_{ij}|^2 (D_{ii} + D_{jj}), \qquad i\neq j. 
    	\end{aligned} \label{eq.dCij_dt}
    \end{equation}
    In order to find the decay of the correlation function $C_{ij}$, we follow Ref. \cite{Chen2020a,Zhang2022universal}, defining $f_n = \frac{1}{L}\sum_x (|D_{x,x+n}|^2+|D_{x,x-n}|^2)$ for $n\geq 1$. Note $\sum_{n\geq 1}f_n \approx \nu(1-\nu) = \frac{1}{4}$, where $\nu=1/2$ at half-filling.
    Therefore, the term $\sum_{k\neq i,j}|D_{ik}|^2|D_{kj}|^2$ is related to $f_n$ by
    \begin{equation}
    	\begin{aligned}
    		\sum_{k=1}^{\infty} f_k f_{n+k} + {1\over 2} \sum_{k=1}^{n-1} f_k f_{n-k} .
    	\end{aligned}
    \end{equation}
    Similarly, the term $|D_{ij}|^2 (D_{ii} + D_{jj})$ gives
    \begin{equation}
    	\begin{aligned}
    		f_n \sum_{k=1}^{\infty} f_k .
    	\end{aligned}
    \end{equation}
    
    Therefore, we can rewrite Eq.~\eqref{eq.dCij_dt} as
    \begin{equation}
    	\begin{aligned}
    		\frac{df_n}{dt} &\approx \frac{\mu}{n^{2\alpha}} + 4\gamma \left [\sum_{k=1}^{\infty} f_k f_{n+k} + {1\over 2} \sum_{k=1}^{n-1} f_k f_{n-k} -f_n \sum_{k=1}^{\infty} f_k \right ]\\
    		&= \frac{\mu}{n^{2\alpha}} + 2\gamma \sum_{k=1}^{\infty} f_k[ f_{n+k} +  f_{n-k} - 2f_n ] ,
    	\end{aligned}\label{eq.dfn_dt}
    \end{equation}
    where $\mu$ is a positive constant $ \sim {1\over \gamma L} \sum_j \left [\overline{(\langle n_j \rangle- \langle n_{j+n} \rangle)^2} + \overline{(\langle n_j \rangle- \langle n_{j-n} \rangle)^2}  \right]$ that comes from the spatial inhomogeneity of the local occupation number. The second term of the above equation is similar to Ref. \cite{Chen2020a,Zhang2022universal}, which contributes to the imaginary Brownian dynamics. When the system reaches the steady dynamical state, we can let $df_n/dt=0$. Considering the first term is power-law, one expect a power-law ansatz $f_n \sim n^{-z}$ will work. Letting $x=k/n$, 
    \begin{equation}
    	\begin{aligned}
    	&\sum_{k=1}^{\infty} f_k \left [f_{n+k} +  f_{n-k} - 2f_n \right ] \sim \\&n^{1-2z} \int_{0}^{\infty} dx [(x+x^2)^{-z} + |x-x^2|^{-z} - 2 x^{-z}] .
    	\end{aligned}
    \end{equation}
    The integral gives a constant, so to balance the terms in the right size of Eq.~\eqref{eq.dfn_dt}, we must have $n^{1-2z} = n^{-2\alpha}$, which lead to $z = \alpha + 1/2$. Since $f_n$ is effectively the average of correlation function over all sites, the decay of the correlation function should be
    \begin{equation}
    	C(r) \propto 1/r^{\alpha + 1/2}. \label{eq.Cr_vs_r} 
    \end{equation}
    In Fig.~\ref{fig.ar_vs_alpha}, we compared the above predictions $\alpha + 1/2$ with the extracted fitting parameters $a_r$ from $C(r) \sim r^{-a_r}$. From the above expression, we can compute the second cumulant of the particle number \cite{poboiko2023}
    \begin{equation}
    	\begin{aligned}
    		C_{\ell_A}^{(2)} = \int_{0}^{\ell_A} dx \int_{\ell_A+1}^{L} dy C(x-y) \propto \ell_A^{3/2-\alpha}.
    	\end{aligned}
    \end{equation}
	The scaling of the EE is then given by \cite{levitov2009},
	\begin{equation}
    	S(\ell_A = L/2)\sim {\pi^2\over 3}C_{\ell_A}^{(2)} \sim L^{3/2-\alpha} , \quad 1/2<\alpha<\alpha_c(\gamma).
	\end{equation}
	In the sub-volume phase, our fits in Fig.~\ref{fig.as_vs_alpha} are indeed well described by $a_s\simeq \frac{3}{2}-\alpha$.
	
	At the MIPT, the correlation function $C(r) \sim r^{-2}$ leading to \cite{poboiko2023}
	\begin{equation}
    	S(\ell_A = L/2)\sim {\pi^2\over 3}C_{\ell_A}^{(2)} \sim \ln L
	\end{equation}
	which agrees with the logarithmic scaling observed at $\alpha=\alpha_c(\gamma)$ in Fig.~\ref{Fig:SA_vs_L_monitoring}. For $\gamma \le 0.5$, $\alpha_c$ lies very close to $3/2$, while for stronger monitoring, $\gamma=2.0$, the transition is shifted to $\alpha_c \simeq 1.36$. In the region $\alpha>3/2$, the system crosses over to classical diffusion and the approximations leading to Eq.~(\ref{eq.Cr_vs_r}) no longer apply. The long-range tail is then irrelevant at long distances, and therefore the monitored dynamics is similar to that of monitored free fermions with short range hopping which results \cite{garcia2026} in the absence of a MIPT.	
    For $\alpha < 1$, the variance of the particle number grows with system size, so there is always a probability of classical jumps, connecting arbitrarily distant points of the sample, that would contribute, independently of the monitoring strength, to a power-law growth of the EE with the system size. This implies the absence of a MIPT for any monitoring strength because the area-laws phase is avoided.
	Finally, for $\alpha < 1/2$, there is divergence of even the first moment of particle number, which physically corresponds with a system \cite{mirlin00} resembling a dense random matrix where we expect that particles can reach distant points of the sample instantaneously, which suppresses any monitoring effect and will lead to the observed volume-law phase even in one dimension.

	\section{Conclusion and outlook}

	In this work, we have investigated numerically and, for some observables analytically, the entanglement dynamics of 1d non-interacting complex fermions with random power-law random hopping, modeled by the decay exponent $\alpha$, subjected to continuous monitoring of strength $\gamma$ via the QSD protocol. By computing the EE, MI and the density-density correlation functions and its singularity spectrum, that characterizes its multifractal features, over a wide range of system sizes, decay rates and monitoring strengths, we have constructed a detailed phase diagram that describes the entanglement properties, including the identification and characterization of MIPT, as a function of $\alpha$ and $\gamma$. Our numerical results agree with analytical arguments.  

	Our main finding is that the combination of long-range hopping and disorder leads to a rich entanglement dynamics that is qualitatively different from previous studies that only studied one of these two features. For $\alpha \leq 1$, and any monitoring strength, the EE scales with the system size with an exponent that range from a volume-law $\propto L$ for $\alpha \lesssim 1/2$, to a sub-volume-law growth $\propto L^{a_s}$ ($0< a_s <1$) for $\alpha > 1/2$ and sufficiently strong $\gamma$. This demonstrates that a sufficiently long-range hopping, a classical feature, dominates the entanglement dynamics preventing an area-law scaling, and therefore a MIPT, in this region. By contrast, for $\alpha > 3/2$ the dynamics is described by a process of normal diffusion and therefore \cite{garcia2026,liao2026} the system is in an area-law phase for any monitoring strength $\gamma$. We believe that small deviations in the numerical results from this prediction are caused by finite size effects in the weak monitoring limit. 
	In the intermediate regime $1 < \alpha \lesssim 3/2$, we identify a MIPT, at a certain $\alpha_c$, that depends on the monitoring strength, separating a sub-volume-law phase from the area-law phase. At $\alpha \approx \alpha_c$, the entanglement entropy scales logarithmically with system size, the MI $\mathcal{I}_2$ becomes scale-invariant, and the density-density correlation function decays as $C(r) \sim r^{-2}$.
	We have further characterized this critical region around the MIPT through the singularity spectrum  $f(\alpha_q)$ obtained from the density-density correlation functions. It displays a parabolic form  with $\alpha_0 > 2d$, typical of weak-multifractal features.

	Several directions remain open for future investigations. For instance, an analytical approach based on mapping the entanglement dynamics onto a NLSM \cite{mirlin1996,poboiko2023,poboiko2023a,fava2023}, the extension of the present results to other universality classes or the role of topology in the entanglement dynamics. Extending this analysis to interacting systems, still with long-range hopping, where the present approach based on the Gaussianity of the state is no longer applicable, is another promising avenue for future research. We plan to address some of them in the near future.

	\acknowledgments
 	We acknowledge support from the National Natural Science Foundation of China (NSFC): Individual Grant No. 12374138. B.F. acknowledges support from the Deutsche Forschungsgemeinschaft (DFG, German Research Foundation) – 553096561. We thank Yin Can for multiple interesting and illuminating discussions. Bo Fan thanks Igor Poboiko and Alexander D. Mirlin for valuable discussions and for pointing out that, in the weak measurement limit, the superdiffusive propagation $G(\omega, q) \sim 1/(q^{2\alpha -1} + \omega^2)$ , implies the second cumulant scales as $C^{(2)}_\ell \sim \ell^{(3/2 - \alpha)}$, in agreement with our findings.
 
\appendix

 \section{Scaling analysis} \label{app:scaling_analysis}
 
 In this appendix, we introduce the algorithm following Ref.~\cite{Skinner2019a} to extract $\alpha_c$ and $\nu$ from the scaling data collapse, and to estimate their statistical uncertainties. The algorithm objective is to minimize a cost function $R(\alpha,\nu)$ which essentially measures the deviation from a data collapse on a universal (unknown) curve. The corresponding optimal values are our estimates for $\alpha_c$ and $\nu$. First, the data are rescaled using the finite-size scaling ansatz $x= (\alpha -\alpha_c)L^{1\over \nu}$ and $y(x)=\mathcal{I}_2(\alpha,L)$ to produce a set of curves $x_i$ and $y(x_i,L)$ by a piece-wise linear interpolation, representing the rescaled $x$ coordinate and rescaled $y$ coordinate, which we wish to collapse on a single curve. Next, we sample from these curves at a discrete set of points $x_i$ and define the cost function:
 \begin{equation}
 	R = \sum_{i,L} [y(x_i,L) - \bar{y}(x_i)]^2 \label{eq.cost_function}
 \end{equation}
 where $\bar{y}(x_i)$ is the average of $y(x_i,L)$ over all system sizes $L$ at each grid point $x_i$. A perfect data collapse yields $R=0$. Minimizing $R$ over $(\alpha,\nu)$ gives our best estimates for $\alpha_c$ and $\nu$.
 
 \begin{figure}[!htbp]
 	\begin{center}
 		\includegraphics[width=7cm]{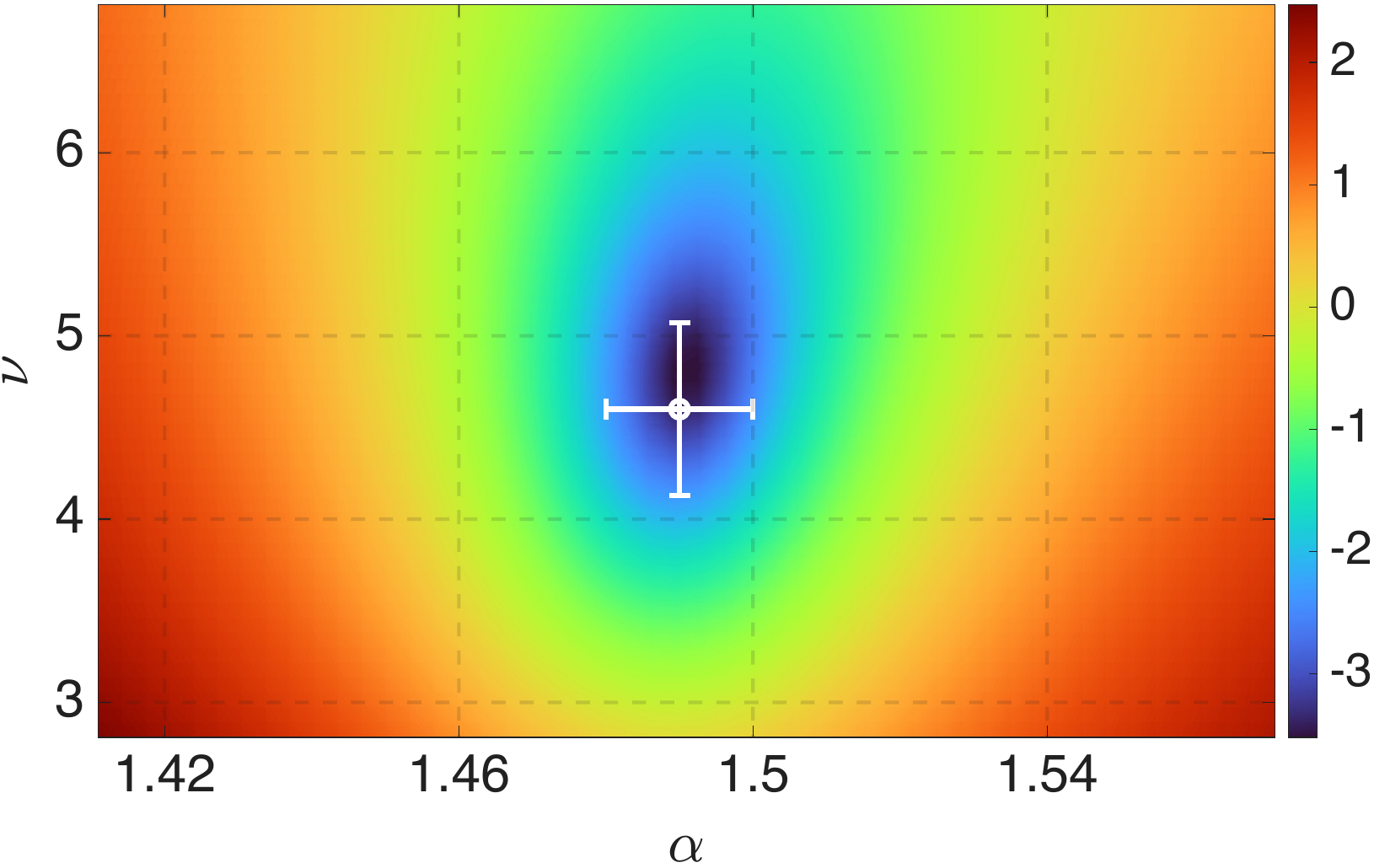} 
 		\caption{ Heatmap of the cost function $\ln{R(\alpha,\nu)}$ Eq.~\eqref{eq.cost_function}. The white circle and the errorbar correspond to the $\nu$ and $\alpha_c$ and the corresponding error in Fig.~\ref{Fig:I2_vs_alpha_QSD}. The corresponding measurement strength is $\gamma = 0.5$.
 		}\label{Fig:cost_function}
 	\end{center}
 	\vspace{-0.5cm}
 \end{figure}
 
 To propagate the statistical errors in the original data into uncertainties on $\alpha_c$ and $\nu$, we then employ the Monte Carlo samplings. We generate $1\times10^4$ synthetic data sets by adding independent Gaussian noise to each $\mathcal{I}_2(\alpha,L)$, with variance equal to the corresponding variance of the original data. For each synthetic data set, we repeat the previous procedures to find the minimum of cost function $R(\alpha,\nu)$, yielding a distribution of estimates $(\alpha_c(k),\nu(k))$, $k=1,2,\ldots, 1\times10^4$. We then obtain the errorbar or uncertainties of $\alpha_c$ and $\nu$ from the standard deviations of the corresponding data $\alpha_c(k)$ and $\nu(k)$.

\vspace{-0.3cm}
\section{Mutual information, density-density correlation function under monitoring and critical $\alpha$ for additional values of the monitoring strength $\gamma$}\label{app:mut}
 
This appendix presents supplementary numerical results supporting the findings of the main text including the validity of Eq.~(\ref{eq.Klich_Levitov}) for the approximation of the EE, the robustness of our results to changes in the time steps and the critical $\alpha$ for other values of $\gamma$. 

\begin{figure}[!htbp]
	\begin{center}
		\subfigure[]{ \label{fig.EE_vs_t}
			\includegraphics[width=3.9cm]{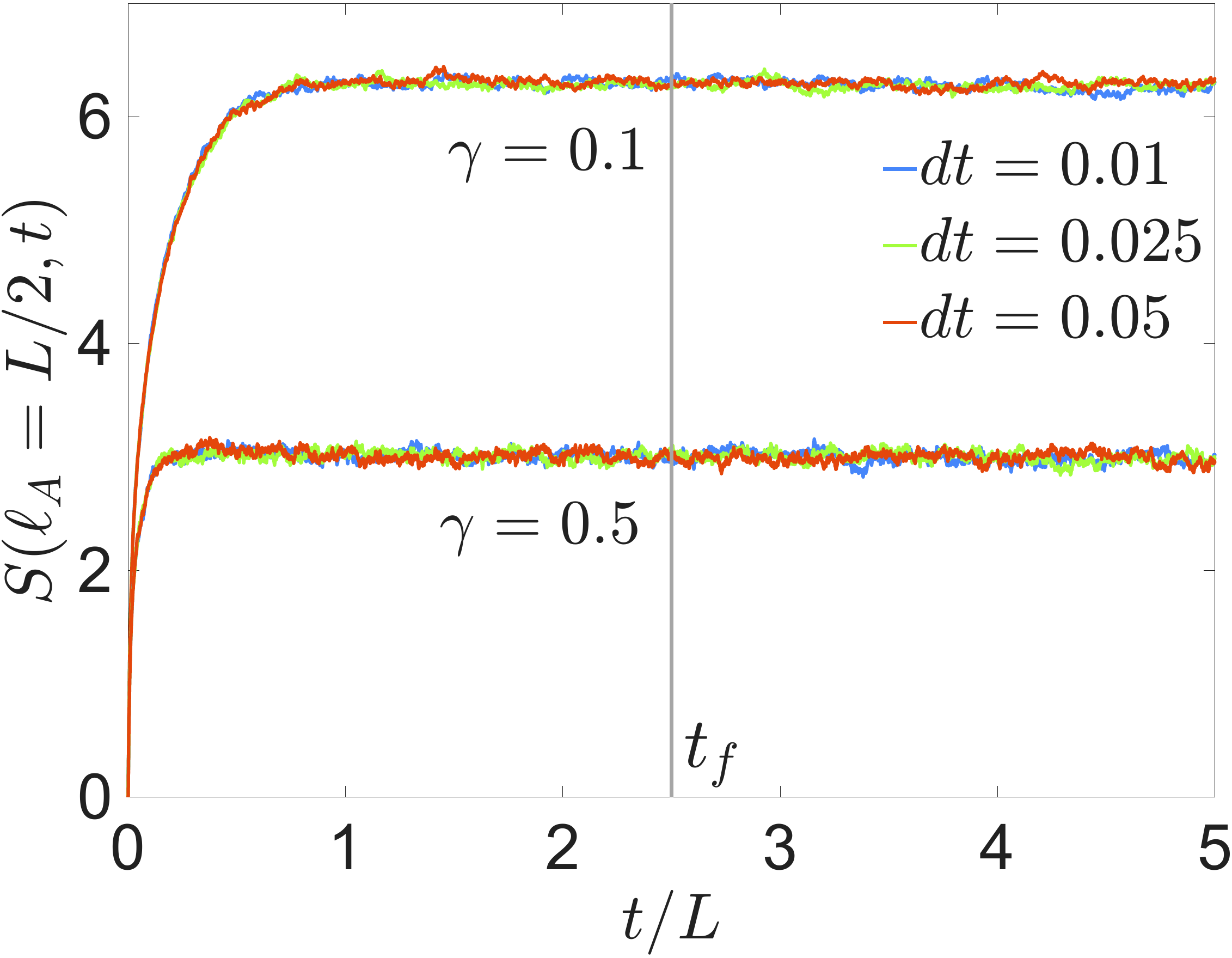} \hspace{0.1cm}
			\includegraphics[width=4.0cm]{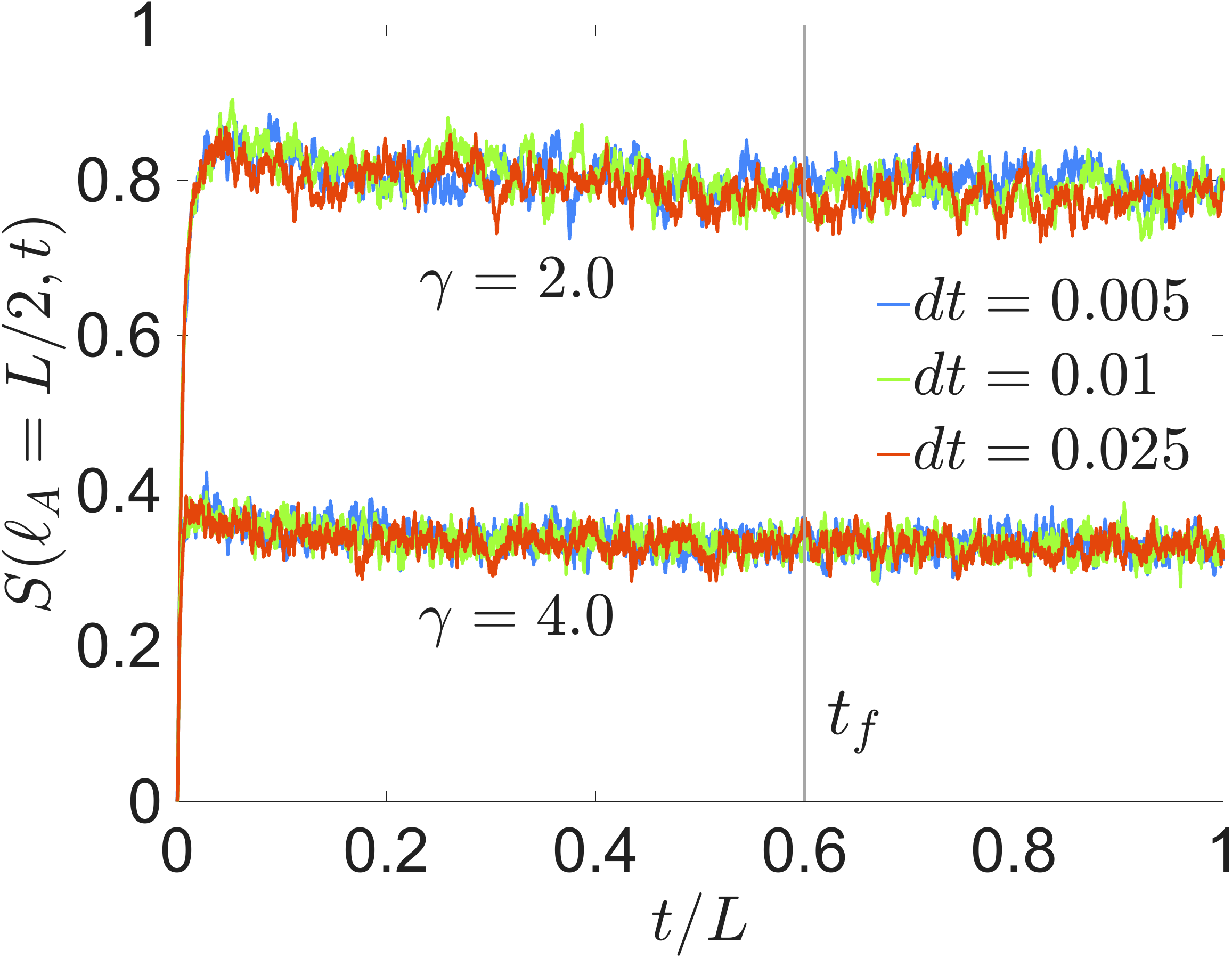} } \vspace{-0.4cm}
		\subfigure[]{ \label{fig.SA_vs_C2}
			\includegraphics[width=4.2cm]{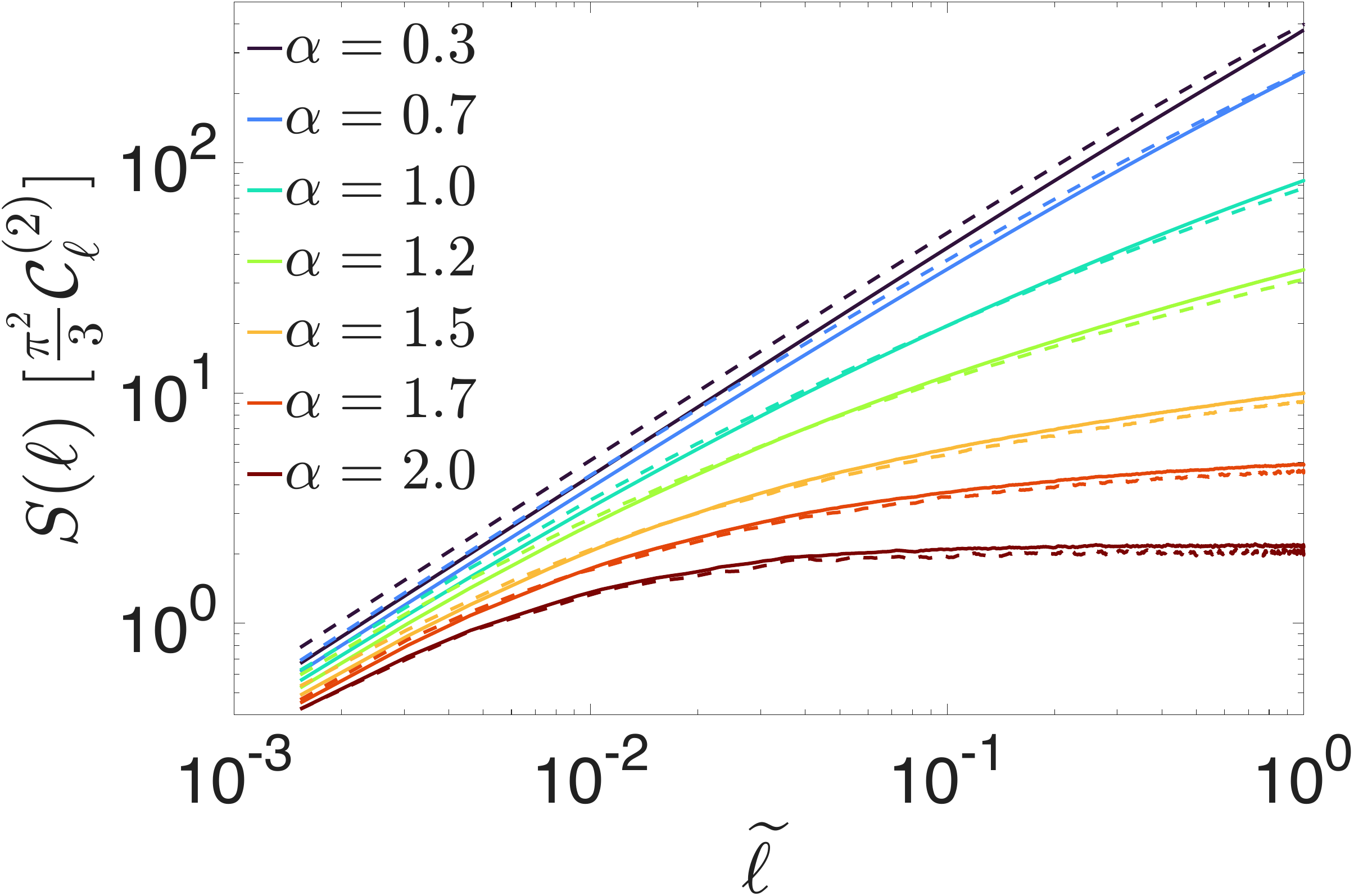} 
			\includegraphics[width=3.9cm]{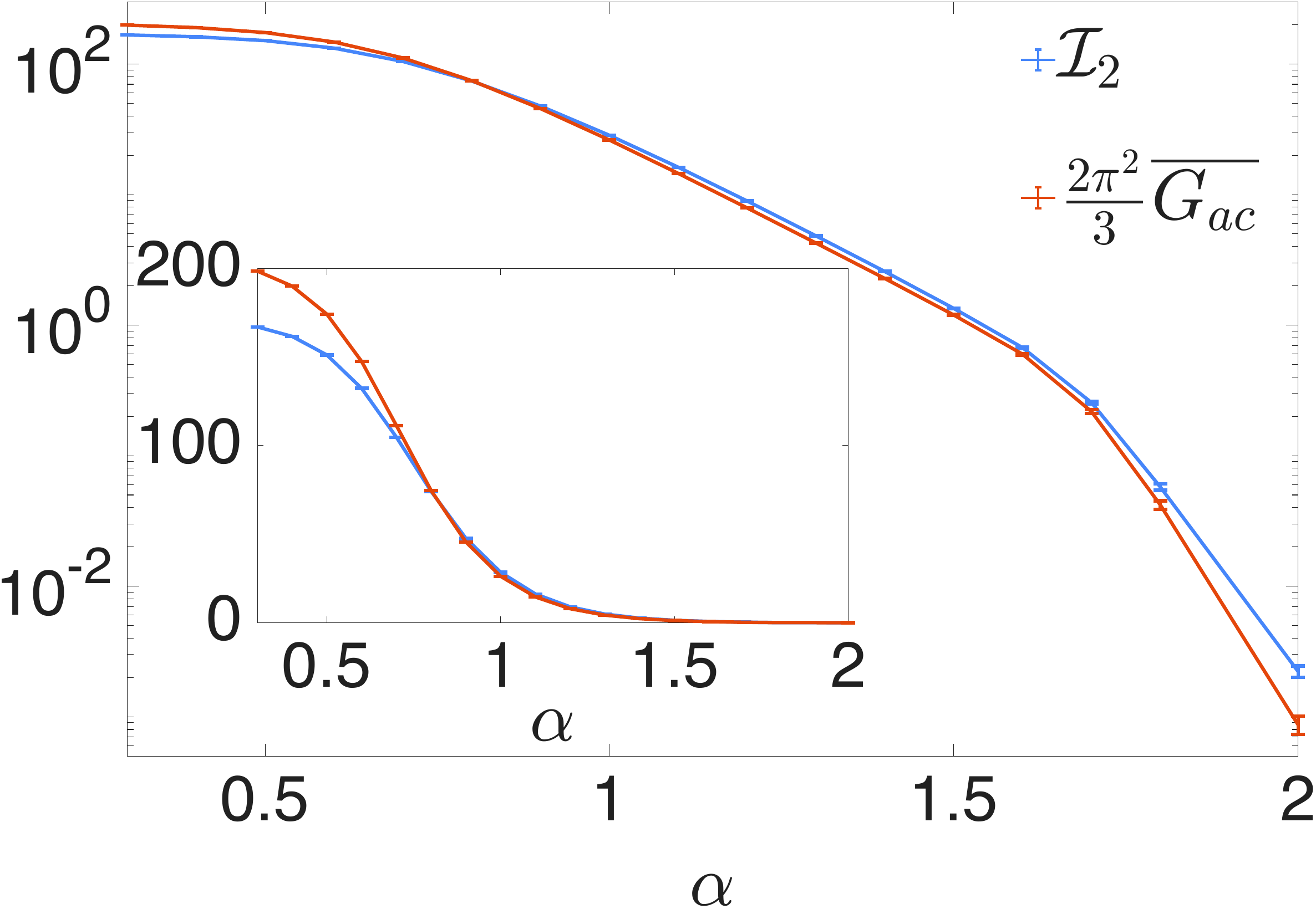} } 
		\caption{
			\subref{fig.EE_vs_t}. Dynamical evolution of the half-chain EE $S(\ell_A = L/2,t)$ for different values of $dt$, at $L = 128, \alpha = 1.8$, for weak monitoring $\gamma = 0.1, 0.5$ (left) and strong monitoring $\gamma = 2.0, 4.0$ (right). The $S(\ell_A = L/2,t)$ is independent of $dt$ for sufficiently small $dt$, confirming the numerical accuracy and convergence of the simulation.
			\subref{fig.SA_vs_C2}. Left: $S(\ell)$ and second cumulant of the particle number ${\pi^2\over 3}C^{(2)}_\ell$ as a function of subsystem size $\widetilde{\ell} = {L\over \pi}\sin{\pi\ell \over L}$ for several values of $\alpha$. Right: MI $\mathcal{I}_2$ and particle-number covariance ${2\pi^2\over3}\overline{G_{ac}}$ of the corresponding regions as a function of $\alpha$. Both panels are at $\gamma = 0.5$ and $L = 2048$. The agreement in both cases validates the theoretical relations Eq.\eqref{eq.Klich_Levitov} between entanglement and particle-number fluctuations discussed in the main text.
		}\label{Fig:EE_vs_t_C2}
	\end{center}
	\vspace{-0.5cm}
\end{figure}

Fig.~\ref{fig.EE_vs_t} shows the dynamical evolution of the half-chain EE $S(\ell_A = L/2, t)$ for several values of the time step $dt$ under weak ($\gamma = 0.1, 0.5$) and strong ($\gamma = 2.0, 4.0$) monitoring strength. The value of the EE is insensitive to the choice of $dt$ across the full range tested, confirming that our simulation is numerically convergent and that the time steps used in the main text introduce no significant systematic errors. Estimated from the numerical results in the system site $L = 128$, we use $dt = 0.05$ for $\gamma \le 0.5$, $dt=0.025$ for $ \gamma = 1$, $dt = 0.01$ for $2 \le \gamma \le 3$, $dt = 5\times10^{-3}$ for $\gamma = 4$, and $dt = 1\times10^{-3}$ for $\gamma=8$ throughout the paper. The saturation time $t_f$, when $S(\ell_A,t) \approx S(\ell_A,t_f)$ for all $t>t_f$, are also estimated from the smaller system size that depends significant on $\alpha$ and $\gamma$.

	Fig.~\ref{fig.SA_vs_C2} compares the EE $S(\ell_A)$ with the second cumulant of the particle number ${\pi^2\over 3}C^{(2)}_\ell$ as a function of subsystem size (left panel), and the MI $\mathcal{I}_2$  with the particle-number covariance ${2\pi^2\over 3}\overline{G_{ac}}$ of the corresponding regions as a function of $\alpha$ (right panel), both at $\gamma = 0.5$ and $L = 2048$. The close agreement between these quantities around the critical values of $\alpha$ confirms the theoretical prediction of Refs.~\cite{levitov2009,Burmistrov2017,poboiko2023,poboiko2023a}, which also means the validity of the close relation between $C(r)$ and the EE in our system.
	
\begin{figure*}[!htbp]
 	\begin{center}
 		\subfigure[]{ \label{fig.I2_vs_a_g0p1}
 			\includegraphics[width=5.2cm]{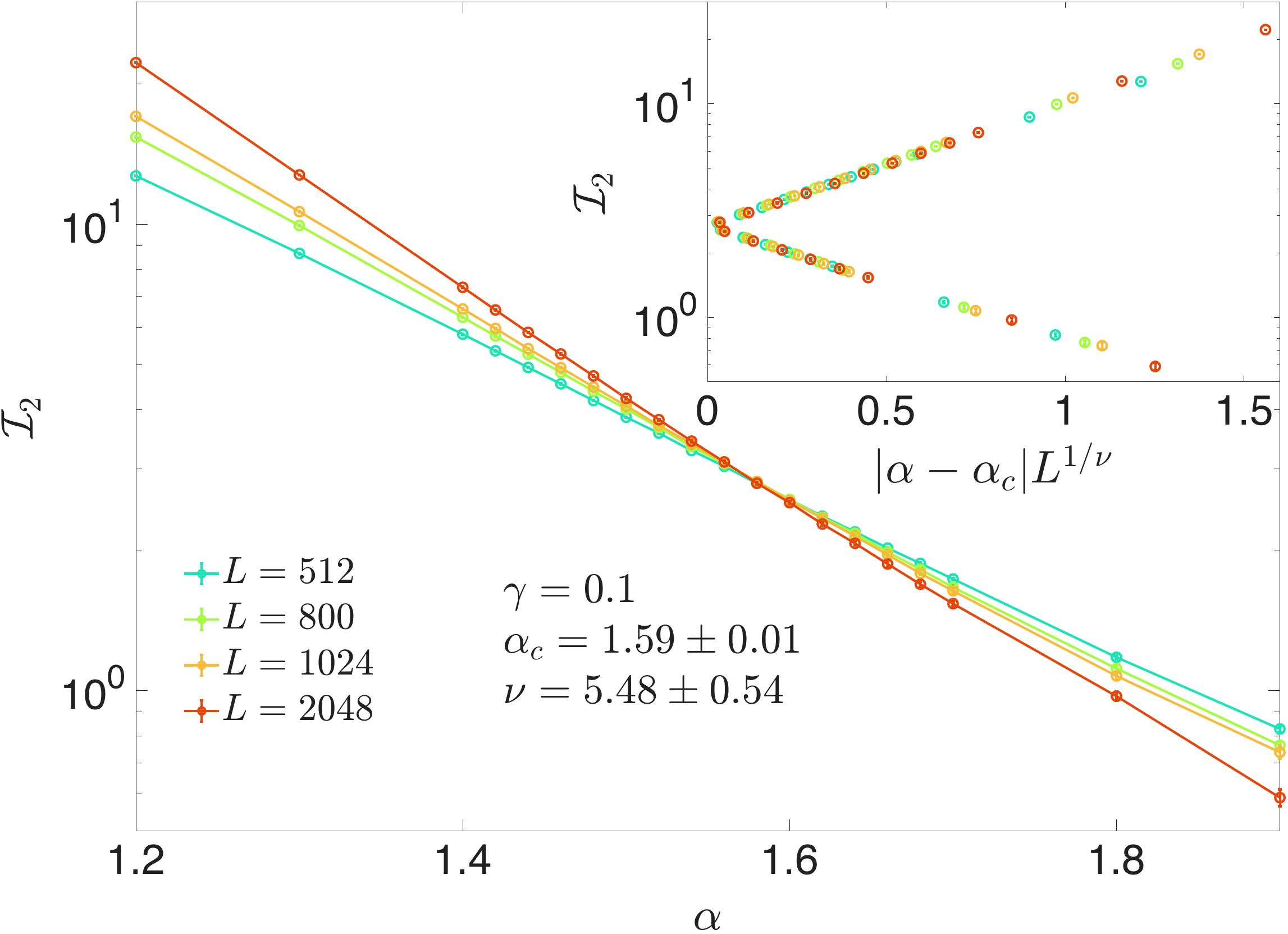} }
 		\subfigure[]{ \label{fig.I2_vs_a_g0p3}
 			\includegraphics[width=5.2cm]{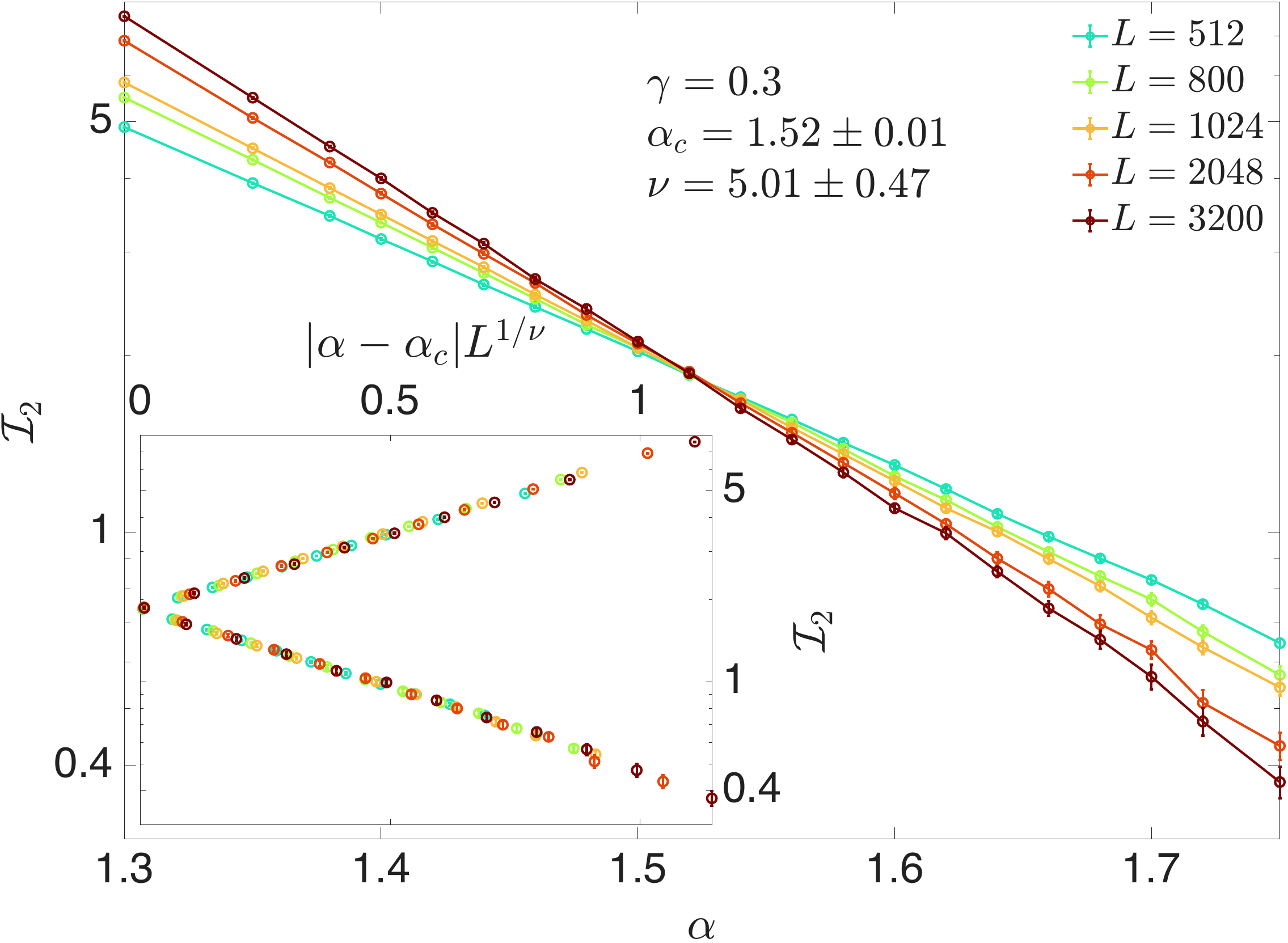} }
 		\subfigure[]{ \label{fig.I2_vs_a_g1p0}
 			\includegraphics[width=5.2cm]{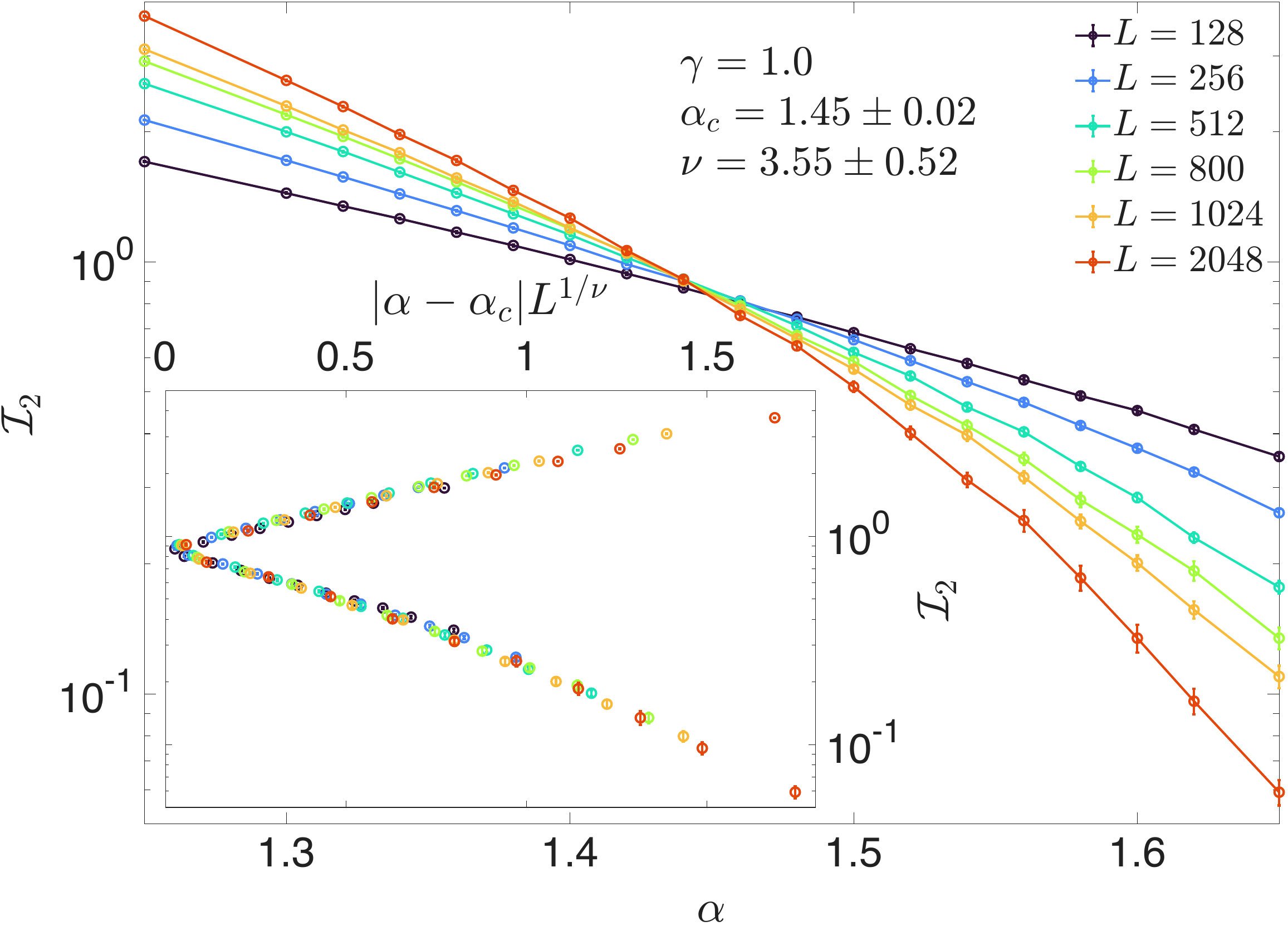} }
 		\subfigure[]{ \label{fig.I2_vs_a_g3p0}
 			\includegraphics[width=5.2cm]{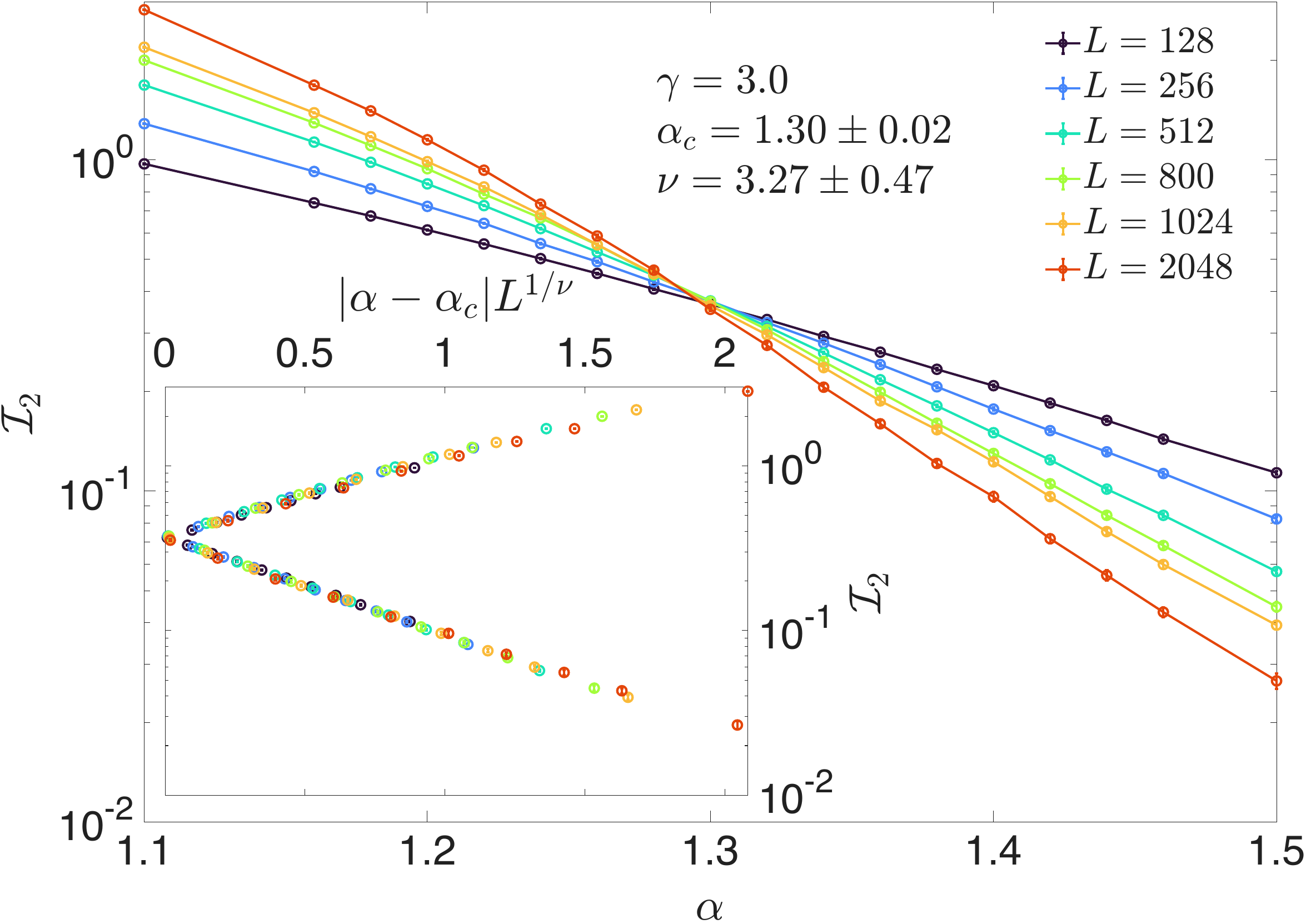} }
 		\subfigure[]{ \label{fig.I2_vs_a_g4p0}
 			\includegraphics[width=5.2cm]{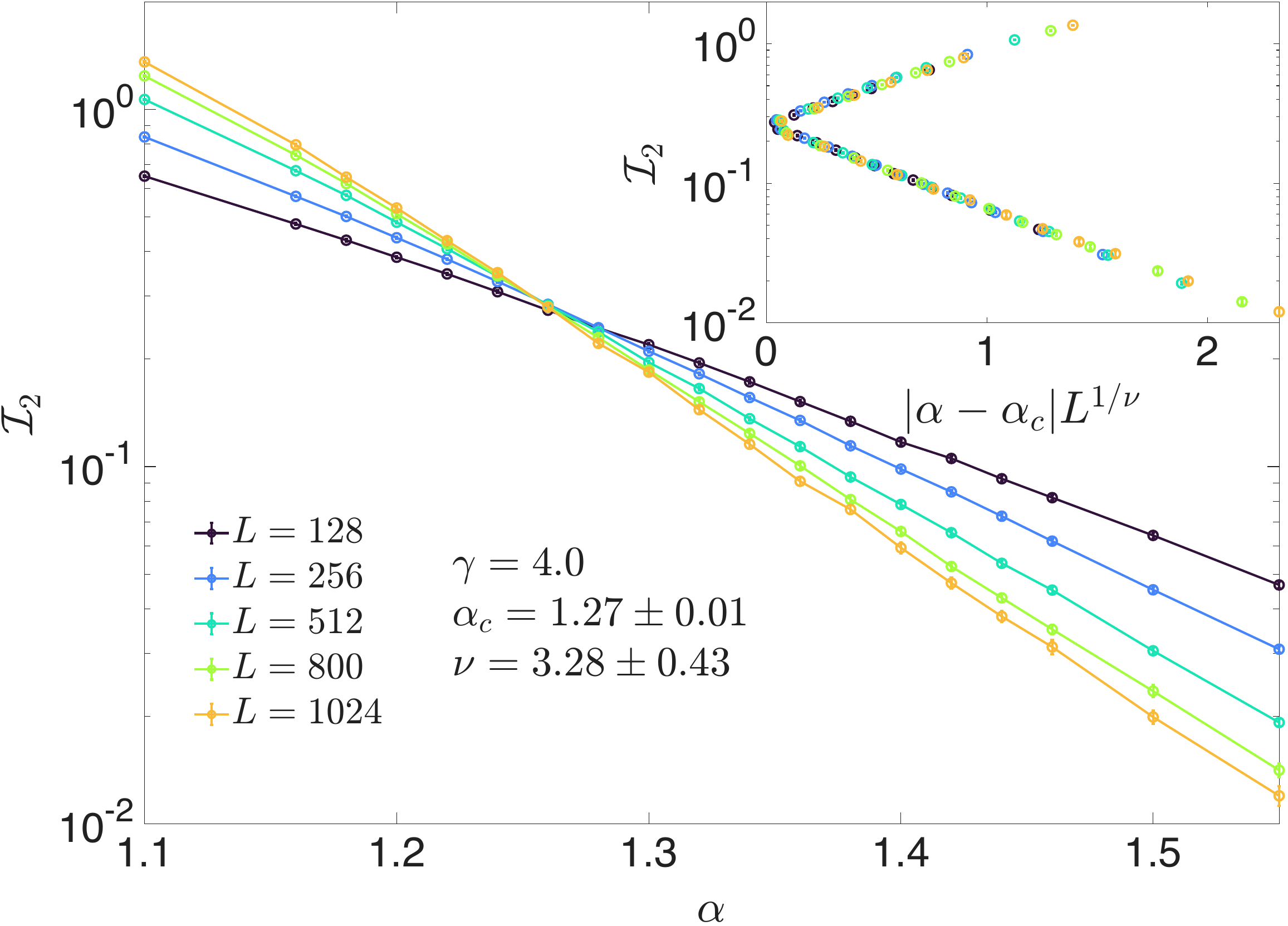} }
 		\subfigure[]{ \label{fig.I2_vs_a_g8p0}
 			\includegraphics[width=5.2cm]{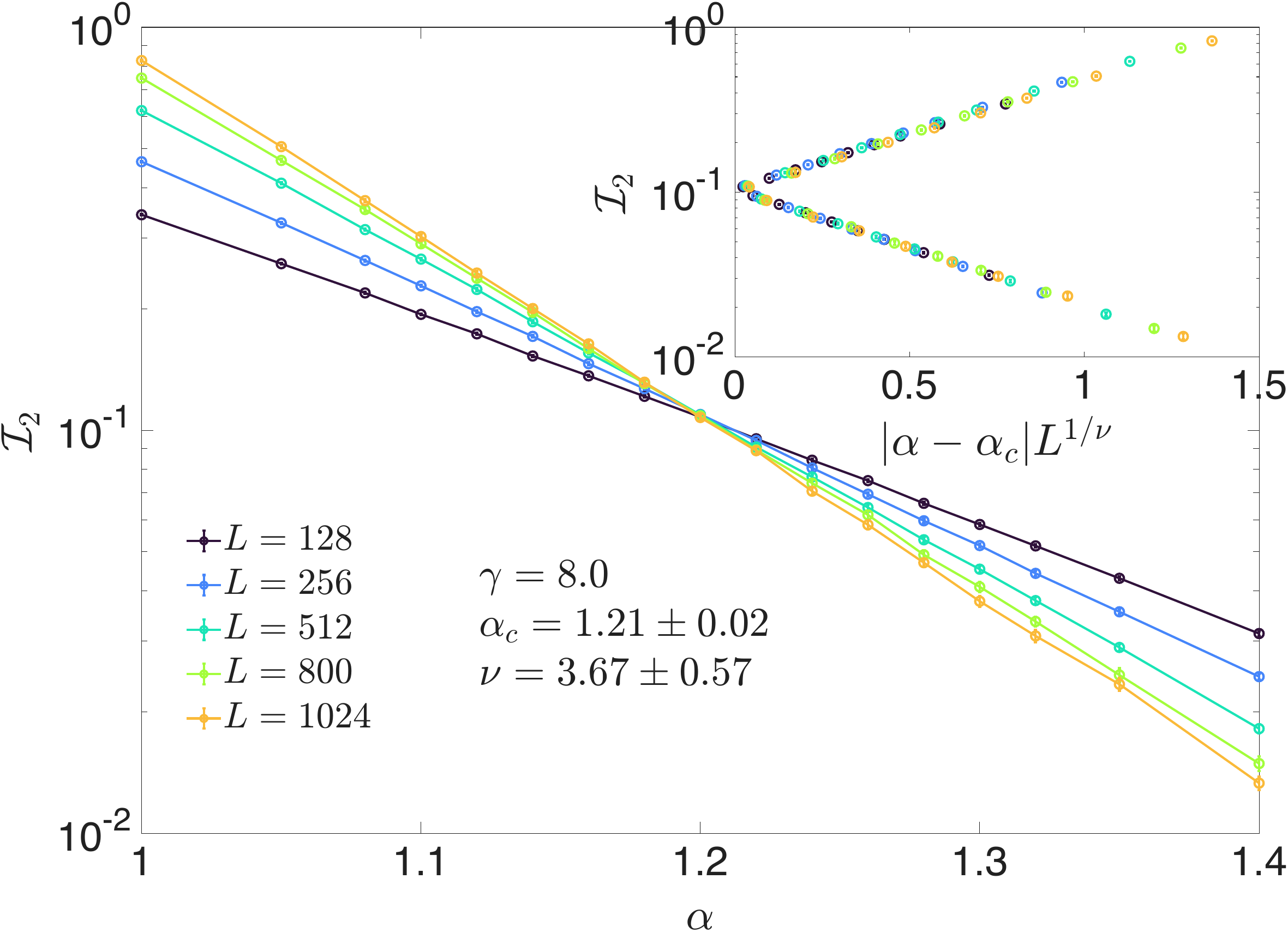} }
 		\caption{The mutual information $\mathcal{I}_2$ as a function of $\alpha$ for different system sizes $L$, and monitoring strengths ranging from weak ($\gamma = 0.1$) to strong ($\gamma = 8.0$). A sharp crossing at $\alpha = \alpha_c$ signals a MIPT, at a certain $\gamma > 0$, characterized by a scale-invariant $\mathcal{I}_2$. The inset shows the optimal data collapse around $\alpha_c$. While the obtained critical point $\alpha_c$ is robust with only small uncertainties, the critical exponent $\nu$ exhibits substantially larger fluctuations likely due to the absence of a well-defined critical exponent in this model.
 		}\label{Fig:I2_vs_alpha_QSD_app}
 	\end{center}
 	\vspace{-0.5cm}
 \end{figure*}

	\begin{figure*}[!htbp]
		\begin{center}
			\subfigure[]{ \label{fig.Cr_g0p1_full}
				\includegraphics[width=5.4cm]{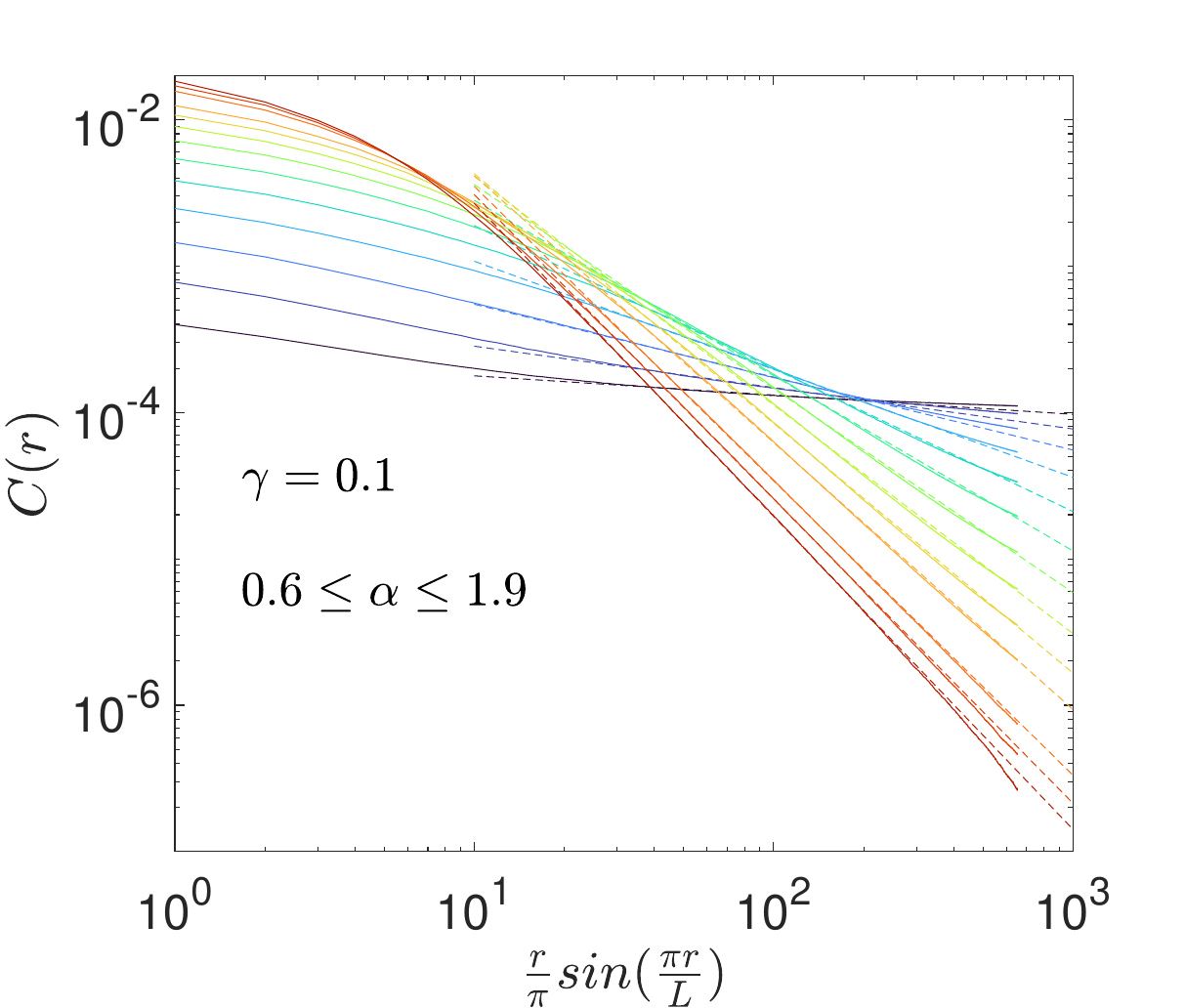} }\hspace{-0.2cm}
			\subfigure[]{ \label{fig.Cr_g0p5_full}
				\includegraphics[width=5.4cm]{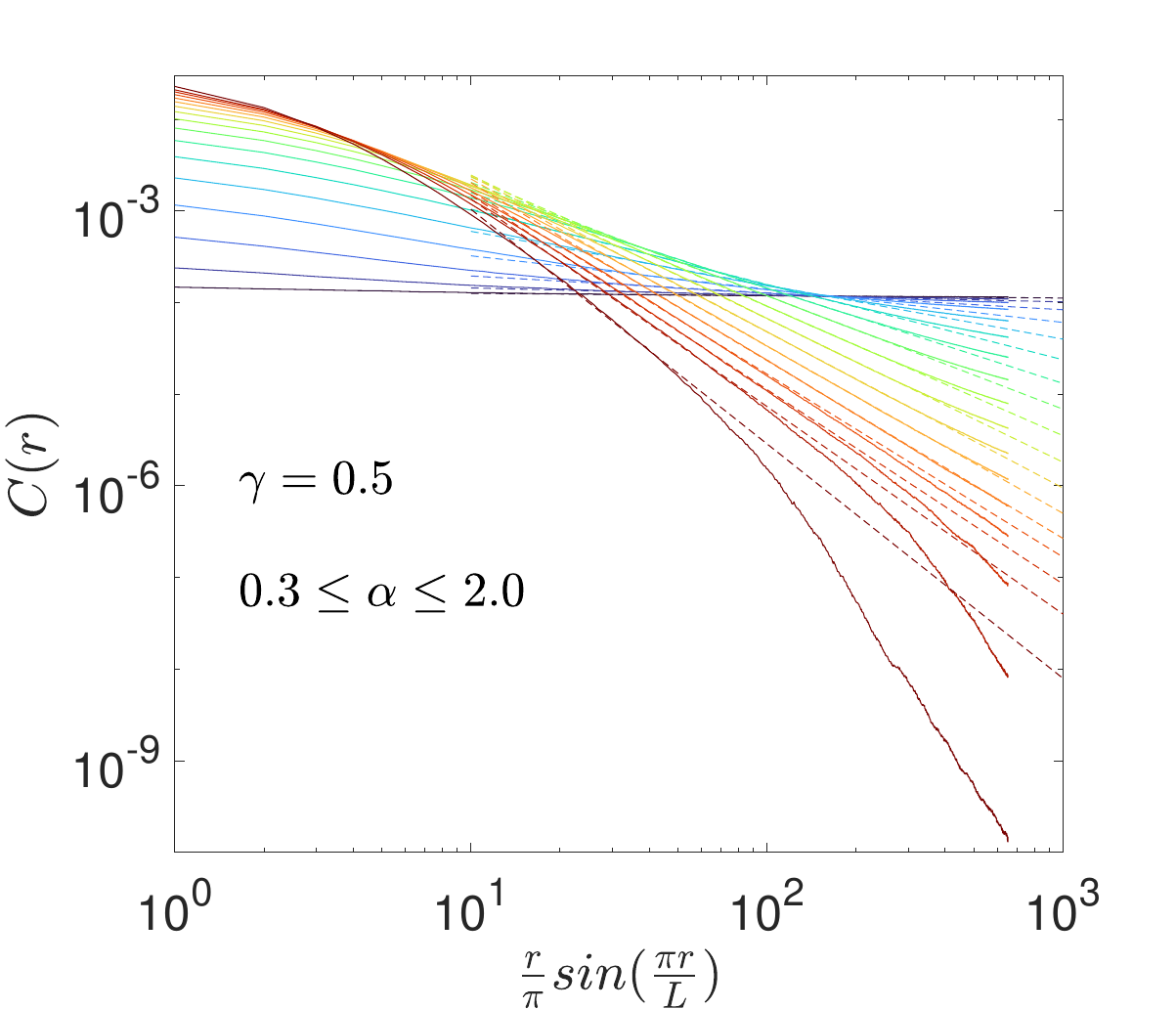} }\hspace{-0.2cm}
			\subfigure[]{ \label{fig.Cr_g2p0_full}
				\includegraphics[width=5.4cm]{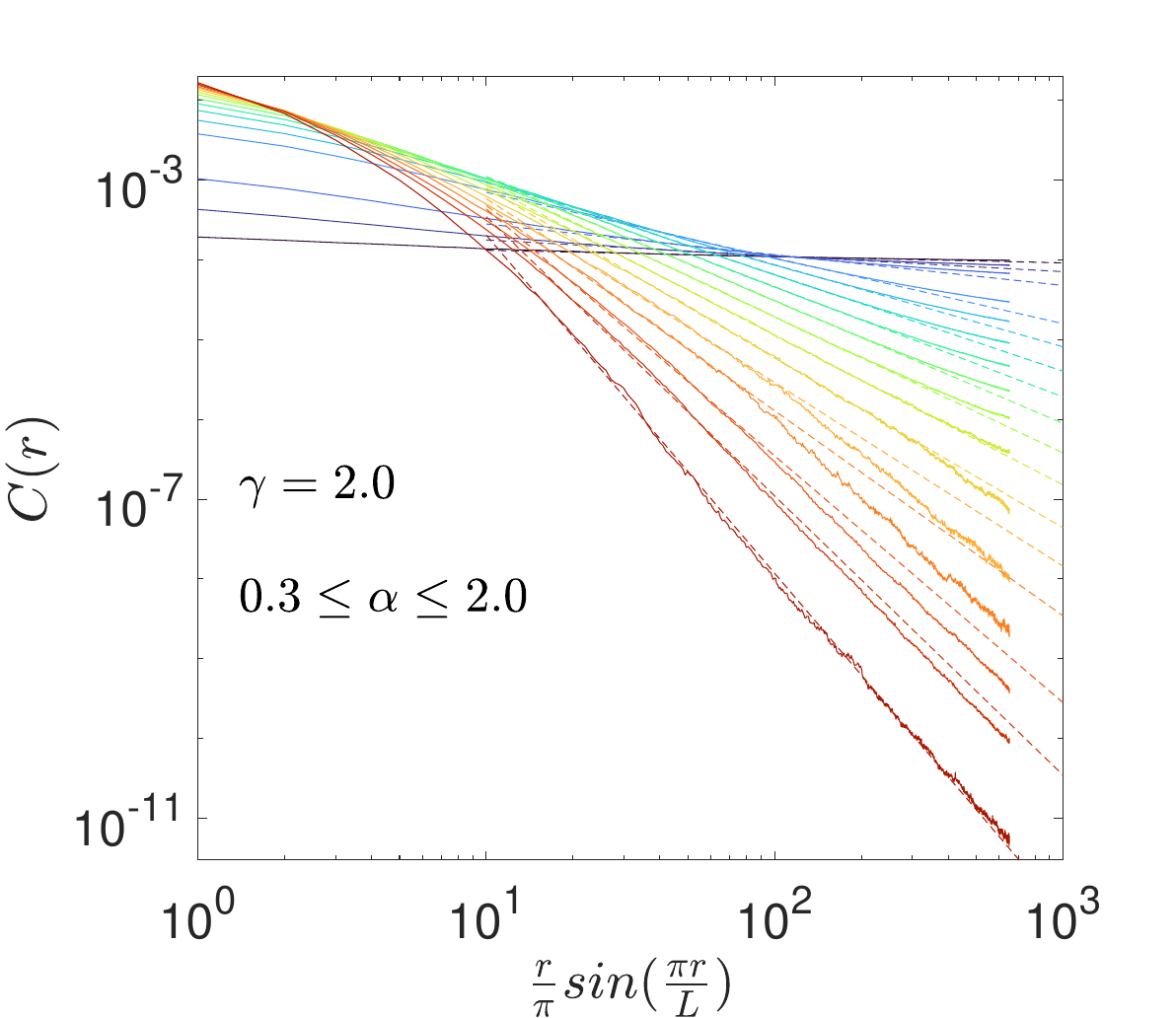} }
			\caption{
				Density-density correlation function $C(r)$ Eq.~(\ref{eq:cr}) as a function of the distance $r$ for different values of $\alpha$ and $\gamma$. The dashed lines show the corresponding power-law fits $C(r) \sim r^{-a_r}$. 
			}\label{Fig:Cr_full_app}
		\end{center}
		\vspace{-0.5cm}
	\end{figure*}
    
    Fig.~\ref{Fig:I2_vs_alpha_QSD_app} extends the finite-size analysis of MI to monitoring strengths ranging from $\gamma = 0.1$ to $\gamma = 8.0$. For all values of $\gamma$ examined, $\mathcal{I}_2$ exhibits a sharp crossing at $\alpha = \alpha_c$, signalling the MIPT, with scale-invariant behavior at criticality confirmed by the data collapse shown in the insets. As in the main text, $\alpha_c$ is extracted with high accuracy, while the critical exponent $\nu$ shows larger uncertainties due to the absence of a well-defined universal exponent in this model.

    Fig.~\ref{Fig:Cr_full_app} shows $C(r)$ over a broader set of $\alpha$ and $\gamma$ values than those in Fig.~\ref{Fig:Cr_QSD} of the main text, at fixed $L = 2048$. The power-law fits $C(r) \sim r^{-a_r}$ confirm that the decay exponent $a_r$ increases continuously with $\alpha$, with $a_r \approx 2$ at the critical point $\alpha \approx \alpha_c$, consistent with the theoretical prediction \cite{poboiko2023} and the logarithmic EE scaling at the MIPT.

	\vspace{-0.3cm}
	\section{Distribution of mutual information in the (sub-)volume-law and area-law phase} \label{app:pro_Iac}

    To complement the analysis of MI $\mathcal{I}_2$ presented in the main text, we examine here the full probability distribution $P(\ln \mathcal{I}_2)$ across the phase diagram. The shape and system-size dependence of the distribution carry information beyond the mean, in particular about the nature of sample-to-sample fluctuations, which are characteristic of each phase and provide independent evidence for the MIPT.

    \begin{figure*}[!htbp]
    	\begin{center}
    		\subfigure[]{ \label{fig.pro_Iac_gm0p5_a1p3}
    			\includegraphics[width=6.0cm]{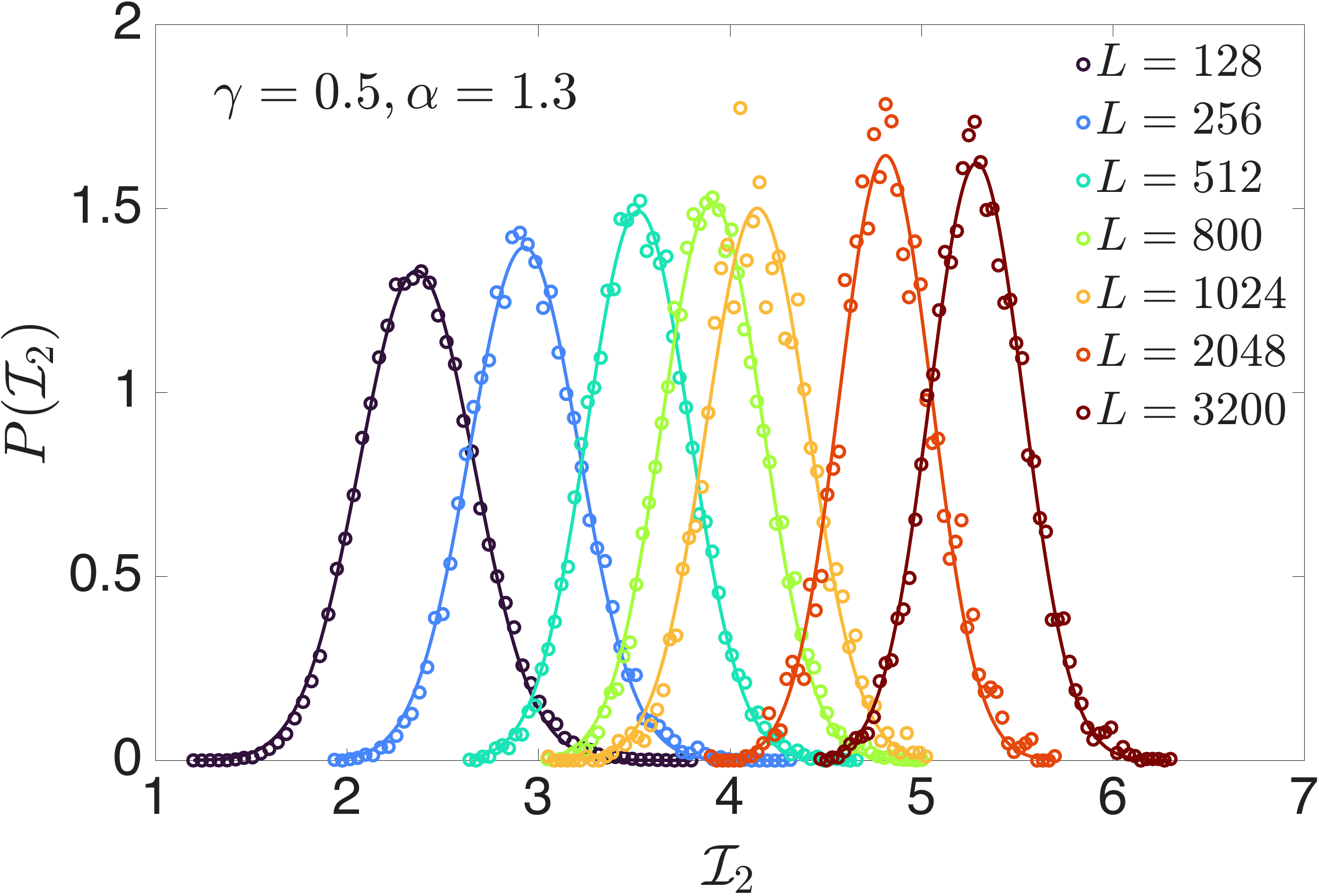} } \hspace{0.3cm}
            \subfigure[]{ \label{fig.pro_Iac_gm0p5_a1p7}
    			\includegraphics[width=6.0cm]{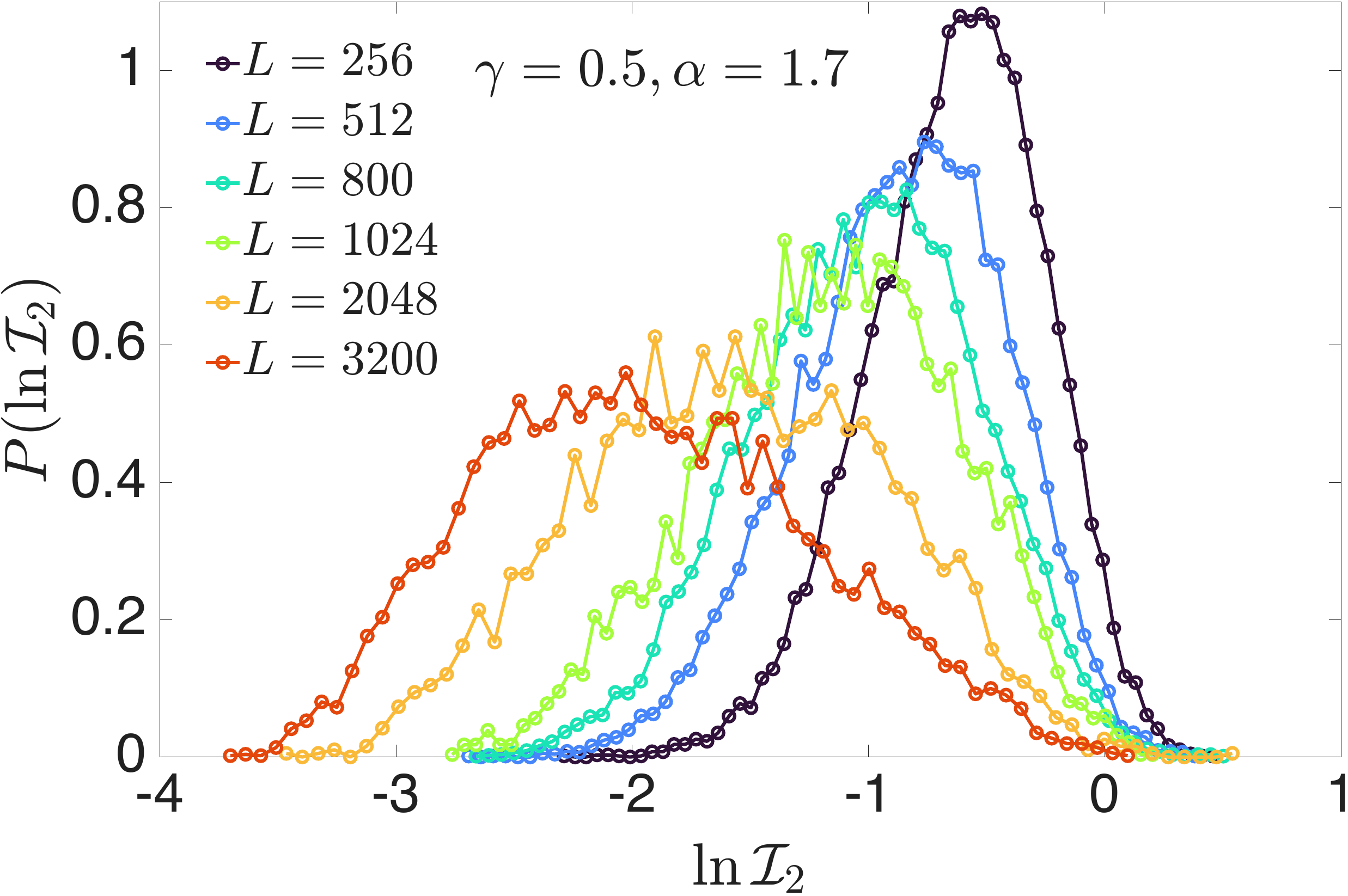} } 
    		\subfigure[]{ \label{fig.pro_Iac_gm2p0_a1p2}
    			\includegraphics[width=6.0cm]{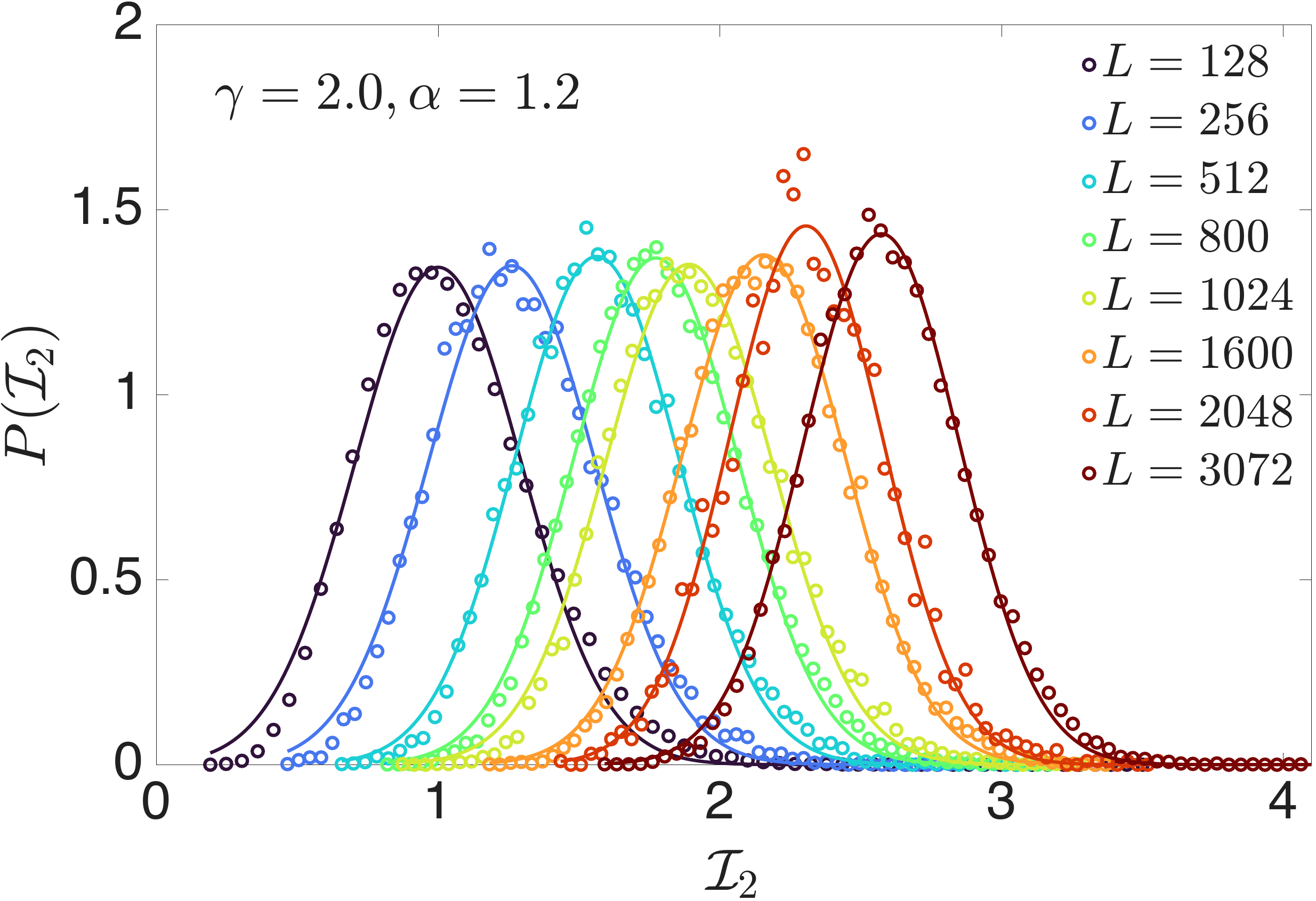} } \hspace{0.3cm}
            \subfigure[]{ \label{fig.pro_Iac_gm2p0_a1p7}
    			\includegraphics[width=6.0cm]{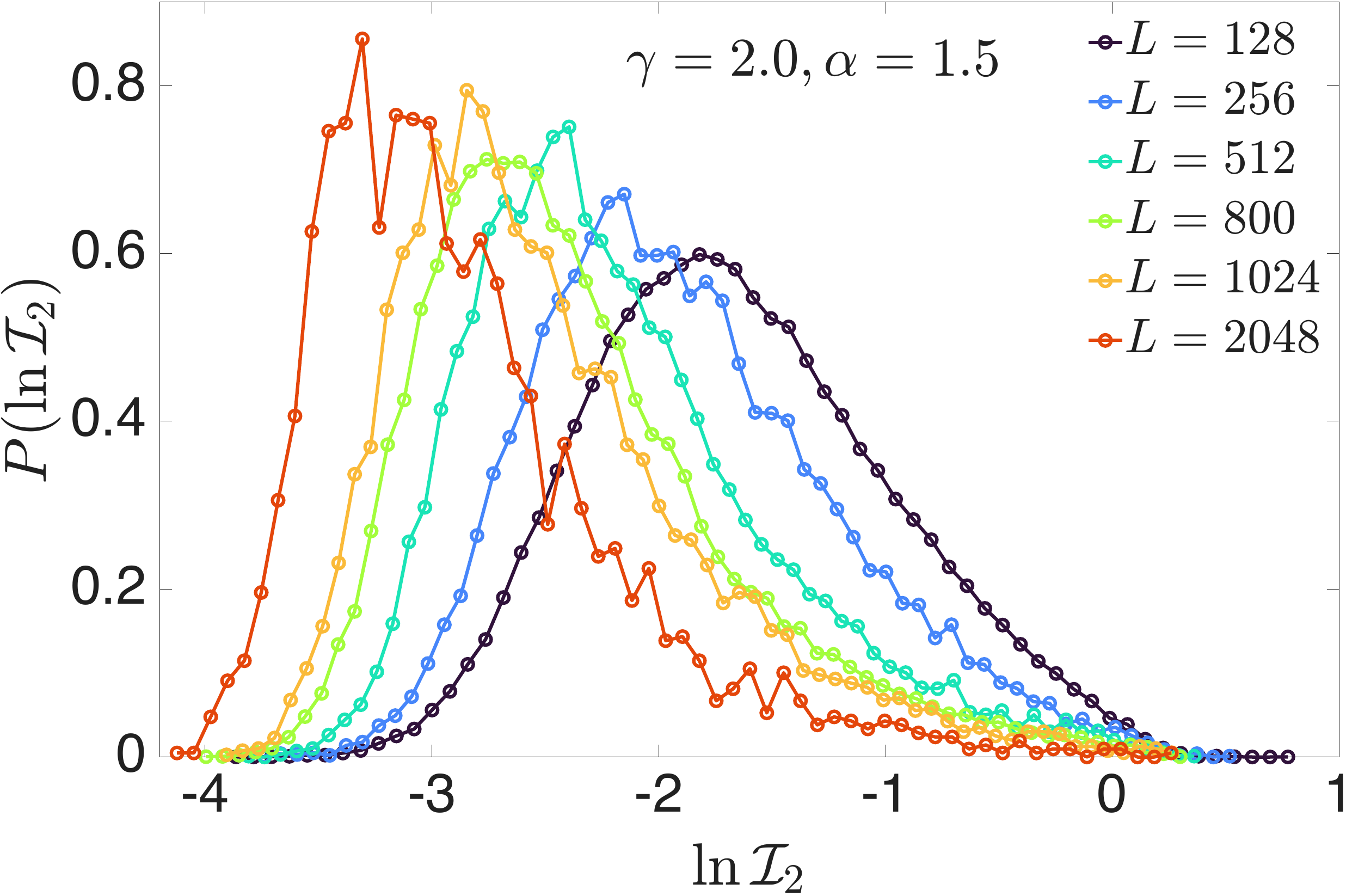} } 
    		\caption{Probability distribution $P(\mathcal{I}_2)$ ($P(\ln \mathcal{I}_2)$) for different system sizes $L$ in the (sub-)volume-law (area-law) phases. In the (sub-)volume-law phase, the distribution $P(\mathcal{I}_2)$ are well approximated by a Gaussian, whose mean grows with system size $L$. In the area-law phase, the peak of $P(\ln{\mathcal{I}_2})$ shifts towards $\ln \mathcal{I}_2 \to -\infty$ with increasing system size, reflecting the vanishing of long-range correlations. Both behaviors are in sharp contrast to the scale-invariant distribution found at criticality, shown in Fig.~\ref{Fig:pro_Iac_critical} of the main text.
    		}\label{Fig:Pro_I2_volume_area}
   	    \end{center}
   	    \vspace{-0.5cm}
    \end{figure*}
    
    As shown in Fig.~\ref{Fig:pro_Iac_critical} of the main text, at the critical point $\alpha = \alpha_c$ the distribution $P(\ln \mathcal{I}_2)$ becomes scale-invariant, i.e., independent of system size L, which is a direct signature of criticality. Fig.~\ref{Fig:Pro_I2_volume_area} shows the behavior away from criticality, in both the sub-volume-law and area-law phases. In the sub-volume-law phase, $P(\mathcal{I}_2)$ is well described by a Gaussian whose mean grows and whose width narrows with increasing $L$, consistent with an averaging MI that grows with system size in the thermodynamic limit. In the area-law phase, the peak of $P(\ln \mathcal{I}_2)$ shifts systematically toward $-\infty $ with increasing $L$, i.e., $\mathcal{I}_2 \to 0$ as $L \to \infty$, reflecting the suppression of long-range correlations. 
    
    \begin{figure*}[!htbp]
		\begin{center}
			\subfigure[]{ \label{fig.fit_q1_gm0p5_a1p5}
				\includegraphics[width=6cm]{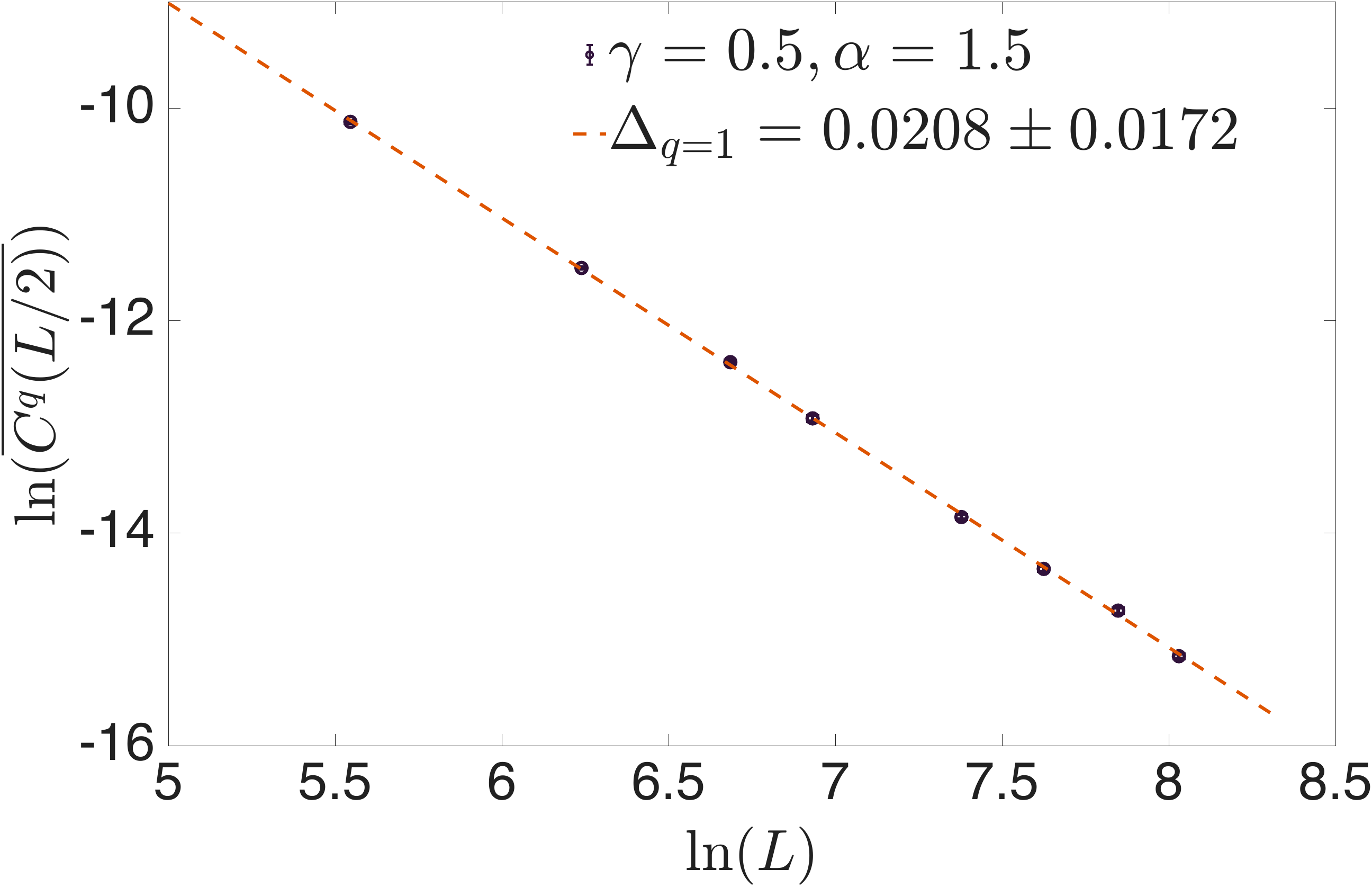}  \hspace{0.5cm}
				\includegraphics[width=6cm]{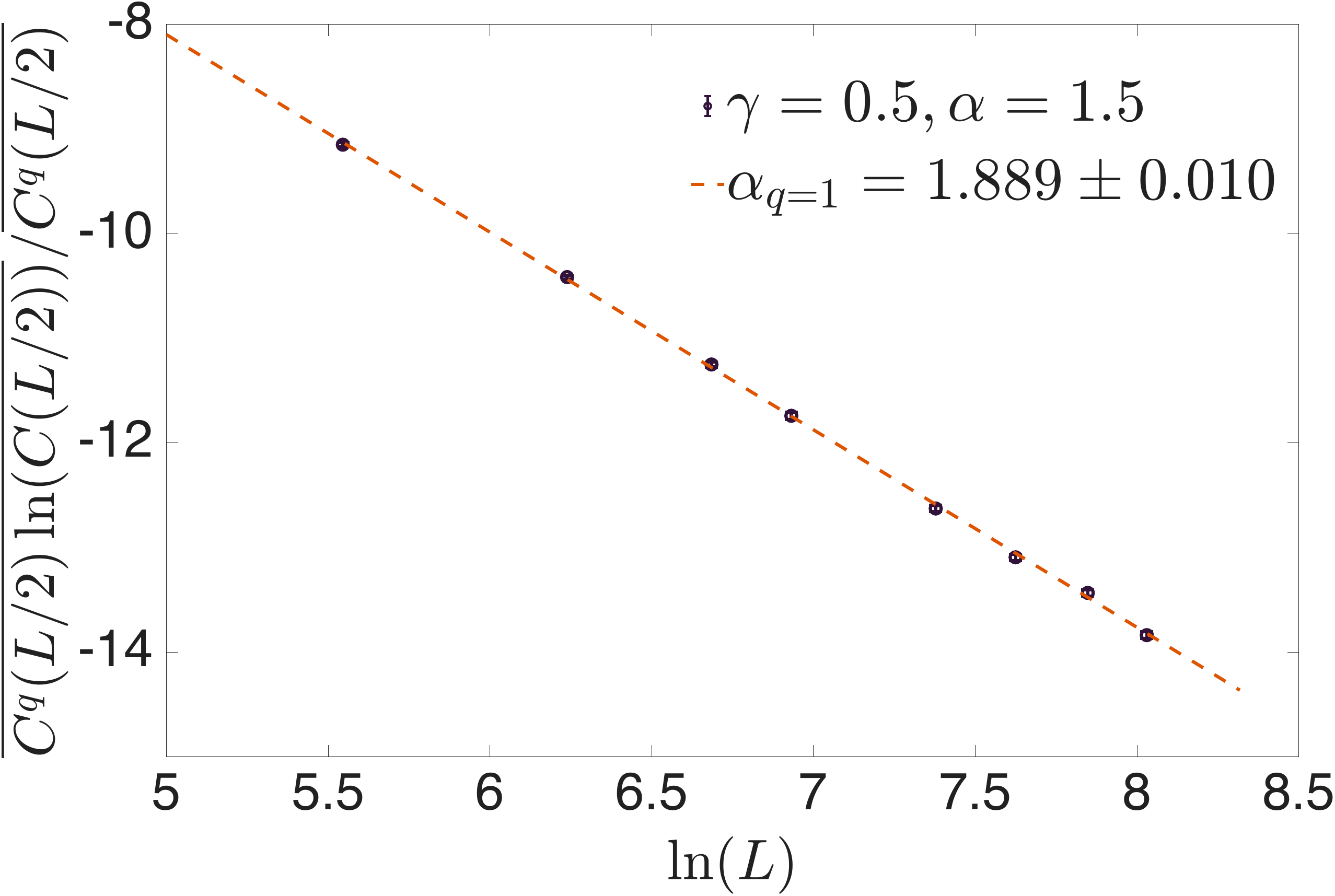} } 
            \subfigure[]{ \label{fig.fit_q1_gm2p0_a1p35}
				\includegraphics[width=6cm]{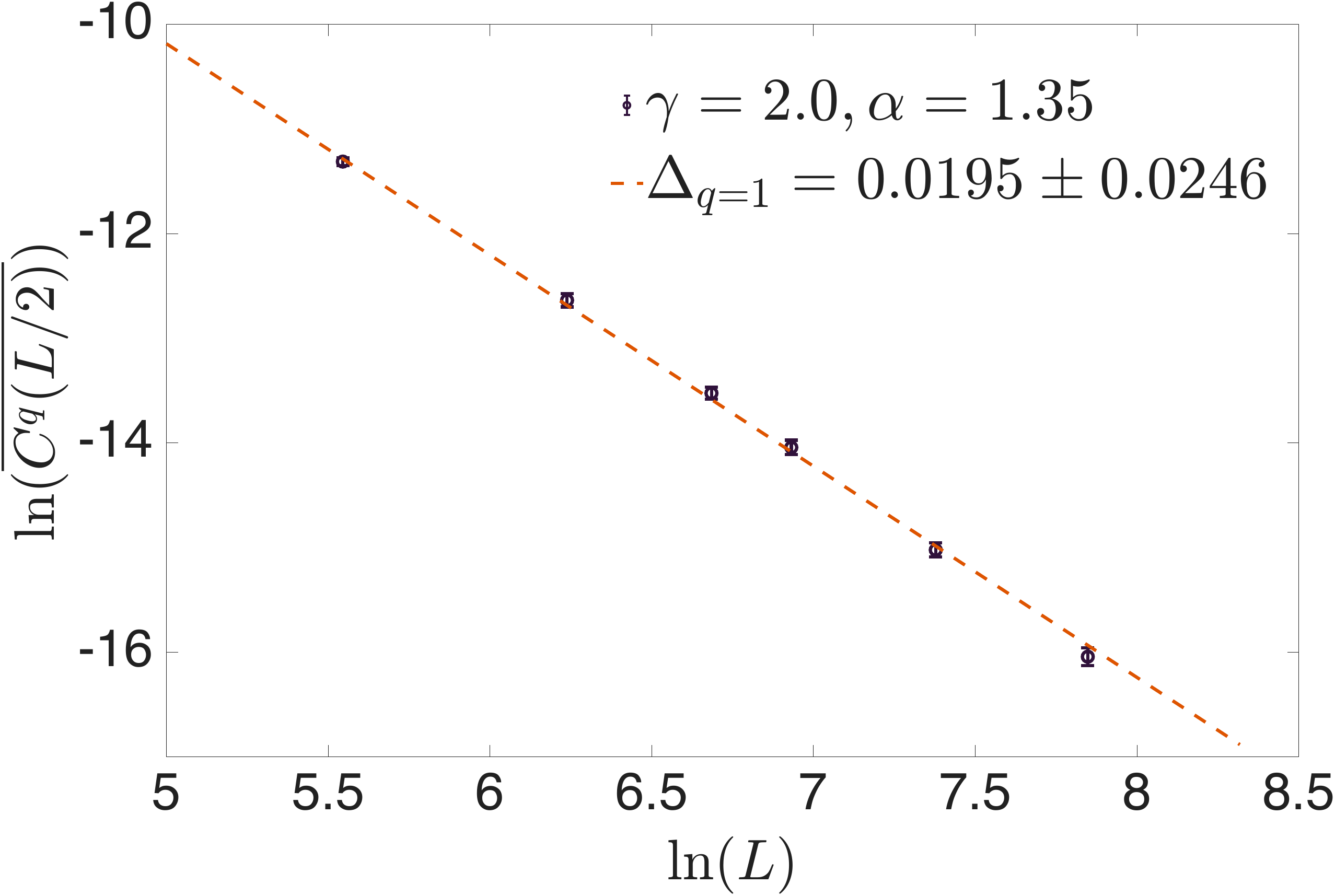} \hspace{0.5cm}
				\includegraphics[width=6cm]{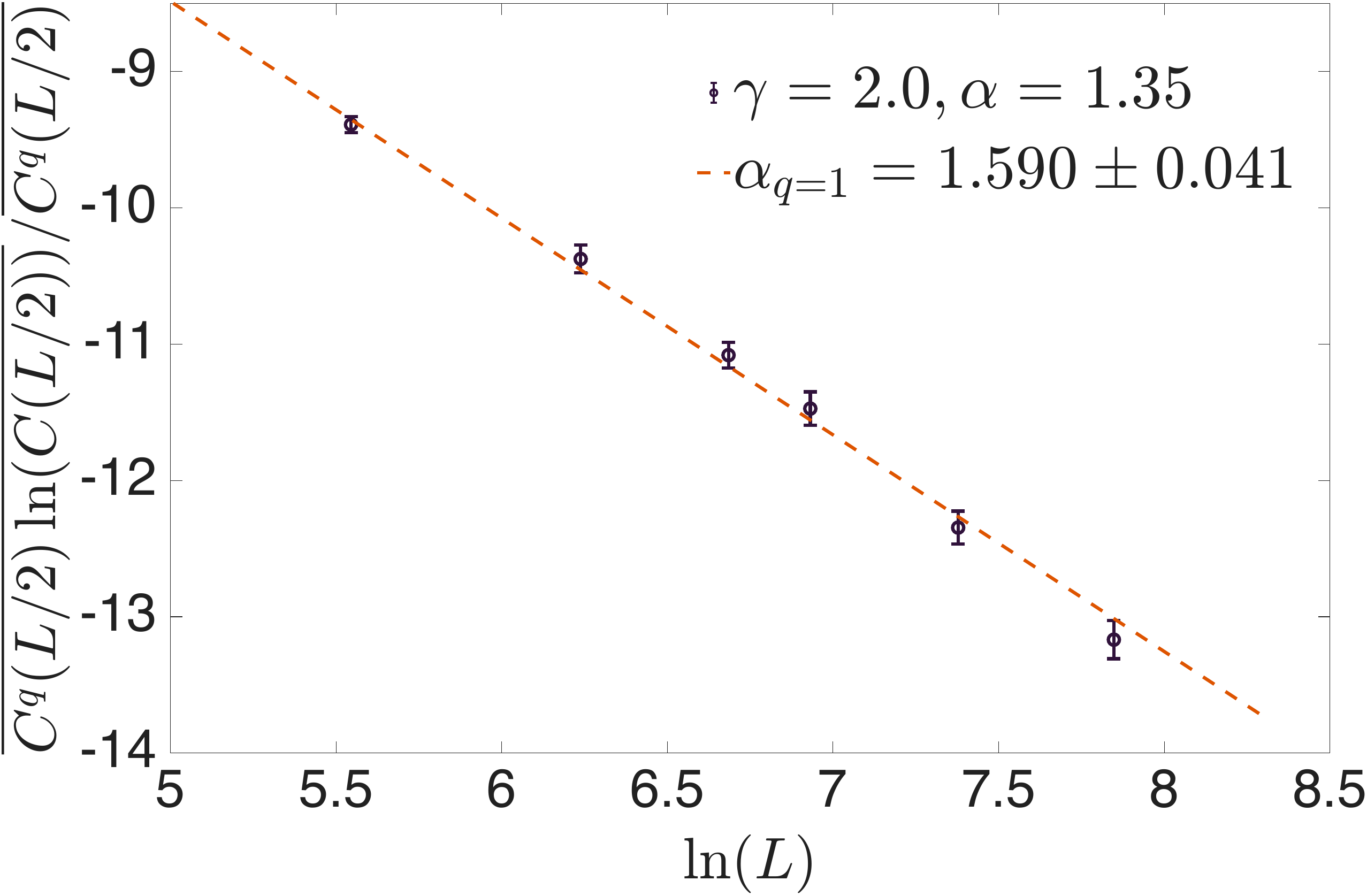} }
			\caption{
            Linear fits used to extract $\Delta_q$ (left panels, Eq.~\eqref{eq:delta_q_fit}) and $\alpha_q$ (right panels, Eq.~\eqref{eq:alpha_q_fit}) at $q = 1$ for the critical points. \subref{fig.fit_q1_gm0p5_a1p5} $\gamma = 0.5, \alpha_c = 1.5$. \subref{fig.fit_q1_gm2p0_a1p35} $\gamma = 2.0, \alpha_c = 1.35$. In both cases, the extracted $\Delta_{q=1} \approx 0$, consistent with the critical scaling $C(r) \sim r^{-2}$.
			}\label{Fig:fit_delta_q_alpha_q}
		\end{center}
		\vspace{-0.5cm}
	\end{figure*}

    \section{Multifractality of the correlation function $C(r)$ at criticality} \label{app:multifractal}
    
    The correlation function $C(r)$ can be determined directly from the correlation matrix via $C(r) = |D_{i+r,i}|^2$. To characterize its multifractal properties at the critical point, we follow the approach of Refs.~\cite{chhabra1989direct,poboiko2025} and analyze the scaling of the $q-$th moments of $C(L/2)$ averaged over disorder realizations and quantum trajectories, which is defined as $\overline{C^q(L/2)}$.
    
    At criticality, the moments are expected to obey the multifractal scaling ansatz \cite{chhabra1989direct,poboiko2025},
    \begin{equation}
        \overline{C^q(L/2)} \sim L^{-\tau(q)},
        \label{eq:scaling_ansatz}
    \end{equation}
    where $\tau(q) = q(d+1) + \Delta_q$. Therefore, we can obtain $\Delta_q = \tau_q - q(d+1)$ through the linear fitting 
    \begin{equation}
        \ln{\overline{C^q(L/2)}} = -\tau_q\ln{L} + a_q.
        \label{eq:delta_q_fit}
    \end{equation} 

    As a self-consistency check, the known critical scaling $C(r) \sim r^{-2}$ \cite{poboiko2023}, which has also been confirmed in the main text, implies $\overline{C(L/2)} \sim L^{-2d} = L^{-2}$ at $q=1$, and hence $\Delta_{q=1} = 0$. The fitting results shown in Fig.~\ref{Fig:fit_delta_q_alpha_q} confirm this expectation, validating the multifractal scaling ansatz.
    
    To extract the singularity spectrum $f(\alpha_q)$, we differentiate Eq.~\eqref{eq:delta_q_fit} with respect to $q$, which yields 
    \begin{equation}
        {\overline{C^q(L/2)\ln{C(L/2)}} \over \overline{C^q(L/2)} } = -\alpha_q \ln{L} - b_q
        \label{eq:alpha_q_fit}
    \end{equation}
    from which $\alpha_q$ is obtained by linear fitting. This quantity is related to the anomalous dimensions through
    \begin{equation}
        \Delta'_q \equiv \frac{d\Delta_q}{dq} = \alpha_q - (d+1).
    \end{equation}
    
    The singularity spectrum then follows from the Legendre transform of $\tau_q$:
    \begin{equation}
        f(\alpha_q) = q\alpha_q - \tau(q) = q\Delta'_q - \Delta_q.
    \end{equation}
    Applying those fittings independently for each value of $q$, we can then extract the corresponding singularity spectrum $f(\alpha_q)$, and $\alpha_q$, which is depicted, and commented in Fig.~\ref{Fig:f_alpha_q} of the main text.

	\bibliography{monitoring.bib} 
	
\end{document}